%% file: main.tex
\documentclass[compsoc, conference, a4paper, 10pt, times]{IEEEtran}

\usepackage{cite}
\usepackage{amsmath,amssymb,amsfonts,amsthm}
\usepackage{graphicx}
\usepackage{textcomp}
\usepackage{xcolor}
\usepackage{booktabs}

\usepackage{url}

\usepackage{breakurl}
\usepackage[breaklinks,hidelinks]{hyperref}
\usepackage{color, colortbl}
\usepackage{capt-of}
\usepackage{enumitem}
\newcommand{\hs}{\hspace{-2mm}}
\usepackage[makeroom]{cancel}
\newcommand{\st}{\scriptscriptstyle}

\usepackage{blindtext}
\usepackage{enumitem}
\usepackage{framed}
\usepackage{esvect}
\usepackage{tikz}
\usepackage{latexsym}
\usepackage{multirow}

\usepackage[mathscr]{eucal}
\usepackage{algpseudocode}

\newcommand{\size}{3.3pt}

\theoremstyle{remark}
\newtheorem{remark}{Remark}
\usepackage{arydshln}

\usepackage{algorithm}

\usepackage{scalerel}
\usepackage{listings}

\usepackage{array}
\usepackage{subcaption}
\usepackage{longtable}

\usepackage{tikz}
\usepackage{pgfplots}
\usepackage{pgfplots, pgfplotstable}

\usepackage{blkarray}
\usepackage{mathtools}
\usepackage{bm}
\usepackage{adjustbox}
\usepackage{blindtext}
\usepackage{multicol}

\definecolor{bubblegum}{rgb}{0.99, 0.76, 0.8}

\usepackage{makecell, cellspace, caption}

\definecolor{Gray}{gray}{0.9}

\usepackage[first=0,last=9]{lcg}

\usepackage{boxedminipage}

\usepackage{multirow}

\usepackage{adjustbox}
\usepackage{blindtext}
\usepackage{multicol}

\usepackage{preliminary}

\theoremstyle{theorem}
\newtheorem{theorem}{Theorem}
\newtheorem{lemma}{Lemma}

\theoremstyle{definition}
\newtheorem{definition}{Definition}
\newtheorem{claim}{Claim}
\theoremstyle{theorem}

\begin{document}

\title{Recurring Contingent Service Payment}


\author{\IEEEauthorblockN{Aydin Abadi}
\IEEEauthorblockA{\textit{University College London} \\
aydin.abadi@ucl.ac.uk}
\and
\IEEEauthorblockN{Steven J. Murdoch}
\IEEEauthorblockA{\textit{University College London} \\
s.murdoch@ucl.ac.uk}
\and
\IEEEauthorblockN{Thomas Zacharias}
\IEEEauthorblockA{\textit{University of Edinburgh} \\
thomas.zacharias@ed.ac.uk}
}

\maketitle

\begin{abstract}
Fair exchange protocols let two mutually distrustful parties exchange digital data in a way that neither party can cheat. They have various applications such as the exchange of digital items, or the exchange of digital coins and digital services between a buyer/client and seller/server. 

In this work, we formally define and propose a \emph{generic} blockchain-based construction called ``\emph{Recurring Contingent Service Payment}'' (RC-S-P). It (i)  lets a fair exchange of digital coins and verifiable service \emph{reoccur securely}  between clients and a server while ensuring that the server is paid if and only if it delivers a valid service, and (ii) ensures the parties' privacy is preserved. RC-S-P supports arbitrary verifiable services, such as ``Proofs of Retrievability'' (PoR) or verifiable computation and imposes low on-chain overheads. Our formal treatment and construction, for the first time, consider the setting where either client or server is malicious. 

We also present a concrete \emph{efficient} instantiation of RC-S-P when the verifiable service is PoR. We implemented the concrete instantiation and analysed its cost. When it deals with a $4$-GB outsourced file, a verifier can check a proof in only $90$ milliseconds, and a dispute between a prover and verifier is resolved in $0.1$ milliseconds.

At CCS 2017, two blockchain-based protocols were proposed to support the fair exchange of digital coins and a certain verifiable service; namely, PoR. In this work, we show that these protocols (i) are susceptible to a \emph{free-riding attack} which enables a client to receive the service without paying the server, and (ii) are not suitable for cases where parties' privacy matters, e.g., when the server's proof status or buyer's file size must remain private from the public. RC-S-P simultaneously mitigates the above attack and preserves the parties' privacy.

\end{abstract}

\input{content.tex}
\input{Conclusion}

\input{Acknowledgments}
\bibliographystyle{plain}

\bibliography{ref}

\appendices

%
\input{Survery-of-Related-work}

\input{notation-Table}

\input{appendix_preliminaries}

\input{proof-status-issue-further-discussion}

\input{VS-definition.tex}

\input{appendix_VSID}

\input{Recurring-Contingent-Service-Payments}

\input{trade-offs}

\input{SAP-discussion}

\input{correctness}

\input{appendix_RC-PoR-P}

\input{RC-PoR-P-without-Arbiter}


\input{RC-PoR-P-complexity-table}
\end{document}

%% file: content.tex
\input{Introduction}

\input{Related-work.tex}

\input{Preliminary}

\input{motivation}
\input{limitations-And-our-solution}

\input{RC-S-P-Definition}

\input{RC-S-P-overview}
\input{PoR-protocols}

\input{RC-PoR-P-Evaluation}

%% file: Introduction.tex
\vspace{-2mm}
\section{Introduction}\label{introduction}

Fair exchange is an interesting problem in which two mutually distrustful parties want to swap digital items such that neither party can cheat the other, in the sense that either each party gets the other's item, or neither party does.  It captures various real-world scenarios; for instance, when two parties want to exchange digital items or when a seller wants to sell a digital verifiable service in exchange for digital coins. Solutions to the problem are usually certain cryptographic schemes, called fair exchange protocols, and have been studied for decades. It has been shown that fairness is unachievable without the aid of a trusted third party \cite{Cleve86}. 

With the advent of decentralised cryptocurrencies and blockchain, it seemed fair exchange protocols can be designed without having to rely on a single trusted third party, in the sense that the third party’s role can be turned into a computer program, i.e., smart contract, which is maintained and executed by the decentralised blockchain.  This ultimately results in a stronger security guarantee, as there would be no need to trust a single entity, anymore. Ever since various fair exchange protocols that rely on blockchain have been proposed.  They mainly support the fair exchange of two digital items (e.g., documents, films) or a digital item and coins (except for the ones in \cite{CampanelliGGN17} that will be discussed shortly).



\noindent\textbf{Our Contributions.} In this work, we:
\begin{enumerate}[leftmargin=4.1mm]

\item define and propose the first generic construction, called ``\emph{recurring contingent service payment}'' (RC-S-P), that (i) supports the fair exchange of digital verifiable services and coins and (ii) preserves the parties’ privacy. RC-S-P makes black-box use of any scheme that offers a verifiable service and remains secure in the recurring setting where the construction is executed many times.


\item propose the first recurring contingent PoR payment (RC-PoR-P). It is a concrete efficient instantiation of the RC-S-P scheme. To achieve efficiency,  it avoids generic cryptographic tools and utilises mainly symmetric-key primitives and smart contracts.  

\item implement   RC-PoR-P  and analyse its cost. Our cost analysis illustrates RC-PoR-P is highly efficient.  When it deals with a $4$-GB outsourced file, in each verification, a verifier can check a proof in only $90$ milliseconds, and a dispute between a prover and a verifier can be resolved in $0.1$ milliseconds. Also, the contracts' computation is constant in file size. We have made the implementation source code publicly available.

\item identify a free-riding attack in the state-of-the-art fair exchange protocols that were designed to support the fair exchange of a verifiable service and coins, i.e., the two  (publicly and privately verifiable)  protocols of Campanelli \textit{et al}.~\cite{CampanelliGGN17}. We show that the attack lets an adversary use a service without paying the fee.

\item show that the protocols of~\cite{CampanelliGGN17} are not suitable for cases where parties’ privacy matters. We argue that the schemes leak in real-time non-trivial fresh information about the seller and buyer to the public, e.g., deposits' actual amount and proof status.

\end{enumerate}

The identified issues in the protocols of~\cite{CampanelliGGN17} indicate a need for a secure mechanism like RC-S-P.  
Our RC-S-P can be used to prevent a variant of \emph{Authorised Push Payment} (APP) fraud, called Purchase fraud, where a service provider may wish to receive a certain amount of coin without delivering the service\footnote{We refer readers to  \cite{2021-Half-Year-Fraud-Update} for further discussion about APP fraud.}. Defining and designing generic RC-S-P is challenging, for three reasons: (i) there exists no generic definition for \emph{verifiable service} (VS) schemes in the literature,  (ii) most of the application-specific VS schemes (e.g., proofs of retrievability \cite{DBLP:conf/asiacrypt/ShachamW08}, or verifiable computation \cite{GennaroGP10}, verifiable searchable encryption \cite{9075374}) assume the client is trusted, while in a fair exchange setting either party can be an active adversary, and (iii) the majority of VS schemes do not (need to) consider the privacy of exchanged messages, as they are in the traditional setting where the client and server directly interact with each other; hence, their messages' privacy can be protected from the public by using secure channels. 

The primary novelties of this work include: (a) RC-S-P, a fair exchange protocol which remains provably secure in the case where either server or client acts malicious while preserving the parties' privacy, (b) RC-PoR-P, which inherits all appealing features of RC-S-P for the special case of PoR without using zero-knowledge proofs, and (c) identifying and addressing the free-riding attack in the state-of-the-art fair exchange protocols, proposed in \cite{CampanelliGGN17}.

%% file: Related-work.tex

\section{Related Work}\label{Related-Work}
In this section, we summarise related work. In Appendix \ref{Survey-of-Related-Work}, we present a  detailed survey. Maxwell~\cite{contingent-payment} proposes  a fair exchange scheme, called ``zero-knowledge contingent payment’’ that supports the  fair exchange of  digital goods and coins. It  is based on  Bitcoin’s smart contracts,   a hash function, and zero-knowledge (zk) proofs. After the advancement of the ``succinct non-interactive argument of knowledge'' (zk-SNARK) \cite{GennaroGP013} that yields more efficient zk proofs, the scheme   was modified to use zk-SNARKs.  Later, Campanelli \textit{et al.}~\cite{CampanelliGGN17} identified an   issue in the above   scheme. The issue  lets a malicious buyer receive the item without paying. To address it, the authors propose the  ``zero-knowledge Contingent Service Payments'' (zkCSP) scheme that also supports contingent payment for  digital  services. It is  based on Bitcoin smart contracts,   hash functions,  and witness indistinguishable proof of knowledge.  To improve  efficiency, they  use zk-SNARKs where  the buyer generates a public parameter, i.e., CRS, and the seller  performs minimal  checks on the CRS.  The authors, as the zkCSP's concrete instantiations,  propose public and private verifiable schemes where the service is ``proofs of retrievability'' (PoR)  \cite{DBLP:conf/asiacrypt/ShachamW08}. To date, they are the only ones  designed for the fair exchange of  digital coins and a digital service. Shortly, we will explain their shortcomings undetected in the literature.


Fuchsbauer~\cite{Fuchsbauer19}  identifies a flaw in  the zkCSP and  shows that the seller's minimal  check  in the zkCSP  does not prevent the buyer from cheating. Later, Nguyen \textit{et al.}~\cite{NguyenAA20} show that by relying on a  stronger assumption, the zkCSP   remains secure. Tramer \textit{et al.}~\cite{TramerZLHJS17} propose a fair exchange scheme that uses  trusted hardware and Ethereum smart contracts.    Dziembowski \textit{et al.}~\cite{DziembowskiEF18} propose FairSwap, a fair exchange scheme  using the Ethereum smart contracts and the notion of proof of misbehaviour \cite{CanettiRR11}.   Later, Eckey \textit{et al.}~\cite{EckeyFS20} propose OPTISWAP that improves FairSwap’s performance. Similar to FairSwap, OPTISWAP uses a smart contract and proof of misbehaviour, but  it relies on an {interactive} dispute resolution protocol.  Recently, outsourced fair PoRs  letting a client delegate the verifications to a smart contract were proposed in \cite{s-PoR,9436004}. The scheme in \cite{s-PoR} uses message authentication codes (MACs) and time-lock puzzles.  The one in \cite{9436004}  uses polynomial commitment and involves a high number of exponentiations. As a result, it imposes higher costs, of proving and verifying,   than the former scheme.  The schemes in  \cite{s-PoR,9436004} assume the client is honest. 

To date, the zkCSP (in \cite{CampanelliGGN17}) remains the only protocol designed to support the fair exchange of digital coins and a verifiable service; accordingly, it is the closest work to ours. The rest of the above schemes are out of the scope of our work because they only support the exchange of two items or the exchange of an item and digital coins.

%% file: Preliminary.tex


\section{Preliminaries} \label{preliminaries}
We use $\lambda$ as the security parameter. We write $x\stackrel{\st\$}\leftarrow X$ to denote that $x$ is chosen uniformly at random from set $X$. We write $\mathsf{negl}(\lambda)$ to denote that a function is negligible in $\lambda$, i.e., asymptotically smaller than the inverse of any polynomial. In the formal definitions in this paper, we use the notation $\Pr\left[\begin{array}{c}\mathsf{Exp}\\
\hline\mathsf{Cond}\\\end{array}\right]$, where $\mathsf{Exp}$ is an experiment that involves an adversary $\mathcal{A}$, and $\mathsf{Cond}$ is the set of the corresponding winning conditions for $\mathcal{A}$. We use $\mathcal{C}$,  $\mathcal{S}$, and $\mathcal{R}$ to denote the client,   server, and arbiter, respectively. We let $pl$ be $\mathcal{S}$'s public price list, $o$ be the amount paid to $\mathcal{S}$ for each valid proof,  and $l$ be the amount (misbehaving)  $\mathcal{C}$ or $\mathcal{S}$  pays  to $\mathcal{R}$ for resolving a  dispute for each verification, $o_{\st max}$ be the maximum amount paid to $\mathcal{S}$ for a valid  proof,  $l_{\st max}$ be the maximum amount  to resolve a potential dispute, and $z$ be the total number of verifications and $(o,l,o_{\st max},l_{\st max})\in pl$.

 We provide a notation table in Appendix \ref{sec:notation-table}. Similar to the \emph{optimistic} fair cryptographic protocols that aim efficiency, e.g., in \cite{AsokanSW97,eurocrypt/AsokanSW98,DongCCR13}, we assume the existence of a trusted third party arbiter  which remains offline most of the time and is only invoked to resolve disputes. 


\subsection{Smart Contract (SC)}\label{subsec:SC} 


Cryptocurrencies, such as Bitcoin \cite{bitcoin} and Ethereum \cite{ethereum}, beyond offering a decentralised currency,  support  computations on  transactions. In this setting, often a certain computation logic is encoded in a computer program, called a \emph{``smart contract''}. To date, Ethereum is the most predominant cryptocurrency framework that enables users to define arbitrary smart contracts. In this framework,  contract code is stored on the blockchain and  executed by all parties maintaining the cryptocurrency. To prevent a denial of service attack, the framework requires a transaction creator to pay a  fee, called \emph{``gas''}. In this work, we require \emph{minimal capabilities} of Ethereum smart contracts, i.e., given a set of addresses, a certain amount of deposit that the contracts hold, certain integer variables, and simple equations registered to the contracts (i.e., linear combinations of the variables), it distributes among the account holders, a portion of the deposit specified by the equations' output. We assume the smart contract and underlying blockchain are secure, i.e., the used signature scheme is unforgeable and the blockchain is immutable.



\subsection{Building Blocks}\label{subsec:blocks}

We outline the main cryptographic primitives that we utilize in our protocols. We provide a detailed description of the said primitives in Appendix~\ref{app:blocks}.
\begin{itemize}[leftmargin=.38cm]
\item \emph{Pseudorandom Fuction} (PRF): we apply a pseudorandom function $\mathtt{PRF}:\{0,1\}^{\st\psi}\times \{0,1\}^{\st \eta}\rightarrow \{0,1\}^{\st  \iota}$ that on input a random $\psi$-bit key and $\eta$-bit message, it outputs a $ \iota$-bit pseudorandom value (cf. Appendix \ref{subsec:PRF}).
\item \emph{Commitment Scheme}: we deploy a binding and hiding commitment scheme. In the \emph{commit} phase, the sender  commits to a message $x$ as $\mathtt{Com}(x,r)=\mathtt{Com}_{\scriptscriptstyle x}$, that involves a secret value,  $r$. In the \emph{open} phase, the sender sends the opening $\ddot{x}:=(x,r)$ to the receiver which verifies its correctness: $\mathtt{Ver}(\mathtt{Com}_{\scriptscriptstyle x},\ddot{x})\stackrel{\scriptscriptstyle ?}=1$ and accepts if the output is $1$  (cf. Appendix~\ref{subsec:commit}).
\item \emph{Publicly Verifiable Non-interactive Zero-knowledge Proof} (NIZK): is a non-interactive proof where a prover $\mathcal{P}$, given a witness $w$ for some statement $x$ in an NP language $L$, wants to convince in zero-knowledge a verifier $\mathcal{V}$ of the validity of $x\in L$. A NIZK is publicly verifiable when any party can verify the validity of $x\in L$ by obtaining the proof (cf. Appendix~\ref{subsec:PV-NIZK}).
\item \emph{Symmetric-key Encryption Scheme}: it consists of a key generation algorithm $\mathtt{SKE.keyGen}$, an encryption algorithm $\mathtt{Enc}$, and a decryption algorithm $\mathtt{Dec}$. We require that the scheme satisfies IND-CPA security (cf. Appendix~\ref{subsec:SKE}).
\item \emph{Digital Signature Scheme}: it consists of a key generation algorithm $\mathtt{Sig.keyGen}$, a signing algorithm  $\mathtt{Sig.sign}$, and a verification algorithm $\mathtt{Sig.ver}$. We require that the digital signature scheme satisfies EUF-CMA security  (cf. Appendix~\ref{subsec:DS}).
\item \emph{Merkle Tree}: A Merkle tree scheme~\cite{DBLP:conf/sp/Merkle80,DBLP:conf/crypto/Merkle89} is a data structure often used for efficiently checking the integrity of an outsourced file. The Merkle tree scheme includes three algorithms; namely, $\mathtt{MT.genTree}$, $ \mathtt{MT.prove}$, and  $\mathtt{MT.verify}$. Briefly, the first algorithm constructs a Merkle tree on file blocks, the second generates a proof of a block's (or set of blocks') membership, and the third one verifies the proof (cf. Appendix~\ref{sec::merkle-tree}).
\end{itemize}

\vspace{-3mm}
\subsection{Proofs of Retrievability (PoR)}\label{PoR-def}
\input{PoR-def}

%% file: PoR-def.tex

A PoR scheme considers the case where an honest client wants to outsource the storage of its file to a potentially malicious server, i.e., an active adversary. It is a challenge-response interactive protocol, where the server proves to the client that its file is intact and retrievable. Informally, a PoR's soundness requires that if a prover convinces the verifier, then the file is stored by the prover. This is formalized via the notion of an extractor algorithm that can extract the file in interaction with the adversary. Appendix \ref{sec::PoR-definition} presents  the PoR's  formal definition.

 We briefly describe the privately verifiable PoR in \cite{DBLP:conf/asiacrypt/ShachamW08} because it was used as a subroutine in zkCSP's concrete instantiation in \cite{CampanelliGGN17} which we are going to expose its vulnerabilities.
In the setup phase, the client splits its file $u$ into fixed size blocks $u=m_{\st 1},..., m_{\st n}$. It generates a tag, i.e., Message Authentication Code (MAC), for each block $m_{\st i}$ as $\sigma_{\st i}=r_{\st i}+\alpha\cdot {m_{\st i}}$, where $r_{\st i}=\mathtt{PRF}(k, i)$, $\alpha$ is a random value and $k$ is the client's secret key. It outsources the storage of all blocks and tags to the server. Later, to check whether the file is still retrievable, it sends $c$ pair of the form $(j, v_{\st j})$ to the server, where $j$ is a block index and $v_{\st j}$ is a random value. Let set $J$ contain all indices that the client sends to the server. The server generates and then sends to the client a proof pair $(\sigma, \beta)$, where $\sigma = \sum\limits_{\st \forall j\in J} v_{\st j}\cdot \sigma_{\st j}$ and $\beta=\sum\limits_{\st \forall j\in J}v_{\st j}\cdot m_{\st j}$.  The client can verify the proof, by checking if $\sigma=\alpha\cdot \beta+ \sum\limits_{\st \forall j\in J} v_{\st j}\cdot r_{\st j}$.

Shacham and Waters \cite{DBLP:conf/asiacrypt/ShachamW08} also propose a publicly verifiable PoR, which requires the client to generate a signature, as a tag, for each block.

%% file: motivation.tex
\vspace{-2mm}
\section{Putting Forth the RC-S-P Concept}\label{sec::motivation}

In this section, we briefly describe the Recurring Contingent Service Payment (RC-S-P) concept and explain why it is needed. RC-S-P concerns fair exchange of a digital verifiable service (offered by a server) and digital coins (offered by a client), and supports any verifiable digital service. Broadly speaking, it ensures that a client pays a predefined amount of digital coins to the server \emph{if and only if} the server (proves that it) provided the promised service. It considers the case where either client or server is potentially an active/malicious adversary. RC-S-P ensures the above security guarantees hold even if the payments reoccur, e.g., the server deals with multiple clients. RC-S-P also preserves the two parties' privacy.    

\vspace{-3mm}
\subsection{Overview of RC-S-P}
 
RC-S-P involves a client $\mathcal{C}$, a server $\mathcal{S}$, an arbiter $\mathcal{R}$, and a smart contract SC. 
It comprises eight phases: (1) \emph{key generation}, in which $\mathcal{C}$ generates the system parameters, (2) \emph{client-side initiation}, where $\mathcal{C}$ encodes the service input, generates metadata, computes a proof asserting that the input and metadata are well-formed, and masks the actual amount of coin it wants to pay for the service, (3) \emph{server-side initiation}, where $\mathcal{S}$ checks $\mathcal{C}$'s proof, ensures the input and metadata are well-formed, and masks the actual deposits it puts on SC, (4) \emph{client-side query generation}, where $\mathcal{C}$ generates an encoded query, (5) \emph{server-side proof generation}, where $\mathcal{S}$ generates an encoded proof asserting that the service was delivered correctly, (6) \emph{client-side proof verification}, where $\mathcal{C}$ checks $\mathcal{S}$'s proof, (7) \emph{dispute resolution},  where $\mathcal{R}$ compiles the validity of a complaint made by $\mathcal{S}$/$\mathcal{C}$, and (8) \emph{coin transfer}, where SC distributes parties' deposits depending on their (mis)behaviour.

Intuitively, RC-S-P is said to be secure if it satisfies
three main security properties: (i) \emph{security against a malicious server}: an adversary corrupting $\mathcal{S}$ wins only with a negligible probability if it does not provide the promised service but persuades $\mathcal{C}$ to accept it, or makes $\mathcal{C}$ or $\mathcal{R}$ withdraw an incorrect amount of coins they deposited in SC; (ii) \emph{security against a malicious client}: an adversary corrupting $\mathcal{C}$ wins only with a negligible probability if it provides an invalid metadata/query but convinces the
$\mathcal{S}$ or $\mathcal{R}$ to accept it, or make $\mathcal{S}$ or $\mathcal{R}$ withdraw an incorrect amount of coins from SC; (iii) privacy: the privacy of (a) the service input (e.g., outsourced file), and (b) the service proof’s status is preserved (for a predefined time period).

RC-S-P offers stronger security guarantees compared to the existing solutions in the real world and literature; below, we explain why that is the case.

\vspace{-2mm}
\subsection{Service Payment in the Real World} 
\vspace{-2mm}

To date, in the real world, each client must pay \emph{in advance} to a server for a digital service that it wishes to use in future, e.g., Dropbox. If a malicious server does not provide the promised service, then the client has to either (i) spend time and effort to follow up the matter (i.e., by arguing the matter with the server or taking legal actions) or (ii) ignore the problem resulting in the loss of the money it paid for the service. Thus, it is important to ensure that an honest client pays only if the server provides the promised service.

\vspace{-2mm}
\subsection{Service Payment in the Literature} 
\vspace{-1mm}

The state-of-the-art protocol in \cite{CampanelliGGN17} that has been designed to support a fair exchange of digital services and coins has an important oversight; namely, it does not protect an honest server from a malicious client. This oversight allows a malicious client to mount an attack which enables it to use the service without paying the server, i.e., a free-riding attack (see Section \ref{sec::Free-Riding-Attack} for further details). So, it is important to ensure that an honest server provides the service only if the client pays for it.

\vspace{-2mm}
\subsection{Privacy} 
\vspace{-1mm}

The existing fair exchange protocol designed for service payment in \cite{CampanelliGGN17} uses a public blockchain to hold clients' deposits and to transfer the deposit to the server if the service is delivered. Nevertheless, the use of blockchain in the fair exchange protocol reveals in real-time non-trivial information about the parties involved, e.g., the deposits’ actual amount or service type (see Section \ref{sec::Lack-of-Privacy} for further discussion). Therefore, it is important to have a fair exchange protocol that could also preserve the parties' privacy. 

%% file: limitations-And-our-solution.tex


\section{The Privacy Issue and Attack}\label{sec::previous-work-limitations}\label{sec:attack-mitigation}


In this section, we elaborate on the lack of privacy of zkCSP and the attack.  
We first explain why the zkCSP schemes do not preserve privacy. Later, in the attack description, we show how the lack of privacy can benefit an attacker.  
%
We focus only on the zkCSP protocols in \cite{CampanelliGGN17}, as they have been specifically designed for a fair exchange of verifiable services and digital coins, whereas the other protocols studied in Section \ref{Related-Work} were designed for a fair exchange of digital items, e.g., a file and coins. 


\input{Limitations-of-zkCSP.tex}

\input{Our-Solution}

%% file: Limitations-of-zkCSP.tex

%


\vspace{-2mm}
\subsection{Lack of Privacy in zkCSP}\label{sec::Lack-of-Privacy}
\vspace{-2mm}

The zkCSP protocols reveal in real-time non-trivial fresh information about the server and clients to the public. The revealed information includes (i) proof status and (ii) deposit amount. We explain them below.


\vspace{-2mm}
\subsubsection{Proof status}\label{sec::Proofs-status} In the traditional setting, the client and server directly interact with each other to verify and prove the  integrity of  agreed-upon services. In this case, the verification's result is only apparent to them. Nevertheless, in the blockchain era,  where a blockchain plays a role in the verification and payment phases (e.g., in the zkCSP schemes)  it becomes visible in \emph{real-time} to \emph{everyone} whether the verification (proof) has been accepted, which reflects whether the server has successfully delivered the service. This issue remains even if the service proofs are not stored (in plaintext) in the blockchain, as the \emph{coins  transfer}  itself  reveals the status of proofs. 

 In certain settings, this leakage might be undesirable and could have \emph{immediate} consequences for both the server and (business) clients, e.g.,   stock value drop \cite{zoom,slack}, or can benefit attackers (as we will explain in Section \ref{sec:attack-mitigation}). For further discussion on proofs status leakage, we refer readers to Appendix \ref{appendix::proof-status-further-discussion}.

\vspace{-2mm}

\subsubsection{Deposit amount}   
 The amount of deposit placed in the smart contract,  swiftly reveals non-trivial  information about the client to the public. In the case of PoR, an observer learns the approximate size of  outsourced data, service type, or in certain cases even the region of clients' outsourced data, by comparing the amount of deposit with the service provider's price list which is often publicly available, e.g., in \cite{amazon,dropbox-,googleOne}. 

\vspace{-2mm}
\subsection{Free-Riding Attack}\label{sec::Free-Riding-Attack}
\vspace{-2mm}

In this section, we describe an attack scenario in which a malicious client (the attacker) is served by an honest server but it would not pay the service fee, i.e., the free-riding attack. We explain the attack for a concrete instantiation of the zkCSP,  when it uses the privately verifiable PoR  of Shacham and Waters \cite{DBLP:conf/asiacrypt/ShachamW08} as a subroutine. This instantiation was presented in \cite[Section 5.2]{CampanelliGGN17}.

Briefly, to mount the attack, the client exploits (1) the lack of privacy in zkCSP and (2) the lack of server-side verification mechanism in the PoR that zkCSP uses.

First, we explain how the lack of privacy in zkCSP benefits the attacker. As we discussed in Section \ref{sec::Proofs-status}, proof status is revealed in real time to everyone including the attacker. Observing proofs' status (when a server deals with multiple clients) over a sufficiently long time (e.g., a few months) allows the attacker to construct  comprehensive background knowledge of the server's  behaviour, e.g., the server has been acting honestly or not suffering from hardware failures that affect its clients' data.  

Next, we explain how the lack of server-side verification mechanism in the PoR that zkCSP uses can benefit the attacker, who acts as follows. At the setup phase in PoR, it generates \textbf{ill-formed tags}. Specifically, instead of honestly generating a tag (MAC) $\sigma_{\st i}$ on a file block $m_{\st i}$ as $\sigma_{\st i}=r_{\st i}+\alpha\cdot {m_{\st i}}$ (cf. Subsection~\ref{PoR-def}), it generates a tag for some arbitrary block $m'_{\st i}$; i.e., $\sigma'_{\st i}=r_{\st i}+\alpha\cdot {m'_{\st i}}$, where  $m_{\st i}\neq m'_{\st i}$. It follows the rest of PoR protocol honestly, with one exception; namely, during the verification, when it sends a query containing some pairs $(j, v_{\st j})$, it also includes pair $(i, v_{\st i})$ in the query, where $i$ is the index of a block whose tag is ill-formed.

 In this case, the server cannot pass the verification; we show why it is the case. For simplicity, we let the attacker send to the server only two pairs $(j, v_{\st j})$ and $(i, v_{\st i})$. Given the pairs, the honest server computes proof pair $(\sigma, \beta)$ as follows  $\sigma=v_{\st j} \cdot \sigma_{\st j} + v_{\st i}\cdot \sigma'_{\st i}=v_{\st j}\cdot (r_{\st j}+\alpha\cdot m_{\st j})+v_{\st i}\cdot (r_{\st i}+\alpha\cdot m'_{\st i})$ and $\beta=v_{\st j}\cdot m_{\st j} + v_{\st i} \cdot m_{\st i}$.  However, this proof is invalid, as if a verifier follows the PoR verification (described in Section \ref{PoR-def}), it  will get the inequality: 
 %
\begin{equation*}
\begin{split}
\sigma&=v_{\st j}\cdot (r_{\st j}+\alpha\cdot m_{\st j})+v_{\st i}\cdot (r_{\st i}+\alpha\cdot {m}'_{\st i})\neq\\
&\neq\alpha\cdot(v_{\st j}\cdot m_{\st j} + v_{\st i} \cdot m_{\st i})+ v_{\st j}\cdot r_{\st j}+ v_{\st i}\cdot r_{\st i}=\\
&=\alpha\cdot\beta+ v_{\st j}\cdot r_{\st j}+ v_{\st i}\cdot r_{\st i}.
\end{split}
\end{equation*}

 This means that, in zkCSP, an honest server cannot generate valid proof when $i$-th block is challenged. In zkCSP, the server cannot detect whether the attacker generated ill-structured tags, as the used PoR does not require a client to prove the tags' correctness. Even after the server fails to generate a valid proof, it cannot tell (or prove to anyone) whether itself (e.g., due to hardware failures) or the malicious client was the source of the problem.

Hence, the malicious client can mount the free-riding attack to avoid paying the server who delivered the services honestly, i.e.,  kept the file intact during the period between the setup and verification phases. To do that, it (i) ensures the server is honest (by exploiting the lack of privacy in zkCSP), (ii) generates ill-formed tags (by exploiting the lack of server-side verification mechanism in the PoR that zkCSP uses), and then (iii) asks the server to generate proof for the blocks related to ill-formed tags. 

In another concrete instantiation of the zkCSK, Campanelli \textit{et al}. used a publicly verifiable PoR as a subroutine. 
 The idea behind the above free-riding attack can be easily applied to that instantiation too.

\noindent\underline{\textit{Main source of the attack}.}  The reason that the attacker can construct ill-structure tags, without being detected, is that the  zkCSP scheme uses a subprotocol that offers a weaker security guarantee than required. Specifically, it assumes \emph{either party} can be potentially corrupted by an active adversary, yet it uses  a certain verifiable service protocol (i.e., PoR) that is secure against \emph{only} a malicious server and assumes the client is fully honest. This \textbf{mismatch of security assumption/requirement} lets a malicious  client misbehave without any consequences. 

%% file: our-solution.tex
\vspace{-2mm}

\section{Overview of our Solution}\label{sec::Overview-of-Our-Solution}

%
 \vspace{-2mm}
\noindent\underline{\textit{Ensuring Security Holds When Either Party is Malicious}}. 
We design the RC-S-P in a modular fashion. First, we formally define the notion of Verifiable Service (VS) and then upgrade VS to a ``Verifiable Service with IDentifiable abort'' (VSID)  inspired by the notion of ``secure multi-party computation with identifiable abort'' \cite{DBLP:conf/crypto/IshaiOZ14}. The latter guarantees that not only the service takes into consideration that the client can be malicious too, but also a third-party arbiter can identify the misbehaving party and resolve any potential disputes between the two. Second, we require a client to deposit its coins to the contract \emph{right before} it starts using the service (similar to the protocol in \cite{s-PoR}) and it is forced to provide correct inputs, via NIZK (similar to the scheme in \cite{armknecht2014outsourced}); otherwise, its deposit is sent to the server. Third, we require parties to post their messages to the contract, to avoid any potential repudiation issue, which is a standard technique. Forth, we let the party which resolves disputes get paid by a corrupt party (similar to the protocol in \cite{DongWAMM17}). The combination of the above techniques allows RC-S-P to deal with the free-riding attack as well.

Now we explain how the solution works. Before using the service, the client deposits a fixed amount of coins in a smart contract, where the deposit amount covers the service payment: $o$ coins, and dispute resolutions' cost: $l$ coins. The server deposits  $l$ coins. Then, the client and server engage in the   VSID protocol such that (the encryption of)  messages exchanged between them are put in the contract.  They perform the verifications locally. When a party detects misbehaviour, it can raise a dispute that invokes the arbiter which  checks the party's claim, off-chain. The arbiter sends the output of the verification to the contract. If the party's claim is  valid, then it can withdraw its coins and the arbiter is paid by the misbehaving party.
%
%
If the party's claim is invalid, that party has to pay the arbiter and the other party can withdraw its deposit. If both the client and server behave honestly, then the arbiter is never invoked;  in this case, the server (after a fixed time) gets its deposit back and is paid for the service, while the client gets $l$ coins back.  
%
%

We provide a formal definition of VS in Appendix \ref{subsec:VS}. Since VSID as a separate notion which might be of independent interest too, we provide its definition, construction, and proof in Appendix \ref{app:VSID}. We note that in the concrete instantiation of our generic solution in which the VS is PoR, we will use a Merkle tree and \emph{proof of misbehaviour} letting us avoid using NIZK and reduce arbiter-side computation (cf. Section~\ref{sec:RC-PoR-P_protocol}).

\noindent\underline{\textit{Preserving Parties' Privacy}}. To preserve the parties' privacy and prevent real-time information leakage,  we use the following ideas. First, to hide proof status, we let the client and server take control of the time of the information release. This lets them keep the information private from the public within a certain period, and release it when it loses its sensitivity.\footnote{\scriptsize The concept of delayed information release has already been used by researchers, e.g., in smart metering in \cite{DelayedInfoRelease}, and in the real world  through the declassification approach taken by most democratic countries  which declassify sensitive information after the information loses its sensitivity.}  Specifically, they agree on the period in which the information must remain hidden, ``private time bubble''. During this period, all messages sent to the contract are encrypted and the parties do not raise any dispute. They raise disputes after the private time bubble ends (or bubble bursts). 

Nevertheless, the client/server can still find out whether a proof is valid when it is provided by its counter-party, because it can locally verify the proof. Second, to hide the amount of deposit, we let each party ``mask'' its coins, by increasing the actual coins amount to the maximum amount of coins in the server's price list.  So, the masked coins hide the actual coins amount from the public.   But, this raises another challenge: \textit{how can the mutually untrustful parties claim back their masking coins (i.e., the difference between the maximum and actual coins amount) after the bubble bursts, while hiding the actual coins amount from the public in the private time bubble?} 

Our third idea, which addresses this challenge, is to let the client and server, at the beginning of the protocol,  agree on a private statement specifying the deposit details (e.g., parties' actual coins amount for the service, dispute resolution, or masking). Later, when they want to claim their coins, they also provide the statement to the contract which checks the statement validity and if it is accepted, it distributes coins according to the statement (and the contract status). We will show how they can efficiently agree on such a statement, by using a statement agreement protocol (SAP). In Appendix \ref{subsec:RCSP_discussion} we also show how they can promise their locked share of coins to a third party.
 
 Our generic framework that offers the above features is called ``Recurring Contingent Service Payment'' (RC-S-P).  Also, as a concrete instantiation of RC-S-P, we present the RC-PoR-P protocol, in which the VS is PoR.

\noindent\textbf{Strawman Solutions.}  
  To address the issue related to the leakage of deposit amount, one may use privacy-preserving cryptocurrency frameworks, e.g., Zerocash \cite{Ben-SassonCG0MTV14} or Hawk \cite{KosbaMSWP16}. Although such frameworks  solve this problem, they impose additional high cost to their users, as each transaction involves a generic proofs system that are computationally expensive.  Also, one might want to let the server pick a fresh address for each verifier to  preserve its pseudonymity with the hope that an observer cannot link clients to a server (so proofs status and deposit amount issues can be addressed). But, for this to work, we have to assume that multiple service providers  use the same protocol on the blockchain and all of them are pseudonymous which  is a strong assumption.

\noindent\textbf{Design Choices.} 
  We make certain design choices that are not present in the zkCSP of \cite{CampanelliGGN17}, to keep our RC-S-P efficient. Specifically, since not only the server but also the clients can be malicious, we (a) involve a third-party arbiter, and (b) require parties to store more messages on a blockchain, for dispute resolution to be feasible. To achieve privacy while avoiding costly cryptographic computations, we require parties to deposit extra coins to mask the actual value of their deposit. To allow fair distribution of the parties' deposit (without having to use the arbiter for the distribution) we use smart contracts. We refer readers to Appendix \ref{sec::Trade-offs} for further discussion.

%% file: RC-S-P-Definition.tex
\vspace{-2mm}

\section{RC-S-P Definition}\label{sec::RC-S-P-Definition}
\vspace{-2mm}

In this section, we introduce a formal definition of RC-S-P.    Before presenting the formal definition of RC-S-P, we outline what a \emph{verifiable service} (VS) is. At a high level, a VS scheme is a two-party protocol in which a client chooses a function, $F$, and provides (an encoding of) $F$, its input $u$, and a query $\bm{q}$ to a server, which is expected to evaluate $F$ on $u$ and $\bm{q}$ (and some public parameters) and respond with the output. Then, the client verifies that the output is indeed the output of the function computed on the provided input. In verifiable services, either the computation (on the input) or both the computation and storage of the input are delegated to the server. We present a full formal definition of a VS scheme in Appendix \ref{subsec:VS}.

%
%

\begin{definition}[RC-S-P Scheme]\label{RC-S-P-def}
A recurring contingent service payment scheme RC-S-P
 %
  %
  involves four parties; namely, a client, server, arbiter, and smart contract (which represents a bulletin board). The scheme is parameterized by five functions:
\begin{itemize}[leftmargin=4.5mm]
\item A function $F$ that will be run on the client's input by the server as a part of the service it provides.
\item A  metadata generator function $M$.
\item A pair of encoding/decoding functions $(E,D)$.
\item  A query generator function $Q$.
\end{itemize}
\noindent The scheme consists of eight algorithms defined below.\\[2pt]
{\small{\underline{$\mathtt{RCSP}.\mathtt{keyGen}(1^{\lambda})\rightarrow \bm{k}$}}}:  It is run by client $\mathcal{C}$. It takes as input security parameter $1^\lambda$. It outputs $\bm{k}:=(k,k')$ that contains a secret and public verification key pair $k:=(sk,pk)$ and a set of secret and public parameters, $k':=(sk',pk')$. It sends $pk$ and $pk'$ to the contract. \\[2pt]
{\small{\underline{$\mathtt{RCSP}.\mathtt{cInit}(1^\lambda, u,\bm{k},z, pl)\rightarrow (u^{\st *},e,T, p_{\st\mathcal S},  \bm{y}, coin^{\st*}_{\st\mathcal C})$}}}: It is run by  $\mathcal{C}$. It takes as input $1^\lambda$,  the service input $u$,   $\bm{k}:=(k, k')$,  the total number of verifications $z$, and  price list $pl$ containing pairs of  actual coin amount for each accepting service proof and the amount for covering each potential dispute resolution's cost.  It represents $u$ as an input of $M$, let  $u^{\st *}$ be this representation. {It  sets $pp$ as (possibly) input dependent parameters, e.g., file size.}
 It computes metadata $\sigma=M(u^{\st *},k,{pp})$ and a  proof $w_{\st\sigma}$ asserting the metadata is well-structured. It sets the value of $p_{\st\mathcal S}$ to the total coins the server should deposit. It picks a private price pair $(o,l)\in pl$. It sets coin secret parameters $cp$ that include $(o,l)$ and parameters of $pl$. It constructs coin encoding token $T_{\st cp}$ containing   $cp$ and   $cp$'s witness, $g_{\st cp}$. It   constructs  encoding token $T_{\st qp}$ that contains secret  parameters $qp$ including {$pp$}, (a  representation of $\sigma$) and parameters (in $sk'$) that will be used to encode the service queries/proofs.   $T_{\st qp}$ contains     $qp$'s witness,  $g_{\st qp}$. 
  Given a valid value and its witness,  anyone can check if they match. It sets a vector of parameters $\bm{y}$ that includes  binary vectors $[\bm{y}_{\st \mathcal  C}, \bm{y}_{\st \mathcal  S},\bm{y}'_{\st \mathcal  C}, \bm{y}'_{\st \mathcal  S}]$ each of which is set to $0$ and its length is $z$.  Note $\bm{y}$ may contain other public parameters, e.g., the contract's address.  It outputs  $u^{\st *}$,  $e:=(\sigma,w_{\st\sigma})$,  $T:=(T_{\st cp},T_{\st qp})$,  $p_{\st\mathcal S}$,    $\bm{y}$,  and the encoded coins amount  $coin^{\st*}_{\st\mathcal C}$ (that contains $o$ and $l$ coins in an encoded form).  $\mathcal{C}$ sends  $u^{\st *}$, $z$,  $e$, $T_{\st cp} \setminus  \{g_{\st cp}\}$, and $T_{\st qp} \setminus  \{g_{\st qp}\}$ to the server  $\mathcal{S}$ and  sends $g_{\st cp},  g_{\st qp}, p_{\st\mathcal S}$, $\bm{y}$,  and $coin^{\st*}_{\st\mathcal C}$  coins to the  contract. \\[2pt]
\underline{$\mathtt{RCSP}.\mathtt{sInit}(u^{\st *},e, pk,z, T, p_{\st\mathcal S},\bm{y})\rightarrow (coin^{\st *}_{\st\mathcal S},a)$}: It is  run by  server $\mathcal{S}$. It takes as input $u^{\st *}$, metadata-proof pair $e:=(\sigma,w_{\st\sigma})$,   $pk$ (read from the contract),  $z$, and $T:=(T_{\st cp},T_{\st qp})$, where $\{g_{\st cp},  g_{\st qp}\}$ are read from the smart contract. It reads $p_{\st\mathcal S}$, and $\bm{y}$ from the smart contract. It  checks the validity of $e$ and  $T$ elements. It checks elements of $\bm{y}$ and ensures each element of $\bm{y}_{\st \mathcal  C}, \bm{y}_{\st \mathcal  S},\bm{y}'_{\st \mathcal  C}, \bm{y}'_{\st \mathcal  S}\in \bm{y}$ has been set to $0$. If all checks pass, then it encodes the amount of its coins that yields $coin^{\st*}_{\st\mathcal S}$, and sets $a=1$. Otherwise, it sets $coin^{\st *}_{\st\mathcal S}=\bot$ and $a=0$. It outputs $coin^{\st *}_{\st\mathcal S}$ and $a$. The smart contract is given $coin^{\st *}_{\st\mathcal S}$ coins and $a$.  \\[2pt]
\underline{$\mathtt{RCSP}.\mathtt{genQuery}(1^\lambda, \text{aux},k,T_{\st qp})\rightarrow c^{\st *}_{\st j}$}: It is run by  $\mathcal{C}$. It takes as input  $1^\lambda$,  auxiliary information $\text{aux}$,  the key pair $k$, and encoding token $T_{\st qp}$.  It computes a pair $c_{\st j}$ containing a query vector $\bm{q}_{\st j}=Q( \text{aux},k,{pp})$,  and proof $\bm{w}_{\st q_{_{\scaleto{j}{\size}}}}$ proving the query is well-structured, where {$pp\in T_{\st qp}$}. It outputs the encoding of the pair, $c^{\st *}_{\st j}=E(c_{\st j},T_{\st qp})$, and sends the output to the contract. \\[2pt]
\underline{$\mathtt{RCSP}.\mathtt{prove}(u^{\st *},  \sigma,  c^{\st *}_{\st j},pk, T_{\st qp})\rightarrow (b_{\st j},m_{\st \mathcal{S},j},\bm\pi^{\st *}_{\st j})$}: It is run by  $\mathcal{S}$. It  takes as input $u^{\st *}$, metadata $\sigma$,   $c^{\st *}_{\st j}$,  $pk$,  and  $T_{\st qp}$.   It checks the validity of decoded query pair $c_{\st j}=D(c^{\st *}_{\st j},T_{\st qp})$.  If it is rejected, then it sets $b_{\st j}=0$ and constructs a complaint $m_{\st \mathcal{S},j}$. Otherwise,  it   sets $b_{\st j}=1$ and $m_{\st \mathcal{S},j}=\bot$. It outputs $b_{\st j},m_{\st \mathcal{S},j}$,   and  encoded proof $\bm\pi^{\st *}_{\st j}=E(\pi_{\st j},T_{\st qp})$, where  $\bm\pi_{\st j}$ contains $h_{\st j}=F(u^{\st *},\bm{q}_{\st j},{pp})$ and a proof $\delta_{\st j}$  asserting the evaluation is performed correctly ($\bm\pi_{\st j}$ may contain dummy values if $b_{\st j}=0$). The smart contract is given $\bm\pi^{\st *}_{\st j}$.\\[2pt]
\underline{$\mathtt{RCSP}.\mathtt{verify}(\bm\pi^{\st *}_{\st j}, c^{\st *}_{\st j},k,T_{\st qp})\hspace{-1mm}\rightarrow (d_{\st j}, \bm m_{\st \mathcal{C},j})$}: \hspace{-2.1mm} A deterministic algorithm run by  $\mathcal{C}$. It takes as input  $\bm\pi^{\st *}_{\st j}$,  query vector $\bm{q}_{\st j}\in c^{\st *}_{\st j}$,  $k$,  and  $T_{\st qp}$. It checks the decoded proof $\bm\pi_{\st j}=D(\bm\pi^{\st *}_{\st j},T_{\st qp})$, if it is rejected,  it outputs $d_{\st j}=0$ and a complaint $\bm m_{\st \mathcal{C},j}$. Else, it outputs  $d_{\st j}=1$ and $\bm m_{\st \mathcal{C},j}=\bot$.\\[2pt]
\underline{$\mathtt{RCSP}.\mathtt{resolve}(\bm m_{\st \mathcal{C}}, \bm m_{\st \mathcal{S}},z, {\bm{\pi}}^{\st *}, {\bm{c}}^{\st *},pk, T_{\st qp})\rightarrow \bm{y}$}: It is run by the arbiter $\mathcal{R}$. It takes as input $\mathcal{C}$'s complaints $\bm m_{\st \mathcal{C}}$, $\mathcal{S}$'s complaints $\bm m_{\st \mathcal{S}}$, $z$, all encoded proofs $ {\bm{\pi}}^{\st *}$, all encoded query pairs $ {\bm{c}}^{\st *}$,  $pk$, and encoding token $T_{\st qp}$. It verifies the token, decoded queries, and proofs.  It reads the  binary vectors $[\bm{y}_{\st \mathcal  C}, \bm{y}_{\st \mathcal  S},\bm{y}'_{\st \mathcal  C}, \bm{y}'_{\st \mathcal  S}]$ from the smart contract. It  updates $\bm{y}_{\st \mathcal P}$ by setting an element of it to one, i.e., $y_{\st \mathcal P,j}=1$, if party $\mathcal{P}\in\{\mathcal{C},\mathcal{S}\}$ has misbehaved in the $j$-th verification (i.e., provided invalid query or service proof). It also updates $\bm{y}'_{\st \mathcal P}$ (by setting an element of it to one) if party $\mathcal{P}$ has provided a complain that does not allow it to identify a misbehaved party, in  the $j$-th verification, i.e., when the arbiter is unnecessarily invoked. \\[2pt]
\underline{$\mathtt{RCSP}.\mathtt{pay}(\bm{y},T_{\st cp},a,p_{\st\mathcal S}, coin^{\st *}_{\st\mathcal C},coin^{\st *}_{\st\mathcal S})\rightarrow ({\bm{coin}}_{\st\mathcal C},{\bm{coin}}_{\st\mathcal S},$}\\ \underline{$\bm{coin}_{\st\mathcal{R}})$}: It is run by the smart contract. It takes as input the binary vectors $[\bm{y}_{\st \mathcal  C}, \bm{y}_{\st \mathcal  S},\bm{y}'_{\st \mathcal  C},$  $\bm{y}'_{\st \mathcal  S}]\in \bm{y}$ that indicate which party misbehaved, or sent invalid complaint  in each verification,   $T_{\st cp}:=\{cp,g_{\st cp}\}$,  $a$, the total coins the server should deposit $p_{\st\mathcal S}$, $coin^{\st *}_{\st\mathcal C}$, and $coin^{\st *}_{\st\mathcal S}$. If $a=1$ and  $coin^{\st *}_{\st\mathcal S}= p_{\st\mathcal S}$, then it verifies the validity of $T_{\st cp}$. If $T_{\st cp}$ is rejected, then it aborts. 
If it is accepted, then it constructs  vector ${\bm{coin}}_{\st\mathcal P}$, where $\mathcal{P}\in\{\mathcal{C},\mathcal{S},\mathcal{R}\}$; It sends ${{coin}}_{\st\mathcal{P},j}\in{\bm{coin}}_{\st\mathcal P}$ coins to party $\mathcal{P}$ for each $j$-th verification. Otherwise (i.e., $a=0$ or $coin^{\st *}_{\st\mathcal S}\neq p_{\st\mathcal S}$)  it  sends $coin^{\st *}_{\st\mathcal C}$ and  $coin^{\st *}_{\st\mathcal S}$ coins to $\mathcal C$ and $\mathcal S$ respectively. 	
\end{definition}
\vspace{-1mm}

The above algorithms $\mathtt{RCSP}.\mathtt{genQuery},\mathtt{RCSP}.\mathtt{prove},$ $\mathtt{RCSP}.\mathtt{verify}$, and $\mathtt{RCSP}.\mathtt{resolve}$ implicitly  take $(a, coin^{\st*}_{\st\mathcal{S}},  p_{\st\mathcal{S}})$ as other inputs and  execute only if $a=1$ and $coin^{\st*}_{\st\mathcal{S}}=p_{\st\mathcal{S}}$; but, for simplicity we  avoided explicitly stating it  in the definition. 
An RC-S-P  scheme must meet correctness and security.  Correctness requires that by the end of the protocol's execution  (that involves  honest client and server), the server accepts an honest client's encoded data and query while the honest client accepts the server's valid service proof (and no one is identified as a misbehaving party). Moreover, the honest client gets back all its deposited coins minus the service payment, the honest server gets back all its deposited coins plus the service payment and the arbiter (that is not involved) receives nothing. Correctness is formally stated below.


\begin{definition}[Correctness] \label{def::RC-S-P-Correctness} An RC-S-P  scheme  with  functions $F,  M, E, D,Q$ is \emph{correct} for auxiliary information
%
 %
 \text{aux} if for any $z$ polynomial in $\lambda$, any price list $pl$, and any service input $u$, it holds that the following probability is equal to $1$: 
\vspace{-1mm}
{\small{
\[
\Pr\left[
\begin{array}{l}
\mathtt{RCSP}.\mathtt{keyGen}(1^\lambda)\rightarrow \bm{k}\\
\mathtt{RCSP}.\mathtt{cInit}(1^\lambda, u,\bm{k},z,pl) \rightarrow (u^{\st *},e,T, p_{\st\mathcal S},  \bm{y},coin^{\st*}_{\st\mathcal C})\\
\mathtt{RCSP}.\mathtt{sInit} (u^{\st *},e, pk,z, T, p_{\st\mathcal S},\bm{y})\rightarrow (coin^{\st *}_{\st\mathcal S},a)\\
\mathsf{For}\; j=1,\ldots,z\;\mathsf{do}: \\

\quad\mathtt{RCSP}.\mathtt{genQuery}(1^\lambda, \text{aux}, k, T_{\st qp})\rightarrow c^{\st *}_{\st j}\\
\quad\mathtt{RCSP}.\mathtt{prove}(u^{\st *},  \sigma,  c^{\st *}_{\st j},pk, T_{\st qp})\rightarrow (b_{\st j},m_{\st\mathcal {S},j}, \bm\pi^{\st *}_{\st j})\\
\quad\mathtt{RCSP}.\mathtt{verify}(\bm\pi^{\st *}_{\st j}, c^{\st *}_{\st j},k,T_{\st qp})\rightarrow (d_{\st j}, \bm m_{\st\mathcal {C},j})\\
\mathtt{RCSP}.\mathtt{resolve}(\bm m_{\st \mathcal{C}}, \bm m_{\st \mathcal{S}}, z, {\bm{\pi}}^{\st *}, {\bm{c}}^{\st *},pk, T_{\st qp})\rightarrow \bm{y}\\
\mathtt{RCSP}.\mathtt{pay}(\bm{y},T_{\st cp},a,p_{\st\mathcal S}, coin^{\st *}_{\st\mathcal C},coin^{\st *}_{\st\mathcal S})\rightarrow (\bm{coin}_{\st\mathcal C},\\\bm{coin}_{\st\mathcal S},\bm{coin}_{\st\mathcal{R}})\\
\hline
(a=1)\wedge (\bigwedge\limits^{\st z}_{\st j=1} b_{\st j}= \bigwedge\limits^{\st z}_{\st j=1} d_{\st j}=1)\ \wedge\\ (\bm{y}_{\st \mathcal C}=\bm{y}_{\st \mathcal S}=\bm{y}'_{\st \mathcal C}=\bm{y}'_{\st \mathcal S}=0)\ \wedge\\
(\sum\limits^{\st z}_{\st j=1}coin_{\st\mathcal C,j}=coin^{\st *}_{\st\mathcal C}-o\cdot z)\ \wedge\\  (\sum\limits^{\st z}_{\st j=1}coin_{\st\mathcal S,j}=coin^{\st *}_{\st\mathcal S}+o\cdot z)\  \wedge (\sum\limits^{\st z}_{\st j=1}coin_{\st\mathcal{R},j}=0)\\
\end{array}
\right]\]
}}
\end{definition}
\noindent where $\bm{y}_{\st \mathcal C},\bm{y}_{\st \mathcal S},\bm{y}'_{\st \mathcal C},\bm{y}'_{\st \mathcal S}\in \bm{y}$.

An RC-S-P scheme is said to be secure if it satisfies  three main properties: (i)  security against  a malicious server, (ii) security against a malicious client, and (cii privacy.  In the following, we formally define each of them.  

Intuitively, security against a malicious server states that, for each $j$-th verification, 
the adversary  wins only with a negligible probability, if it  provides either (a)  correct evaluation of the function on the service input but it either makes the client  withdraw an incorrect amount of coins (i.e., something other than its deposit minus service payment) or makes the arbiter withdraw an incorrect amount of coins if it unnecessarily invokes the arbiter, or (b) incorrect evaluation of the function on the service input,  but either persuades the client or the arbiter to accept it  or makes them  withdraw an incorrect amount of coins (i.e., $coin_{\st\mathcal{C},j}\neq \frac{coin_{\st\mathcal C}^{\st*}}{z}$ or  $coin_{\st\mathcal{R},j}\neq l$ coins).  Below, we formalize this intuition.

\begin{definition}[Security Against Malicious Server]\label{deff::RC-S-P-SecurityAgainstMaliciousServer}  An RC-S-P scheme with functions $F,M,E,D,Q$ is \emph{secure against a malicious server} for auxiliary information $\text{aux}$, if for any $z$ polynomial in $\lambda$,  any price list $pl$, every $j$ (where $1\leq j\leq z$), and any PPT adversary $\mathcal{A}$,  it holds that the following probability is $\mathsf{negl}(\lambda)$:
\vspace{-1mm}
\[ \Pr\left[
{\small{
  \begin{array}{l}
\hs\mathtt{RCSP.keyGen}(1^{\lambda})\rightarrow \bm{k}\\
  \hs \mathcal{A}(1^\lambda,pk, F,M,E,D,Q,z,pl)\rightarrow u\\
  \hs \mathtt{RCSP}.\mathtt{cInit}(1^\lambda, u,\bm{k},z,pl)\rightarrow (u^{\st *},e,T, p_{\st\mathcal S},  \bm{y}, coin^{\st*}_{\st\mathcal C})\\
 \hs  \mathcal{A}(u^{\st *}, e, pk, z, T, p_{\st\mathcal S},\bm{y})\rightarrow (coin^{\st *}_{\st\mathcal S},a)\\
 \hs	\mathtt{RCSP}.\mathtt{genQuery}(1^\lambda, \text{aux},k, T_{\st qp})\rightarrow c^{\st *}_{\st j}\\
   \hs  \mathcal{A}(c^{\st *}_{\st j},\sigma, u^{\st *},a)\rightarrow 
     (b_{\st j},m_{\st \mathcal{S},j},h^{\st *}_{\st j},\delta^{\st *}_{\st j})\\ 
\hs \mathtt{RCSP}.\mathtt{verify}(\bm\pi^{\st *}_{\st j}, c^{\st *}_{\st j}, k,T_{\st qp})\rightarrow (d_{\st j}, \bm m_{\st \mathcal{C},j})\\
  \hs \mathtt{RCSP}.\mathtt{resolve}(\bm m_{\st \mathcal{C}}, \bm m_{\st \mathcal{S}},z, {\bm{\pi}}^{\st *}, {\bm{c}}^{\st *},pk, T_{\st qp})\rightarrow \bm{y}\\
 \hs  \mathtt{RCSP}.\mathtt{pay}(\bm{y},T_{\st cp},a,p_{\st\mathcal S}, coin^{\st *}_{\st\mathcal C},coin^{\st *}_{\st\mathcal S})\rightarrow ({\bm{coin}}_{\st\mathcal C}, \\{\bm{coin}}_{\st\mathcal S},{\bm{coin}}_{\st\mathcal{R}})\\  
\hline  
  \Big(F(u^{\st *}, \bm{q}_{\st j}, {pp})= h_{\st j} \ \wedge (coin_{\st\mathcal{C},j}\neq  \frac{coin_{\st\mathcal C}^{\st*}}{z}-o \vee\\  (coin_{\st\mathcal{R},j}\neq l\ \wedge\  y'_{\st \mathcal {S},j}=1))\Big)\ \vee\\ 
\Big(
F(u^{\st *},\bm{q}_{\st j}, {pp})\neq h_{\st j} \ \wedge (d_{\st j}=1	\vee y_{\st \mathcal{S},j}=0  \vee\\ \ coin_{\st\mathcal{C},j}\neq \frac{coin_{\st\mathcal C}^{\st*}}{z} \ \vee  coin_{\st\mathcal{R},j}\neq l) \Big)\\
\end{array}
}  }  \right]\]

\noindent where  $\bm{q}_{\st j}\in D(c^{\st *}_{\st j},T_{\st qp})$, $\bm\pi^{\st *}_{\st j}=[h^{\st *}_{\st j}, \delta^{\st *}_{\st j}], h_{\st j}= D(h^{\st *}_{\st j},T_{\st qp})$, $\sigma\in e$, $\bm m_{\st \mathcal{C},j}\in \bm m_{\st \mathcal{C}}, m_{\st \mathcal{S},j}\in \bm m_{\st \mathcal{S}}, y'_{\st \mathcal{S},j}\in \bm{y}'_{\st \mathcal{S}}\in\bm{y}$,   $y_{\st \mathcal{S},j}\in \bm{y}_{\st \mathcal{S}}\in\bm{y}$, and $ {pp}\in T_{\st qp}$. 

\end{definition}

Informally, security against a malicious client requires that, for each  $j$-th verification, a malicious client with a negligible probability  wins if it provides either (a) valid metadata and query but either makes the server receive an incorrect amount of coins (something other than its deposit plus the service payment), or makes the arbiter withdraw incorrect amounts of coin if it unnecessarily invokes the arbiter  or (b) invalid metadata or query but convinces the server to accept either of them (i.e., the invalid metadata or query), or (c) invalid query but persuades the arbiter to accept it, or  makes them withdraw an incorrect amount of coins (i.e., $coin_{\st\mathcal{S},j}\neq \frac{coin_{\st\mathcal S}^{\st*}}{z}+o$ or  $coin_{\st\mathcal{R},j}\neq l$ coins). Below, we formally state the property. Note that in the following  definition, an honest server either does not deposit (e.g., when $a=0$) or if it deposits (i.e., agrees to serve) ultimately receives its deposit \emph{plus the service payment} (with high probability).
 \begin{definition}[Security Against Malicious Client]\label{deff::RC-S-P-Security-Against Malicious-Client}  An RC-S-P scheme with functions $F,M,E,D,Q$  is \emph{secure against a malicious client} for auxiliary information \text{aux}, if for any $z$ polynomial in $\lambda$, any price list $pl$, every $j$ (where $1\leq j\leq z$), and any PPT adversary $\mathcal{A}$, it holds that the following probability is $\mathsf{negl}(\lambda)$:
\[\Pr\left[{\small{
  \begin{array}{l}
\hs\mathcal{A}(1^{\lambda}, F,M,E,D,Q,z,pl)\rightarrow (u^{\st *},\bm{k},e,T,  p_{\st\mathcal{S}}, coin^{\st*}_{\st\mathcal C},\\ \bm{y},pk)\\
 \hs \mathtt{RCSP}.\mathtt{sInit}(u^{\st *}, e, pk, z, T, p_{\st\mathcal S},\bm{y})\rightarrow (coin^{\st *}_{\st\mathcal S},a)\\
\hs	\mathcal{A}(coin^{\st *}_{\st\mathcal S},a, 1^\lambda, \text{aux},k, T_{\st qp})\rightarrow c^{\st *}_{\st j}\\
 \hs    \mathtt{RCSP}.\mathtt{prove}(u^{\st *},  \sigma,  c^{\st *}_{\st j}, pk, T_{\st qp})\rightarrow (b_{\st j},m_{\st \mathcal{S},j}, \bm\pi^{\st *}_{\st j})\\
 \hs \mathcal{A}(\bm\pi^{\st *}_{\st j}, c^{\st *}_{\st j}, k, T_{\st qp})\rightarrow (d_{\st j}, \bm {m}_{\st {\mathcal{C}},j})\\
 \hs  \mathtt{RCSP}.\mathtt{resolve}(\bm m_{\st \mathcal{C}}, \bm m_{\st \mathcal{S}}, z, {\bm{\pi}}^{\st *}, {\bm{c}}^{\st *}, pk, T_{\st qp})\rightarrow \bm{y}\\
\hs \mathtt{RCSP}.\mathtt{pay}(\bm{y},T_{\st cp},a,p_{\st\mathcal S}, coin^{\st *}_{\st\mathcal C}, coin^{\st *}_{\st\mathcal S})\rightarrow ({\bm{coin}}_{\st\mathcal C},\\ {\bm{coin}}_{\st\mathcal S},{\bm{coin}}_{\st\mathcal{R}})\\
 
 \hline
 
\hs \Big((M(u^{\st *},k, {pp})= \sigma \ \wedge Q(\text{aux},k, {pp})= \bm{q}_{\st j})\ \wedge\\
\hs  (coin_{\st\mathcal{S},j}\neq  \frac{coin_{\st\mathcal S}^{\st*}}{z}+o  \ \vee
 coin_{\st\mathcal{R},j}\neq l\ \wedge\  y'_{\st \mathcal {C},j}=1)\Big) \vee\hs
  \\
  
\hs  \Big(M(u^{\st *}, k,  {pp})\neq \sigma \wedge a=1\Big) \ \vee\\

 \hs \Big(Q(\text{aux},k,  {pp})\neq \bm{q}_{\st j}\ \wedge
  (b_{\st j}=1 \ \vee\\ \ y_{\st\mathcal{C},j}=0 \ \vee  coin_{\st\mathcal{S},j}\neq \frac{coin_{\st\mathcal S}^{\st*}}{z}+o \ \vee  coin_{\st\mathcal{R},j}\neq l) 
  \Big)
 \\
 
\end{array}    }}\right]\]

\noindent where $\bm{q}_{\st j}\in D(c^{\st *}_{\st j},t_{\st qp})$, $\sigma\in e$, $y'_{\st \mathcal{C},j}\in \bm{y}'_{\st \mathcal{C}}\in\bm{y}$,   $y_{\st \mathcal{C},j}\in \bm{y}_{\st \mathcal{C}}\in\bm{y}$, and ${pp}\in T_{\st qp}$. 


\end{definition}

  Informally, RC-S-P is privacy-preserving if it guarantees the privacy of (1) the service input (e.g., outsourced file) and (2) the service proof's status during the private time bubble. In the following, we formally define privacy.

  \begin{definition}[Privacy]\label{deff::RC-S-P-Privacy}  An RC-S-P scheme  with functions $F, M, E, $ $D,Q$ preserves \emph{privacy} for  auxiliary information
  %
   %
   \text{aux} if for any $z$ polynomial in $\lambda$ and any price list $pl$,  the following hold:
  \begin{enumerate}
\item For any PPT adversary $\mathcal{A}_{\st 1}$, it holds that the following probability is no more than $\frac{1}{2}+\mathsf{negl}(\lambda)$.
\[\Pr\left[{\small{
  \begin{array}{l}
  \hs \mathtt{RCSP.keyGen}(1^{\lambda})\rightarrow \bm{k}\\
   \hs      \mathcal{A}_{\st 1}(1^\lambda, pk, F,M,E,D,Q,z,pl)\rightarrow (u_{\st 0},u_{\st 1})\\
   \hs  \beta\stackrel{\st\$}\leftarrow\{0,1\}\\
  \hs  \mathtt{RCSP}.\mathtt{cInit}(1^\lambda, u_{_{\st\beta}},\bm{k},z,pl)\rightarrow (u^{\st *}_{_{\st\beta}},e,T, p_{\st\mathcal S},  \bm{y},\\\hs  coin^{\st*}_{\st\mathcal C})\\
 \hs \mathtt{RCSP}.\mathtt{sInit}(u_{_{\st\beta}}^{\st *}, e, pk, z, T, p_{\st\mathcal S},\bm{y})\rightarrow (coin^{\st *}_{\st\mathcal S},a)\\
\hs\mathsf{For}\; j=1,\ldots,z\;\mathsf{do}: \\
 \hspace{-3.5mm}   \quad\mathtt{RCSP}.\mathtt{genQuery}(1^\lambda, \text{aux},k, T_{\st qp})\rightarrow c^{\st *}_{\st j}\\
 \hspace{-3.5mm}  \quad  \mathtt{RCSP}.\mathtt{prove}(u^{\st *}_{\st\beta},   \sigma,  c^{\st *}_{\st j}, pk, T_{\st qp})\rightarrow (b_{\st j},m_{\st \mathcal{S},j}, \bm\pi^{\st *}_{\st j})\\  
 \hspace{-3.5mm}   \quad\mathtt{RCSP}.\mathtt{verify}(\bm\pi^{\st *}_{\st j}, c^{\st *}_{\st j}, k,T_{\st qp})\rightarrow (d_{\st j}, \bm m_{\st \mathcal{C},j})\\ 
 \hline
\hs  \mathcal{A}_{\st 1}(\bm{c}^{\st *},coin^{\st *}_{\st\mathcal S},coin^{\st *}_{\st\mathcal  C}, g_{\st cp}, g_{\st qp}, \bm{\pi}^{\st *}, pl, a)\rightarrow\beta\\
\end{array}}}\right]\]
\noindent where   $\bm{c}^{\st *}=[{c}^{\st *}_{\st 1},...,{c}^{\st *}_{\st z}]$ and $\bm{\pi}^{\st *}=[\bm\pi^{\st *}_{\st 1},...,\bm\pi^{\st *}_{\st z}]$.

\item For any PPT adversaries $\mathcal{A}_{\st 2}$, $\mathcal{A}_{\st 3}$, and  $\mathcal{A}_{\st 4}$ the following probability is no more than $Pr_\mathrm{max}+\mathsf{negl}(\lambda)$:
\[\Pr\left[{\small{
  \begin{array}{l}
\hs  \mathtt{RCSP.keyGen}(1^{\lambda})\rightarrow \bm{k}\\
 \hs \mathcal{A}_{\st 2}(1^\lambda, pk, F,M,E,D,Q,z,pl)\rightarrow u\\
 \hs   \mathtt{RCSP}.\mathtt{cInit}(1^\lambda, u,\bm{k}, M,z,pl,enc)\rightarrow (u^{\st *},e,T,\\ \hs p_{\st\mathcal S},  \bm{y}, coin^{\st*}_{\st\mathcal C})\\
\hs \mathtt{RCSP}.\mathtt{sInit}(u^{\st *}, e, pk, z, T, p_{\st\mathcal S},\bm{y})\rightarrow (coin^{\st *}_{\st\mathcal S},a)\\
\hs\mathsf{For}\; j=1,\ldots,z\;\mathsf{do}: \\

 \hspace{-3.5mm}   \quad\mathcal{A}_{\st 2}(1^\lambda, \text{aux}, k, T_{\st qp})\rightarrow c^{\st *}_{\st j}\\
   \hspace{-3.5mm}    \quad \mathcal{A}_{\st 3}(u^{\st *},   \sigma,  c^{\st *}_{\st j}, pk, T_{\st qp})\rightarrow (b_{\st j}, m_{\st \mathcal{S},j}, \bm\pi^{\st *}_{\st j})\\  
 \hspace{-3.5mm}    \quad \mathtt{RCSP}.\mathtt{verify}(\bm\pi^{\st *}_{\st j}, c^{\st *}_{\st j}, k,T_{\st qp})\rightarrow (d_{\st j}, \bm m_{\st \mathcal{C},j})\\
  \hline
 \hs \mathcal{A}_{\st 4}(F,M,E,D,Q,\bm{c}^{\st *},coin^{\st *}_{\st\mathcal S},coin^{\st *}_{\st\mathcal  C}, g_{\st cp}, g_{\st qp}, \bm{\pi}^{\st *},\\ pl, a)\rightarrow(d_{\st j},j)\\
\end{array} }}\right]\]

%
\noindent where  $\bm\pi^{\st *}_{\st j}$ has been encoded correctly, $\bm\pi^{\st *}_{\st j}=[h^{\st *}_{\st j}, \delta^{\st *}_{\st j}], h_{\st j}= D(h^{\st *}_{\st j},T_{\st qp})$, and $Pr_\mathrm{max}$ is defined  as follows. Let  $Exp_{\st\mathsf{priv}}^{\st\mathcal{A}_2, \mathcal{A}_3}(1^{\st \lambda})$ be the above experiment. Let  $\bm{q}_{\st j}\in D(c^{\st *}_{\st j},T_{\st qp})$, $ {pp}\in T_{\st qp}$. We define the events $Con_{\st 0,j}^{\st (1)}:Q(\text{aux}, k, pp)\neq \bm{q}_{\st j}$, $ Con_{\st 0,j}^{\st (2)}:b_{\st j}=0, Con_{\st 1,j}^{\st (1)}:Q(\text{aux}, k, pp)= \bm{q}_{\st j}$, $ Con_{\st 1,j}^{\st (2)}:b_{\st j}=1$, $\overline{Con}_{\st 0,j}^{\st (1)}: F(u^{\st *},\bm{q}_{\st j}, {pp})\neq h_{\st j}$, $ \overline{Con}_{\st 0,j}^{\st (2)}: d_{\st j}=0$, $\overline{Con}_{\st 1,j}^{\st (1)}: F(u^{\st *},\bm{q}_{\st j}, {pp})= h_{\st j}$, and $ \overline{Con}_{\st 1,j}^{\st (2)}: d_{\st j}=1$. For $i\in\{0,1\}$ and $j\in[z]$, let

\vspace{-2mm}

\[Pr_{\st i,j}:=\Pr\left[{\small{
\begin{array}{l}
\hspace{3mm}Exp_{\st\mathsf{priv}}^{\st \mathcal{A}_2,\mathcal{A}_3}(1^{\st \lambda})\\
%
  \hline
\Big(Con_{\st i,j}^{\st (1)} \ \wedge \ Con_{\st i,j}^{\st (2)} \Big)\vee\\ \Big(\overline{Con}_{\st i,j}^{\st (1)} \ \wedge \ \overline{Con}_{\st i,j}^{\st (2)} \Big) \end{array}}}\right]. \mbox{ Then,}\]
%
%
 $Pr_\mathrm{max}:=\mathrm{max}\{Pr_{\st 0,1},Pr_{\st 1,1},..., Pr_{\st 0,z},Pr_{\st 1,z}\}$.

\end{enumerate}
\end{definition}
 
In the above definition, for each $j$-th verification, the adversary $\mathcal{A}_{\st 2}$ or $\mathcal{A}_{\st 3}$ produces an invalid query or invalid proof, respectively, with probability $Pr_{\st 0,j}$  and a valid query or valid proof, respectively, with probability $Pr_{\st 1,j}$. It is required that  privacy is preserved regardless of the queries and proofs  status, i.e., whether they are valid/invalid, as long as they are correctly encoded and provided. In the above definitions, the private time bubble is a time period from the point when $\mathtt{RCSP.keyGen}(\cdot)$ is executed up to the time  when $\mathtt{RCSP}.\mathtt{resolve}(\cdot)$  is run. In other words, the privacy holds up to the point where $\mathtt{RCSP}.\mathtt{resolve}(.)$  is run. This is why the latter algorithm is excluded from the  experiments in Definition \ref{deff::RC-S-P-Privacy}.

 \begin{definition}[RC-S-P Security]\label{def::RC-S-P-Security} An RC-S-P with functions $F,M, $ $E,D,Q$ is \emph{secure} for auxiliary information \text{aux}, if it satisfies security against malicious server, security against malicious client, and preserves privacy for \text{aux}, w.r.t. Definitions \ref{deff::RC-S-P-SecurityAgainstMaliciousServer},~\ref{deff::RC-S-P-Security-Against Malicious-Client}, and~\ref{deff::RC-S-P-Privacy}, respectively. 
 \end{definition}


%% file: RC-S-P-overview.tex

\vspace{-4mm}

\section{Generic RC-S-P Protocol}\label{sec:RCSP-Overview}
\vspace{-2mm}
In this section, we outline the designs of the RC-S-P protocol that realises Definition \ref{def::RC-S-P-Security}. As we stated in Section \ref{sec::Overview-of-Our-Solution}, this protocol relies on the idea that the server and client can efficiently agree on private statements at the beginning of the protocol. We first present a primitive, called \emph{statement agreement protocol} (SAP), that satisfies the above requirement, and then present the RC-S-P.

\input{statement-agreement}

\vspace{-2mm}

\subsection{Overview of RC-S-P Protocol}
\vspace{-2mm}

We have built the RC-S-P protocol using a novel combination of VSID, SAP, the private time bubble notion,  symmetric-key encryption schemes, the coin masking and padding techniques. At a high level, it works as follows.

The client and server use SAP to provably agree on two private statements; the first statement includes payment details, while another one specifies a secret key, k, and the pads’ length. They also agree on public parameters such as (a) the private time bubble’s length, and (b) a smart contract specifying the parameters and the total amount of masked coins each party should deposit. The client deploys the contract. Each party deposits its masked coins in it. To start using the service, they invoke those VSID algorithms which let the server check if the client has generated its metadata correctly (via NIZK) and the server aborts if it decides not to serve.


At the end of each billing cycle, the client generates an encrypted query using $k$. It pads the encrypted query and sends the result to the contract. 
%
%
In the same cycle, the server retrieves the query and checks its validity (via NIZK). If the query is rejected, the server locally stores the index of the billing cycle and then generates a dummy proof. Otherwise, if the server accepts the query, it generates a proof of service. In either case, the server encrypts the proof, pads it, and sends the result to the contract. After the server sends the messages to the contract, the client extracts and locally verifies the proof. If the verification passes, the client knows the server has delivered the service honestly. If the proof is rejected, it waits until the private time bubble passes and dispute resolution time arrives. 

During the dispute resolution period, if the client or server rejects a proof, it invokes the arbiter, refers it to the invalid encrypted proofs in the contract, and sends to it the decryption key and the pads' detail. The arbiter checks the key and pads validity. If they are accepted, then the arbiter extracts the related proofs and checks the validity of the party's claim. The arbiter sends to the contract a report of its findings that includes the total number of times the server and client provided invalid proofs. To distribute the coins, the client or server sends: (a) ``pay'' message, (b) the agreed statement that specifies the payment details, and (c) the statement’s proof to the contract which verifies the statement and if approved it distributes the coins according to the statement's
detail, and the arbiter's report. Appendix \ref{app::RC-S-P} presents the RC-S-P protocol in detail.

%% file: statement-agreement.tex

\vspace{-2mm}
\subsection{Statement Agreement Protocol (SAP)}\label{SAP}

\vspace{-2mm}


An SAP is secure if it meets four security properties:
\begin{enumerate}[leftmargin=5mm]
\item Neither party can persuade  a third party  verifier that it agreed with its counter-party on an invalid statement, i.e., a statement that not both parties have agreed on.
\item After they agree on the statement,  an honest party can always prove to the verifier that it has the agreement. 
\item The privacy of the statement should be preserved (from the public) before either of the two parties attempts to prove the agreement on the statement. 
\item After both parties reach an agreement, neither can later deny the agreement. 
\end{enumerate}
To that end, we use a combination of a smart contract (including digital signatures involved) and a commitment scheme. The idea is as follows. Let $x$ be the statement. The client picks a random value and uses it to commit to $x$. It sends the commitment to the contract and the commitment opening (i.e., statement and the random value) to the server. The server checks if the opening matches the commitment and if so, it commits to the statement using the same random value and sends its commitment to the contract. Later, for a party to prove to the contract/verifier that it has agreed on the statement with the other party, it only sends the opening of the commitment.  The contract/verifier checks if the opening matches both commitments and accepts if it matches. The SAP protocol is provided below. It assumes that each party $\mathcal{P} \in\{\mathcal{C,S}\}$ already has a blockchain public address $adr_{\st\mathcal{P}}$.

%
%
%
 \begin{enumerate}
 
 \item\textbf{Initiate}. $\mathtt{SAP.init}(1^{\lambda}, adr_{\st\mathcal{C}}, adr_{\st\mathcal{S}},x )$

 The following steps are taken by  $\mathcal C$.
 
  \begin{enumerate}
  \item\label{SAP::deploy-contract}  Deploys a smart contract, SAP, that  states both $adr_{\st\mathcal{C}}$ and  $adr_{\st\mathcal{S}}$. Let $adr_{\st\text{SAP}}$ be the SC's address. 

   \item  Picks a random value $r$, and commits to the statement as $\mathtt{Com}(x,r)=g_{\st \mathcal{C}}$. It sends $adr_{\st\text{SAP}}$ and $\ddot{x}:=(x,r)$  to  $\mathcal S$ and sends $g_{\st\mathcal C}$ to the contract. 
    \end{enumerate}

    \item\textbf{Agreement}. $\mathtt{SAP.agree}(x,r,g_{\st \mathcal{C}},adr_{\st\mathcal{C}},adr_{\st\text{SAP}})$

     The following steps are taken   by  $\mathcal S$.
     
     \begin{enumerate}
 
   \item Checks if $g_{\st \mathcal{C}}$ was from $adr_{\st \mathcal{C}}$ and  $\mathtt{Ver}(g_{\st\mathcal C}, \ddot{x})=1$. 
   \item If the checks pass, it sets $b=1$,   computes $\mathtt{Com}(x,r)=g_{\st\mathcal S}$, and sends $g_{\st\mathcal S}$ to the contract. Else, it sets $b=0$ and $g_{\st\mathcal S}=\bot$.
   
    \end{enumerate}

   \item\textbf{Prove}. For  $\mathcal C$ (resp. $\mathcal S$) to prove that it has an agreement on $x$ with $\mathcal S$ (resp. $\mathcal C$),  it sends $\ddot{x}:=(x, r)$  to the contract.

 \item\textbf{Verify}. $\mathtt{SAP.verify}(\ddot{x}, g_{\st\mathcal C},g_{\st\mathcal S},adr_{\st\mathcal{C}}, adr_{\st\mathcal{S}})$

 The following steps are taken   by  the contract.
   \begin{enumerate}
   
\item\label{SAP::check-adr} Ensures $g_{\st\mathcal C}$ and $g_{\st\mathcal S}$ were sent from   $adr_{\st \mathcal{C}}$ and  $adr_{\st \mathcal{S}}$. 
  
   \item\label{SAP::check-commit} Ensures $\mathtt{Ver}(g_{\st\mathcal C},\ddot{x})=\mathtt{Ver}(g_{\st\mathcal S},\ddot{x}) =1$.
   
   \item Outputs $d=1$, if the checks in  steps  \ref{SAP::check-adr} and \ref{SAP::check-commit}  pass. Otherwise, it outputs $d=0$.
    \end{enumerate}
 \end{enumerate}

\vspace{-.6mm}
   In Appendix \ref{sec:Discussion-on-the-SAP}, we discuss the SAP's security and explain why naive solutions are not suitable.

%% file: PoR-protocols.tex

\section{Recurring Contingent PoR Payment}\label{sec:RC-PoR-P_protocol}

\vspace{-2mm}

In this section, we present recurring contingent PoR payment (RC-PoR-P) that is a concrete instantiation of the RC-S-P, when the verifiable service is PoR. We use PoR in the concrete instantiation of RC-S-P because PoR it is an active area of research and it will let us compare our solution to the-state-of-the-art, i.e., concrete instantiation of zkCSP of Campanelli \textit{et al}. \cite{CampanelliGGN17}.

In RC-PoR-P, instead of the function $F$, we have $F_{\st \mathsf{PoR}}$ which is an algorithm that takes as input $\mathcal{C}$'s encoded file $u^*$ and  $\mathcal{C}$'s query $\mathbf{q}$  and outputs a proof asserting the outsourced data $u$ is retrievable. For instance, if a PoR utilises a Merkle tree, then $F_{\st \mathsf{PoR}}$ is the algorithm that generates the Merkle tree’s proofs.
As a concrete instantiation, RC-PoR-P offers two primary added features. Specifically, unlike the generic RC-S-P construction (cf. Appendix~\ref{app::RC-S-P}), it (a) does not use any zk proofs (even though either $\mathcal{C}$ or $\mathcal{S}$ can be malicious) which significantly improves costs, and (b)  has a much lower  arbiter-side computation cost; as we will show later, this also allows for a smart contract to efficiently play the arbiter's role. Below, we first explain how the features are satisfied.

\noindent\underline{\emph{Avoiding the use of zk proofs}.} The majority of  PoRs assume that only $\mathcal{S}$ is potentially malicious while $\mathcal{C}$ is honest. To ensure a file's availability, they rely on metadata that is either a set of tags (e.g., MACs or signatures) or a root of a Merkle tree,  built on the  file blocks. In the case where $\mathcal{C}$ can also be malicious,  if tags are used then using zk proofs seem  an obvious choice, as it allows $\mathcal{C}$ to guarantee to $\mathcal{S}$ that the tags have been constructed correctly (similar to the PoR in \cite{armknecht2014outsourced}). But, this imposes significant computation and communication costs. We observed that using a Merkle tree would benefit our protocol from a couple of perspectives; in short, it removes the need for zk proofs and it supports proof of misbehaviour. Our first observation is that if a Merkle tree is used, then $\mathcal{S}$ can efficiently check the metadata's correctness by reconstructing this tree on the file blocks, without involving  zk proofs.

\noindent\underline{\emph{Low arbiter-side cost}.}  In a Merkle tree-based PoR, in each verification, the number of proofs (or paths) are linear with the number of blocks that are probed, say $\phi$. In this scheme, the verifier  checks all given proofs and rejects them if only one of them is invalid.   We observed that if this scheme is used in the RC-PoR-P, then once $\mathcal{C}$ finds an invalid proof, it can send only that single invalid proof  as a \emph{proof of misbehaviour} to the arbiter  $\mathcal{R}$\footnote{This idea is akin to the proof of misbehaviour proposed in  \cite{CanettiRR11}.}. This technique significantly reduces $\mathcal{R}$'s computation cost from $\phi \log_{\st 2}(n)$ to $\log_{\st 2}(n)$, where $n$ is the number of file blocks.

The RC-PoR-P scheme (cf. Subsection~\ref{sec::RC-PoR-P-protocol}) deploys the following two building blocks:
\begin{enumerate}[leftmargin=5mm]
\item A PoR scheme, presented in Subsection~\ref{protocol::PoR}, that can be seen as a variant of the standard Merkle tree-based PoR \cite{DBLP:conf/ccs/HaleviHPS11,MillerPermacoin,DBLP:journals/iacr/JuelsK07}.  The security of the construction relies on the security of the underlying Merkle tree and  pseudorandom function (cf. Subsection~\ref{subsec:blocks}).
\item A \emph{statement agreement protocol} (SAP), introduced in Subsection~\ref{SAP}, that lets $\mathcal{S}$ and $\mathcal{C}$ efficiently agree on private statements at the beginning of the RC-PoR-P scheme. The SAP is built upon a binding and hiding commitment scheme, a smart contract, and a secure digital signature scheme used to sign transactions on the blockchain (cf. Subsections~\ref{subsec:blocks} and~\ref{subsec:SC}). 
\end{enumerate}

\input{Modified-Merkle-tree-based-PoR}

\input{Recurring-Contingent-PoR-Payment-protocol-With-Arbiter}

\vspace{-3mm}
\subsection{RC-PoR-P Without Arbiter's Involvement}
\vspace{-2mm}

Due to the efficiency of the arbiter-side algorithm, i.e., $\mathtt{RCSPoR.resolve}(\cdot)$, we can delegate the arbiter’s role to the smart contract, SC.  In this case, the third-party arbiter's involvement is no longer needed.  But, to have the new variant of RC-PoR-P,   we need to adjust the original RC-PoR-P’s protocol and definition, primarily from two perspectives. First, how a party pays to resolve a dispute would change, which ultimately affects the amount of coins each party receives in the coin transfer phase (see below for more details).  
%
%
Second, there would be no need to keep track of the number of times a party unnecessarily raises a dispute, as it pays the contract when it sends a query, before the contract processes its claim. In Appendix \ref{sec::RC-S-P-Definition2}, we provide a generic definition for RC-S-P for the case where the arbiter's role can be played by a smart contract. The generic definition also captures the new variant of RC-PoR-P.

Next, we elaborate on how the original RC-PoR-P protocol can be adjusted such that the third-party arbiter's role is totally delegated to the smart contract, SC.  Briefly, Phases \ref{RC-PoR-P::KeyGen}--\ref{Client-sideProofVerification} remain  unchanged, with an exception. Namely,   in step \ref{setcounters}, only two counters $y_{\st\mathcal{C}}$ and $y_{\st\mathcal{S}}$ are created, instead of four counters; accordingly, in the same step,  vector $\bm{y}$ is now   $\bm{y}: [y_{\st\mathcal C},y_{\st\mathcal S}]$, so counters $y'_{\st\mathcal{C}}$ and $y'_{\st\mathcal{S}}$  are excluded from the vector. At a high level,  the changes applied to phase \ref{RCPoRP::Dispute-Resolution} are as follows:  the parties send their complaints to SC now, SC does not maintain $y'_{\st\mathcal{C}}$ and $y'_{\st\mathcal{S}}$ anymore, SC takes the related steps  (on the arbiter's behalf), and it reads its internal state any time it needs to read data already stored on the contract. Moreover, the main adjustment to phase  \ref{RCPoRP::CoinTransfer} is that the amount of coins each party receives changes. In the RC-PoR-P and RC-S-P (presented in sections \ref{sec::RC-PoR-P-protocol} and \ref{sec::RC-S-P-protocol}, respectively),  the party which raises a dispute does not pay the arbiter when it sends to it the dispute query. Instead, loosely speaking, the arbiter in the coin transfer phase is paid by a misbehaving party. In contrast, when the arbiter’s role is played by a smart contract, the party which raises a dispute and sends the dispute query to the contract (due to the nature of Ethereum smart contracts) has to pay the contract \emph{before} the contract processes its query. This means that an honest party which sends a complaint to the contract needs to be compensated (by the corrupt party) for the amount of coins it sent to the contract to resolve the dispute. 
%
%
In Appendix \ref{sec::R-PoR-P-protocol-Withohut-Arbiter}, we present the modified RC-PoR-P protocol in more detail.

%% file: Modified-Merkle-tree-based-PoR.tex


\vspace{-2mm}
\subsection{Modified Merkle tree-based PoR}\label{protocol::PoR}
\vspace{-2mm}

In this section, we first present a modified version of the standard Merkle tree-based PoR and then explain the applied modifications. At a high level, $\mathcal{C}$ encodes its input file using an error-correcting code, splits the result into blocks, and builds a Merkle tree on the blocks. Then, it locally stores the tree's root and  sends the blocks to $\mathcal{S}$ which rebuilds the tree. At the  verification time, $\mathcal{C}$ sends a $\mathtt{PRF}$'s key to $\mathcal{S}$ which derives a  number of blocks'   indices showing which blocks are probed. For each probed block, $\mathcal{S}$  generates a proof. It sends all proofs to $\mathcal{C}$ which  checks them. If $\mathcal{C}$ accepts all proofs, then it concludes that its file is retrievable. Otherwise, if it rejects some proofs, it stores only one index of the  blocks whose proofs were rejected. Below, we  present the PoR protocol. 

\vspace{.8mm}
\begin{enumerate}[leftmargin=5mm]
\item\textbf{{\small{Client-side Setup}}}. $\mathtt{PoR.setup}(1^{\st\lambda},u)$
 
\begin{enumerate}
\item $\mathcal C$ uses an error-correcting code, to encode  the input file, $u$. Let $u'$ be the encoded file. Then, it splits $u'$  into  $m$ blocks as follows, $u^{\st *}=u^{\st '}_{\st 1}||1,...,u^{\st '}_{\st m}||m$.
\item  $\mathcal C$ constructs a Merkle tree on $u^{\st *}$'s blocks, i.e., $\mathtt{MT.genTree}(u^{\st *})$. Let $\sigma$ be the root of the  tree, and $\phi$ be the number of blocks that will be probed.   It sets public  parameters as $pp:= (\sigma, \phi, m, \zeta)$, where $\zeta$ is a $\mathtt{PRF}$'s description, as  defined in Subsection~\ref{subsec:blocks}. It sends $pp$ and $u^{\st *}$ to $\mathcal{S}$.
\end{enumerate}
\item\label{key-chalenge}\textbf{{\small{Client-side Query Generation}}}. $\mathtt{PoR.genQuery}(1^{\st\lambda}, pp)$
\begin{enumerate}
\item  $\mathcal C$ picks a  key $\hat{k}$ for  $\mathtt{PRF}$ and sends $\hat{k}$ to $\mathcal{S}$.
\end{enumerate}
 %
 %
%
%
%
%
\item\label{PoR-server-prove}\textbf{{\small{Server-side Proof Generation}}}. $\mathtt{PoR.prove}(u^{\st *},\hat{k},pp)$

 \begin{enumerate}
\item $\mathcal S$ derives $\phi$ pseudorandom indices  as follows. 

$\forall i,1\leq i\leq \phi: q_{\st i}=\big(\mathtt{PRF}(\hat{k},i)\bmod m\big)+1$. Note that $1\leq q_{\st i}\leq m$. Let ${\bm{q}}=[q_{\st 1},..., q_{\st \phi}]$. 


\item $\mathcal S$ generates a  proof $\pi_{\st{q_{_{\st i}}}}=\mathtt{MT.prove}(u^{\st*},q_{\st i})$, for each random index $q_{\st i}$. Let the final result  be $ {\bm{\pi}}=[(u^{\st *}_{\st q_{_{\st i}}}, $ $\pi_{\st q_{_{\st i}}})]_{\st q_{\st i}\in {\bm{q}}}$, where $i$-th element in $ {\bm{\pi}}$ corresponds to  $q_{\st i}$, and the probed block is   $u^{\st *}_{\st q_{_{\st i}}}$. It  sends $ {\bm{\pi}}$ to $\mathcal{C}$.
 \end{enumerate}

\item\label{PoR-verify}\textbf{{\small{Client-side Proof Verification}}}. $\mathtt{PoR.verify}( {\bm{\pi}},\bm{q},pp)$ 
\begin{enumerate}

\item If $|\bm{\pi}|=|\bm{q}|=1$, then $\mathcal{C}$ sets $\phi=1$. This step is only for the case where a single proof and query is provided (e.g., in the proof of misbehaviour). 
\item $\mathcal C$ checks if  $\mathcal S$  sent all proofs, by parsing each  element of $\bm\pi$ as:  $\mathtt{parse}(u^{\st *}_{\st q_{_{\st i}}})=u^{\st '}_{\st q_{_{\st i}}}||q_{_{\st i}}$, and  checking if its index $q_{_{\st i}}$ equals to  $ {\bm{q}}$'s  $i$-th element (note, $\mathcal C$ recomputes $ {\bm{q}}$ given $\hat k$). If all checks pass,  it takes the next step. Else,  it outputs $ {\bm{d}}=[0,i]$, where $i$ is the index of  ${\bm{\pi}}$'s   element  that did not pass the check. 

\item $\mathcal C$ checks if every path in  $ {\bm{\pi}}$ is valid, by calling $\mathtt{MT.verify}(u^{\st *}_{\st q_{_{\st i}}}, $ $\pi_{\st q_{_{\st i}}},\sigma)$. If all checks pass, it outputs $ {\bm{d}}=[1,\bot]$; otherwise, it outputs $ {\bm{d}}=[0,i]$, where $i$ refers to the index of the first element in $ {\bm{\pi}}$ that does not pass the check. 
\end{enumerate}
\end{enumerate}



 The above protocol differs from the standard Merkle tree-based PoR from two perspectives; First,  in step \ref{PoR-verify}, $\mathcal C$ also outputs one of the rejected proofs' indices. Given that index (and vectors of proofs and challenges), this will let a third party \emph{efficiently} verify that $\mathcal S$ did not pass the verification. Second,   in step \ref{key-chalenge}, instead of sending $\phi$ challenges, we let $\mathcal C$ send only a key/seed of the $\mathtt{PRF}$  to $\mathcal S$ which can derive a set of challenges from it, such a technique has been used before,  e.g., in \cite{GhoshN19,CramerDI05,DamgardI05}. This will lead to a decrease in the $\mathcal C$'s communication and smart contract's storage costs.

%% file: Recurring-Contingent-PoR-Payment-protocol-With-Arbiter.tex
\vspace{-2mm}

\subsection{RC-PoR-P Protocol}\label{sec::RC-PoR-P-protocol}

\vspace{-2mm}

In this section, we present our RC-PoR-P construction. The RC-PoR-P and the generic RC-S-P design share some ideas, yet as already mentioned, the two constructions have several differences. We provide the overview of the RC-PoR-P scheme and its detailed description below.

In the beginning, $\mathcal{C}$ generates a symmetric encryption key $\bar{k}$ and sets the number of dummy values to pad encrypted proofs, $pad_{\st \pi}$.
  In its setup step, $\mathcal{C}$ runs $\mathtt{PoR.setup}(1^{\lambda}, u)$ to obtain the encoding $u^*$ and the parameters $pp:= (\sigma, \phi, m, \zeta)$. The query/proof secret parameters $qp$ include $(\bar{k},pad_{\st \pi}, pp)$. $\mathcal{C}$ sets the coin secret  parameters $cp:=(o,o_{\st max}, l,  l_{\st max},z)$  (cf. Section~\ref{preliminaries}) that determine $coin^{\st *}_{\st\mathcal C}$ and $p_{\st\mathcal{S}}$, i.e.  the total number of masked coins $\mathcal C$ and $\mathcal S$ must deposit. It initiates two SAP sessions for agreements on $qp$ and $cp$ with $\mathcal{S}$ and deploys a smart contract, SC. It completes setup by providing $\mathcal{S}$ with $u^*$, the SAP parameters (including $qp$ and $cp$), and the number of verifications, $z$, and depositing  $coin^{\st *}_{\st\mathcal C}$ coins in SC.
 In server setup, $\mathcal{S}$  checks whether a sufficient amount of coins has been deposited by $\mathcal C$ and runs the agreement step of the two SAP sessions initiated by $\mathcal{C}$. If agreement is successful and the public parameters $(\sigma, \phi, m)$ verify, it sends  $coin^{\st *}_{\st\mathcal S}=p_{\st\mathcal{S}}$ coins to SC.
 
  After their setup is complete, $\mathcal{C}$ and $\mathcal{S}$ engage in the billing cycles phase for a number of $z$ verifications as follows. During the $j$-th verification, $\mathcal{C}$ runs $\mathtt{PoR.genQuery}$ and sends the output query, $\hat{k}_{\st j}$, encrypted to SC. In turn, $\mathcal{S}$ reads SC and decrypts the encrypted query. If $\hat{k}_{\st j}$ is invalid, it creates a complaint  $m_{\st\mathcal{S},j}$. Else, it runs $\mathtt{PoR.prove}$ to generate a proof  $ {\bm{\pi}}_{\st j}$ for $\hat{k}_{\st j}$. Next, it sends $ {\bm{\pi}}_{\st j}$ encrypted and padded to SC. In order to verify, $\mathcal{C}$ removes the pads and decrypts as  $ {\bm{\pi}}_{\st j}$ and runs  $\mathtt{PoR.verify}$ for $ {\bm{\pi}}_{\st j}$ and $\hat{k}_{\st j}$.  If $ {\bm{\pi}}_{\st j}$ does not pass verification, it creates a complaint $\bm m_{\st\mathcal{C},j}$.
 
Dispute resolution takes place when $\mathcal{C}$ rejects service proofs or $\mathcal{S}$ rejects the queries. The arbiter $\mathcal{R}$ receives the complaint vectors $\bm m_{\st\mathcal{C}}$ and $\bm m_{\st\mathcal{S}}$ from  $\mathcal{C}$ and $\mathcal{S}$ along with each party's ``views'' of the two SAP sessions. Given $\bm m_{\st\mathcal{S}}$ and the view of $\mathcal{S}$, if $\mathcal{S}$'s view is valid, then $\mathcal{R}$ decides for every complaint in  $\bm m_{\st\mathcal{S}}$ by decrypting the corresponding query and executing $\mathcal{S}$'s steps for that query in the billing cycles phase described above. Given $\bm m_{\st\mathcal{C}}$ and the view of $\mathcal{C}$, if $\mathcal{C}$'s view is valid, then $\mathcal{R}$ decides for every complaint in  $\bm m_{\st\mathcal{C}}$ by retrieving the rejected proof's details (included in the complaint), decrypting the related query  and (i) executing $\mathcal{S}$'s steps for that query, (ii) executing $\mathcal{C}$'s verification for the rejected proof and the related query. The arbiter updates SC's state based upon its decisions. Finally, coin transfer is carried out according to the state of SC, as updated by $\mathcal{R}$.

 Before we present the protocol, we discuss how metadata generator function $M_{\st \mathsf{PoR}}$, the pair of encoding/decoding functions $(E_{\st \mathsf{PoR}}, $ $D_{\st \mathsf{PoR}})$ and the query generator function $Q_{\st \mathsf{PoR}}$ (involved in the RC-S-P Definition~\ref{RC-S-P-def}) are defined in the PoR context, as they are often implicit in the original definition of  PoR. Briefly, $M_{\st \mathsf{PoR}}$ is a function that processes a file and generates metadata. For instance, when PoR uses a Merkle tree, then $M_{\st \mathsf{PoR}}$ refers to
 %
  %
   $\mathtt{MT.genTree}(w)\rightarrow(tr, \sigma)$, where $tr$ is the tree constructed on in file $w$  and $\sigma$ is the root of the tree. Encoding by $E_{\st \mathsf{PoR}}$ refers to encrypting with a symmetric key and then adding an appropriate number of pads, while decoding by $D_{\st \mathsf{PoR}}$ refers to removing the pads and then decrypting with the symmetric key.
   Furthermore, $Q_{\st \mathsf{PoR}}$ can be a $\mathtt{PRF}$ that generates a set of pseudorandom strings in a certain range, e.g., file block's indices.

\vspace{-.4mm}

\begin{enumerate}[ leftmargin=.44cm]
\item\label{RC-PoR-P::KeyGen} \textbf{{{Key Generation}}}. $\mathtt{RCPoRP}.\mathtt{keyGen}(1^{\lambda})$
\begin{enumerate}[ leftmargin=.11cm]

\item  $\mathcal C$  picks a fresh symmetric encryption key $\bar{k}\leftarrow\mathtt{SKE.keyGen}(1^{\lambda})$. 

\item  $\mathcal C$  sets  parameter $pad_{\st \pi}$: the number of  dummy values to pad encrypted proofs.  Let $sk':=(pad_{\st \pi}, \bar{k})$. The key's size is part of the security parameter. Let  ${k'}:=(sk',pk')$, where    $pk':=(adr_{\st\mathcal{C}},adr_{\st\mathcal{S}})$. The values of  $pad_{\st \pi}$ is determined as $pad_{\st \pi}=\pi_{\st max}-\pi_{\st act}$, where $\pi_{\st max}$ and $\pi_{\st act}$ refer to the maximum and actual PoR's proof size.

\end{enumerate}

\item\label{RCPoRP::Client-side-Initiation} \textbf{{{Client-side Initiation}}}. $\mathtt{RCPoRP}.\mathtt{cInit}(1^\lambda, u,{k'}, z,pl)$

\begin{enumerate}
%
%
\item\label{RCPoRP::setup} 
Calls  $\mathtt{PoR.setup}(1^{\lambda}, u)\rightarrow (u^{\st *},pp)$ to encode $u$. It   appends $pp:= (\sigma, \phi, m, \zeta)$ and $sk'$ to secret parameters $qp$.

%

\item\label{RC-PoR-P::set-cp}  Sets  coin secret  parameters $cp:=(o,o_{\st max}, l,  l_{\st max},z)$,   then $coin^{\st *}_{\st\mathcal C}=z\cdot (o_{\st max}+l_{\st max})$ and $p_{\st\mathcal{S}}=z\cdot l_{\st max}$, given the price list $pl$, where $coin^{\st *}_{\st\mathcal C}$ and $p_{\st\mathcal{S}}$ are   the total number of masked coins $\mathcal C$ and $\mathcal S$ should deposit. The parameters  $pl,o,o_{\st max}, l,  l_{\st max},z$ are explained in Section \ref{preliminaries}.

\item\label{RCPoRP::set-qp}  Calls $\mathtt{SAP.init}(1^{\st\lambda}, adr_{\st\mathcal{C}}, adr_{\st\mathcal{S}},qp )\rightarrow(r_{\st qp}, g_{\st qp}, adr_{\st\text{SAP}_{\st 1}})$ and $\mathtt{SAP.init}(1^{\st\lambda}, $ $adr_{\st\mathcal{C}}, adr_{\st\mathcal{S}},cp )\rightarrow(r_{\st cp},g_{\st cp},adr_{\st\text{SAP}_{\st 2}})$ to initiate  agreements on $qp$ and $cp$ with $\mathcal{S}$.  Let $T_{\st qp}:=(\ddot{x}_{\st qp}, g_{\st qp})$ and $T_{\st cp}:=(\ddot{x}_{\st cp},g_{\st cp})$,  s.t.  $\ddot{x}_{\st qp}:=(qp,r_{\st qp})$ and $\ddot{x}_{\st cp}:=(cp,r_{\st cp})$ are the openings of $g_{\st qp}$ and $g_{\st cp}$.  Let $T:=\{T_{\st qp},T_{\st cp}\}$.

\item\label{setcounters}  Sets a smart contract, SC, that explicitly specifies  parameters $z$,  $coin^{\st *}_{\st\mathcal C}$, $p_{\st\mathcal{S}}$,  $adr_{\st\text{SAP}_{\st 1}}$,  $adr_{\st\text{SAP}_{\st 2}}$, $pk'$,  including time values $\texttt{Time}:=\{ \texttt{T}_{\st 0},..., \texttt{T}_{\st 2},\texttt{G}_{\st 1,1},...,\texttt{G}_{\st z,2},\texttt{J}, $ $ \texttt{K}_{\st 1},...,\texttt{K}_{\st 6}, \texttt{L}\}$ and a vector $[y_{\st\mathcal C},y'_{\st\mathcal C},y_{\st\mathcal S},  y'_{\st\mathcal S}]$ initialized as $[0,0,0,0]$. It deploys SC. Let $adr_{\st \text{SC}}$ be the address of the deployed SC and $\bm{y}:= [y_{\st\mathcal C},y'_{\st\mathcal C},y_{\st\mathcal S},  y'_{\st\mathcal S}]$.

\item\label{RC-PoR-P::client-deposit} Deposits $coin^{\st *}_{\st\mathcal C}$ coins in the contract. It sends $u^{\st *}, z,  \ddot{x}_{\st qp},$ and $\ddot{x}_{\st cp}$ (along with $adr_{\st \text{SC}}$) to $\mathcal S$. Let $\texttt{T}_{\st 0}$ be the time that the above process finishes.

\end{enumerate}

\item \textbf{{{Server-side Initiation}}}\label{RCPoRP::Server-side-Initiation}. $\mathtt{RCPoRP}.\mathtt{sInit}(u^{\st *},  z, T, p_{\st\mathcal{S}} ,\bm{y})$

\begin{enumerate}

\item Checks the parameters  in $T$ (e.g., $qp$ and $cp$) and  in SC (e.g., $p_{\st\mathcal{S}}, \bm{y}$) and ensures sufficient amount of coins has been deposited by $\mathcal C$.  

\item\label{RCPoRP::Server-side-agree} Calls $\mathtt{SAP.agree}(qp,r_{\st qp},g_{\st qp},adr_{\st\mathcal{C}}, adr_{\st\text{SAP}_{1}})\rightarrow (g'_{\st qp},\\ b_{\st 1})$ and $\mathtt{SAP.agree}(cp,r_{\st cp},g_{\st cp},adr_{\st\mathcal{C}},adr_{\st\text{SAP}_{2}})\hspace{-1mm}\rightarrow (g'_{\st cp}, b_{\st 2})$, to check and agree on $qp$ and $cp$.



 \item\label{RCPoRP::Server-side-serve}  If  $b_{\st 1}=0$ or $b_{\st 2}=0$,  it sets $a=0$. Otherwise, it verifies the  public parameters correctness as follows (i) rebuilds the Merkle tree on $u^{\st *}$ and checks the resulting root equals $\sigma$, and (ii) checks $|u^{*}|=m$ and  $\phi\leq m$, where $(m, \phi)\in T$, and $\sigma\in pp\in T$. If the checks pass,  it sets $a = 1$; else, it sets $a = 0$. It sends $a$ and $coin^{\st *}_{\st\mathcal S}=p_{\st\mathcal{S}}$ coins to SC at time $\texttt{T}_{\st 1}$, where  $coin^{\st *}_{\st\mathcal S}=\bot$ if $a=0$.
\end{enumerate}
$\mathcal S$  and  $\mathcal C$ can withdraw their coins at time $\texttt{T}_{\st 2}$, if  $\mathcal S$ sends $a=0$, fewer coins than $p_{\st\mathcal{S}}$, or nothing to the SC. To withdraw, $\mathcal S$  or  $\mathcal C$  sends a ``pay'' message to $\mathtt{RCPoRP}.\mathtt{pay}(.)$ at time $\texttt{T}_{\st 2}$.


\textit{Billing-cycles Onset}. $\mathcal{C}$ and $\mathcal{S}$ engage in  phases \ref{Billing-cycles-genQuery}-\ref{Client-sideProofVerification},  at the end of every $j$-th billing cycle, where $1\leq j\leq z$. Each $j$-th cycle includes two  time points, $\texttt{G}_{\st j,1}$ and $\texttt{G}_{\st j,2}$, where $\texttt{G}_{\st j,2}>\texttt{G}_{\st j,1}$, and $\texttt{G}_{\st 1,1}>\texttt{T}_{\st 2}$.

\item\textbf{{{Client-side Query Generation}}}.\label{Billing-cycles-genQuery} \\$\mathtt{RCPoRP}.\mathtt{genQuery}(1^\lambda,  T_{\st qp})$

\begin{enumerate}


\item\label{RC-PoR-P::genquery} Calls $\mathtt{PoR.genQuery}(1^{\lambda}, pp)\rightarrow \hat{k}_{\st j}$, where $pp\in T_{\st qp}$.

\item\label{RC-PoR-P::enc-query} Sends encryption $c^{\st *}_{\st j}=\mathtt{Enc}(\bar{k},\hat{k}_{\st j})$ to SC at time $\texttt{G}_{\st j,1}$.

\end{enumerate}

\item\label{RC-PoR-P::Server-side-Proof-Generation}\textbf{Server-side Proof Generation}. \\$\mathtt{RCPoRP}.\mathtt{prove}(u^{\st *},  c^{\st *}_{\st j}, T_{\st qp})$

\begin{enumerate}


\item\label{RC-PoR-P::server-decrypt} Decrypts the query, $\hat{k}_{\st j}=\mathtt{Dec}(\bar{k},c^{\st *}_{\st j})$, where $\bar{k}\in T_{\st qp}$.


\item\label{RC-PoR-P::checkquery} Checks the query's correctness by ensuring $\hat{k}_{\st j}$ is not empty, and is in the key's universe, i.e., $\hat{k}_{\st j}\in \{0,1\}^{\st\psi}$.  If the checks pass,  it sets $b_{\st j}=1$; else, it sets $b_{\st j}=0$.

\begin{itemize}[ leftmargin=.25cm]
\item[$\bullet$] if $b_{\st j}=1$, it sets $m_{\st\mathcal{S},j}=\bot$. It generates proofs vector by calling  $\mathtt{PoR.prove}(u^{\st *}, \hat{k}_{\st j},pp)\rightarrow  {\bm{\pi}}_{\st j}$. Then, it encrypts the proofs, i.e., for $1\leq g\leq| {\bm{\pi}}_{\st j}|: \mathtt{Enc}(\bar{k}, {\bm{\pi}}_{\st j}[g])= {\bm{\pi}}'_{\st j}[g]$. Let $ {\bm{\pi}}'_{\st j}$ contain the encrypted proofs. It pads every encrypted proof in $ {\bm{\pi}}'_{\st j}$ with ${pad}_{\st \pi}\in T_{\st qp}$ random values picked from the encryption's output range, $U$. Let $\bm{\pi}^{\st *}_{\st j}$ be the result. It sends $\bm{\pi}^{\st *}_{\st j}$  to SC at time $\texttt{G}_{\st j,2}$. 
\item[$\bullet$]  if $b_{\st j}=0$, it sets the complaint $m_{\st\mathcal{S},j}=j$. It  constructs a dummy proof $ {\bm{\pi}}'_{\st j}$ with elements randomly picked from $U$,  pads the result as above, and sends the result, $\bm{\pi}^{\st *}_{\st j}$, to SC at time $\texttt{G}_{\st j,2}$.


\end{itemize}
It outputs $b_{\st j}$ and $m_{\st\mathcal{S},j}$. 


\end{enumerate}

\item\label{RC-PoR-P::Client-side-Proof-Verification} \textbf{Client-side Proof Verification}.\label{Client-sideProofVerification} \\$\mathtt{RCPoRP}.\mathtt{verify}(\bm{\pi}^{\st *}_{\st j}, c^{\st *}_{\st j},T_{\st qp})$
\begin{enumerate}


\item\label{RC-PoR-P::Client-side-Proof-Verification-dec-ver} Removes the pads from $\bm{\pi}^{\st *}_{\st j}$, yielding   $\bm{\pi}'_{\st j}$. It decrypts the service proofs $\mathtt{Dec}(\bar{k},\bm{\pi}'_{\st j})=\bm{\pi}_{\st j}$ and then verifies the proof by calling  $\mathtt{PoR.verify}(\bm{\pi}_{\st j}, $ $\hat{k}_{\st j}, {pp})\rightarrow \bm{d}_{\st j}$,  where $\hat{k}_{\st j}=\mathtt{Dec}(\bar{k},c^{\st *}_{\st j})$.  

\begin{itemize}
\item[$\bullet$] if $ {\bm{\pi}}_{\st j}$ passes the verification, i.e., $ {\bm{d}}_{\st j}[0]=1$, it sets $\bm m_{\st\mathcal{C},j}=\bot$ and  concludes that the service for this verification was delivered.  

\item[$\bullet$] otherwise (i.e., $ {\bm{d}}_{\st j}[0]=0$),  it sets $g= {\bm{d}}_{\st j}[1]$ and the complaint $\bm m_{\st\mathcal{C},j}=[j,g]$.
%
%
Recall, $ {\bm{d}}_{\st j}[1]$ refers to a rejected proof's index in proof vector $ {\bm{\pi}}_{\st j}$. 

\end{itemize}
\item It outputs ${\bm{d}}_{\st j}$ and $\bm m_{\st\mathcal{C},j}$. 

\end{enumerate}

\item\textbf{Dispute Resolution}. \label{RCPoRP::Dispute-Resolution} \\$\mathtt{RCPoRP}. \mathtt{resolve}(\bm m_{\st \mathcal{C}}, \bm m_{\st \mathcal{S}}, z, {\bm{\pi}}^{\st *}, {\bm{c}^{\st *}},  T_{\st qp})$

This phase takes place only in  case of dispute, i.e., when $\mathcal C$ rejects service proofs or $\mathcal S$ rejects the queries. 

\begin{enumerate}

\item The arbiter $\mathcal{R}$ ensures  counters: $y_{\st\mathcal C}, y'_{\st\mathcal C},y_{\st\mathcal S}$ and  $y'_{\st\mathcal S}$ are set to $0$,  before time $\texttt{K}_{\st 1}$,  where $\texttt{K}_{\st 1}>\texttt{G}_{\st z,2}+\texttt{J}$. 

\item $\mathcal S$  sends complaints $\bm{m}_{\st\mathcal S}$ and $\ddot{x}_{\st qp}$ to $\mathcal{R}$ at time $\texttt{K}_{\st 1}$.

\item Upon receiving $\bm{m}_{\st\mathcal S}$ and $\ddot{x}_{\st qp}$ , $\mathcal{R}$  takes the following steps at time $\texttt{K}_{\st 2}$. 
\begin{enumerate}

\item\label{RCPoRP::Dispute-Resolution-check-qp} checks $\ddot{x}_{\st qp}$'s validity, by calling the SAP’s verification which returns $d$. If the output is $d=0$,  it discards $\bm m_{\st\mathcal S}$ and does not take steps \ref{RCPoRP::check-v} and \ref{sc-check-server-claim}. Otherwise, it proceeds to the next step.

\item\label{RCPoRP::check-v} removes from $ {\bm{m}}_{\st\mathcal{S}}$ any element  duplicated or not in range $[1,z]$. It  constructs an empty vector $ {\bm{v}}$.

\item\label{sc-check-server-claim}  for any element $i\in {\bm{m}}_{\st\mathcal{S}}$: fetches the related encrypted query $c^{\st *}_{\st i}\in \bm{c}^{\st *}$   from SC and   decrypts it as $\hat{k}_{\st i}=\mathtt{Dec}(\bar{k},c^{\st *}_{\st i})$; it checks  the query by doing the same checks performed in step \ref{RC-PoR-P::checkquery}. If the query is rejected, it increments $y_{\st\mathcal C}$ by $1$ and appends $i$ to $ {\bm{v}}$. If the query is accepted, it increments $y'_{\st\mathcal S}$  by $1$. Let $\texttt{K}_{\st 3}$ be the time the above checks are complete.

\end{enumerate}


\item\label{RC-PoR-P::client-send-complaints} $\mathcal C$  sends complaints $\bm m_{\st\mathcal C}$ and $\ddot{x}_{\st qp}$ to $\mathcal{R}$ at time $\texttt{K}_{\st 4}$.

\item Upon receiving $\bm m_{\st\mathcal C}$ and $\ddot{x}_{\st qp}$, $\mathcal{R}$  takes the below steps at time  $\texttt{K}_{\st 5}$. 
\begin{enumerate}
\item\label{RCPoRP::client-complaint-check-qp} checks $\ddot{x}_{\st qp}$'s validity, by calling the SAP’s verification which returns $d'$. If $d'=0$,  it discards $\bm{m}_{\st\mathcal C}$, and does not take steps \ref{sc-checks-client-claim}-\ref{the-contract-loop}. Otherwise, it proceeds to the next step. 

 \item\label{sc-checks-client-claim} ensures each vector   $ {\bm{m}}\in  {\bm{m}}_{\st\mathcal{C}}$ is valid. Specifically, it checks there exist no two vectors: $ {\bm{m}}, {\bm{m}}'\in  {\bm{m}}_{\st\mathcal{C}}$ such that $ {\bm{m}}[0]= {\bm{m}}'[0]$.  If such vectors  exist, it deletes the redundant ones from $ {\bm{m}}_{\st\mathcal{C}}$. This ensures no two claims refer to the same verification.  It removes any vector $ {\bm{m}}$ from $ {\bm{m}}_{\st\mathcal{C}}$ if $ {\bm{m}}[0]$ is not in the range $[1,z]$ or if $ {\bm{m}}[0]\in  {\bm{v}}$. This check ensures $\mathcal C$ cannot hold $\mathcal S$ accountable if $\mathcal C$  generated an invalid query for the same verification.

\item\label{the-contract-loop} for every  vector $ {\bm{m}}\in {\bm{m}}_{\st \mathcal{C}}$:
\begin{enumerate}[ leftmargin=.22cm]
\item retrieves a rejected proof's details by setting   $j= {\bm{m}}[0]$ and $g= {\bm{m}}[1]$. Recall that $g$ refers to the index of a rejected proof in the proof vector  generated for $j$-th verification, i.e., $ {\bm{\pi}}_{\st j}$.

\item fetches the related encrypted query $c^{\st *}_{\st j}\in \bm{c}^{\st *}$   from SC and   decrypts it: $\hat{k}_{\st j}=\mathtt{Dec}(\bar{k},c^{\st *}_{\st j})$. It removes the pads from $g$-th padded encrypted proof. Let $ {\bm{\pi}}'_{\st j}[g]$ be the result. It decrypts the  encrypted proof, $\mathtt{Dec}(\bar{k}, {\bm{\pi}}'_{\st j}[g]))= {\bm{\pi}}_{\st j}[g]$.

\item identifies the misbehaving party as follows.

\begin{itemize}[ leftmargin=.25cm]

\item verifies $\hat{k}_{\st j}$ by doing the same checks done in step \ref{RC-PoR-P::checkquery}. If the checks do not pass,   it sets  $I_{\st j}=\mathcal C$ and skips the next two steps; otherwise, it proceeds to the next step.

\item\label{RC-PoR-P::derive_key} derives the related challenged block index from $\hat{k}_{\st j}$:  $q_{\st g}=\big(\mathtt{PRF}(\hat{k}_{\st j},g)\bmod m\big)+1$.

\item\label{RC-PoR-P::verify-proof} verifies only $g$-th proof, by calling $\mathtt{PoR.verify}( {\bm{\pi}}_{\st j}[g] , $ $q_{\st g},$ $pp)\rightarrow  {\bm{d}}'$. If $ {\bm{d}}'[0]=0$, it sets $I_{\st j}=\mathcal S$. Else, it sets $I_{\st j}=\bot$.

\item  if $I_{\st j}=\mathcal C$ and $y_{\st\mathcal {C}}$ or $y'_{\st\mathcal {C}}$ was not incremented for $j$-th verification,  it increments $y_{\st\mathcal C}$  by $1$. If $I_{\st j}=\mathcal S$ and $y'_{\st\mathcal {S}}$ was not incremented for $j$-th verification, it increments $y_{\st\mathcal S}$  by $1$.  If  $I_{\st j}=\bot$ and $y_{\st\mathcal {C}}$ was not incremented for $j$-th verification,  it increments $y'_{\st\mathcal {C}}$ by $1$.

\end{itemize}

\end{enumerate}

\end{enumerate}

\item\label{RCPORP::arbiter-to-SC} The arbiter at time $\texttt{K}_{\st 6}$ sends $[y_{\st\mathcal C},y_{\st\mathcal S},  y'_{\st\mathcal C},y'_{\st\mathcal S}]$  to SC which accordingly adds them to $\bm{y}$. 

\end{enumerate} 

\item \textbf{Coin Transfer}. \label{RCPoRP::CoinTransfer} $\mathtt{RCPoRP}.\mathtt{pay}(\bm{y},T_{\st cp},a,p_{\st\mathcal S}, coin^{\st *}_{\st\mathcal C},coin^{\st *}_{\st\mathcal S})$
\begin{enumerate}

 \item\label{RC-PoR-P::pay-t2} If SC  receives  ``pay'' message  at time $\texttt{T}_{\st 2}$, where $a=0$ or $coins^{\st *}_{\st\mathcal{S}}<p_{\st\mathcal{S}}$, then it sends $coin^{\st *}_{\st\mathcal C}$ coins to $\mathcal C$ and $coin^{\st *}_{\st\mathcal S}$ coins to $\mathcal S$. Otherwise (i.e., they reach an agreement), the following step is executed.

\item\label{RC-PoR-P::SC-checks-cp} If SC receives ``pay'' message and  $\ddot{x}_{\st cp}\in T_{\st cp}$ at time $\texttt{L}>\texttt{K}_{\st 6}$, it checks $\ddot{x}_{\st cp}$'s validity by calling the SAP's verification which returns $d''$. 


\item If $d''=1$,  SC distributes the coins to the parties as follows:

\begin{enumerate}
\item   $coin_{\st\mathcal C}=coin^{\st *}_{\st\mathcal C}-o\cdot(z-y_{\st\mathcal S})-l\cdot(y_{\st\mathcal C}+y'_{\st\mathcal C})$ coins  to $\mathcal C$.
\item $coin_{\st\mathcal S}=coin^{\st *}_{\st\mathcal S}+o\cdot(z-y_{\st\mathcal S})-l\cdot(y_{\st\mathcal S}+y'_{\st\mathcal S})$ coins to $\mathcal S$.
\item $coin_{\st\mathcal R}=l\cdot(y_{\st\mathcal S}+y_{\st\mathcal C}+y'_{\st\mathcal S}+y'_{\st\mathcal C})$ coins to $\mathcal R$.

\end{enumerate}

\end{enumerate}
\end{enumerate}

Briefly, the RC-PoR-P protocol’s correctness holds dues to the correctness of PoR, symmetric key encryption, SAP, and smart contract. Appendix \ref{sec::RC-PoR-P-Correctness} presents a more detailed discussion. Below, we state our main theorem on the security of the RC-PoR-P scheme. Appendix \ref{app:RC-PoR-P_proof} presents the theorem's proof. 
\begin{theorem}\label{thm:RC-PoR-P}
The RC-PoR-P scheme with functions $F_{\st \mathsf{PoR}},M_{\st \mathsf{PoR}},E_{\st \mathsf{PoR}},$ $D_{\st \mathsf{PoR}},Q_{\st \mathsf{PoR}}$ described in Subsections~\ref{protocol::PoR} and ~\ref{sec::RC-PoR-P-protocol} is secure (cf. Definition \ref{def::RC-S-P-Security}), if the underlying Merkle tree, pseudorandom function, commitment scheme, and digital signature scheme are secure, and the  symmetric-key encryption scheme is IND-CPA secure.
\end{theorem}

%% file: RC-PoR-P-Evaluation.tex

\vspace{-1mm}

\section{Performance Evaluation of RC-PoR-P}\label{sec:evaluation}
\vspace{-2mm}

In this section, we provide an analysis of the RC-PoR-P protocol. 
%
%
    Table \ref{table::RC-PoR-P-runTime} summarises the protocol's concrete cost (we also provide a table for its asymptotic cost in Appendix \ref{sec::RC-PoR-P-Assymptotic-Cost}). 
Also, we compare RC-PoR-P  with  (a) the zero-knowledge contingent (publicly verifiable) PoR payment  in \cite{CampanelliGGN17} and  the fair PoR payment scheme in \cite{s-PoR} that are more efficient than the state-of-the-art and closest to our work. Table \ref{table::Cost-comp} summarises the comparison. The analysis of RC-PoR-P covers both asymptotic and concrete overheads. To conduct the concrete cost study, we have implemented  RC-PoR-P. The protocol's off-chain and on-chain parts have been implemented in  C++ and Solidity programming languages respectively. To conduct the off-chain experiment, we used  a server with dual Intel Xeon Gold 5118,   2.30 GHz CPU and 256 GB RAM. To carry out the on-chain experiment, we used a MacBook Pro laptop with quad-core Intel Core i5, 2 GHz CPU  and 16 GB RAM that interacts with the Ethereum private blockchain. We ran the experiment $10$ times. In the experiment, we used the SHA-2 hash function and set its output length and the security parameter  to $128$ bits.  We  set the size of every block to $128$ bits, as in \cite{DBLP:conf/asiacrypt/ShachamW08}.   We used a random file whose  size is in the range $[64\ \text{MB},4\ \text{GB}]$. This results in the number of file blocks in the range $[2^{\st 22},2^{\st 28}]$. Since in the  experiment we used  relatively large file sizes, to lower on-chain transaction costs, we allow  the parties to use the technique  explained in Section \ref{sec::RC-S-P-protocol}, which lets the server and client exchange  the (PoR) proofs off-chain in an irrefutable fashion\footnote{\scriptsize For each $j$-th verification, $\mathcal{S}$ sends each related path to $\mathcal{C}$, via an authenticated channel. If $\mathcal{C}$ rejects a path, then it inserts into its complaint  $\mathcal{S}$'s message that includes one of the invalid paths for $j$-th  verification. Our analysis excludes signature generation and verification processes as they can be efficiently incorporated by using standard authenticated channels (e.g., PKI-based XML signatures).}. The prototype implementation utilises the Cryptopp \cite{cryptopp} and  GMP \cite{gmp} libraries. The protocol's off-chain and on-chain source code are publicly available in \cite{RC-S-P-off-chain} and \cite{RC-S-P-on-chain} respectively. 



\input{RC-PoR-P-run-time}
\input{table-PoR-cost}

\vspace{-2mm}

\subsection{Computation Cost}\label{subsec:comp_cost}
\vspace{-2mm}

In our analysis,  the cost of erasure-coding a file is not taken into consideration,  as it is identical in all PoR schemes. We first analyse the computation cost of RC-PoR-P. $\mathcal{C}$'s cost is as follows. In phase \ref{RCPoRP::Client-side-Initiation}, its cost in step \ref{RCPoRP::setup} involves $m\cdot\sum\limits^{\st\log_{\st 2}(m)}_{\st i=1}\frac{1}{2^{\st i}}$ invocations of a hash function. So its complexity in this step is $O(m)$. Its total cost in steps \ref{RC-PoR-P::set-cp} and \ref{RCPoRP::set-qp} involves two invocations of the hash function.  Therefore, the  client-side total complexity in this phase is $O(m)$. In this  phase, its off-chain \emph{run-time} increases about $2\times$ (i.e.,  from $23.1$ to  $45.5,...,$ from $732.1$ to $1596.6$ seconds) when $m$ increases (i.e., from $2^{\st 22}$ to $2^{\st 23},...,$ from $2^{\st 27}$ to $2^{\st 28}$ blocks). This phase also  costs it $123\cdot 10^{\st -5}$ ether.  In phase \ref{Billing-cycles-genQuery}, $\mathcal{C}$ invokes $\mathtt{PRF}$ and symmetric-key encryption $\phi$ times and once respectively.  So, for $z$ verifications its total computation cost is $O(z\cdot \phi)$. Its off-chain run-time in this phase is negligibly small. This phase also costs it $6\cdot 10^{\st -5} \cdot z$ ether.  In phase \ref{RC-PoR-P::Client-side-Proof-Verification}, $\mathcal{C}$ for each verification  decrypts and verifies proofs which mainly involves $\phi\cdot (\log_{\st 2}(m)+1)$ invocations of the symmetric key encryption and $\phi\cdot \log_{\st 2}(m)$ invocations of the hash function. So,  its total complexity in this phase is $O(z\cdot\phi\cdot \log_{\st 2}(m))$. Its off-chain run-time in this phase is very low and  grows  almost $1.1\times$   (i.e., from $0.09 \cdot z$ to $0.11 \cdot z, ...,$ from $0.21\cdot z$ to $0.24\cdot z$ seconds) when $m$ increases.

Now, we analyse $\mathcal{S}$'s computation cost. In phase \ref{RCPoRP::Server-side-Initiation}, $\mathcal{S}$'s complexity is $O(m)$. Its off-chain run-time in this phase grows $2\times$  (i.e., from $8.9$ to $16.5, ...,$ from $248.8$ to $548.8$ seconds)  when $m$ increases. This phase costs it  $9\cdot 10^{\st -5}$ ether. In phase \ref{RC-PoR-P::Server-side-Proof-Generation}, $\mathcal{S}$ decrypts a value for each verification, generates and encrypts proofs  that require $\phi\cdot \log_{\st2}(m)$ invocations of the hash function and $\phi\cdot (\log_{\st2}(m)+1)$ invocations of symmetric key encryption, for each verification. So,   its total complexity in phase \ref{RC-PoR-P::Server-side-Proof-Generation} is $O(z\cdot\phi\cdot \log_{\st 2}(m))$.  In this phase, its off-chain run-time grows  about  $2.1\times$ (i.e., from $22.4\cdot z$ to  $30.4 \cdot z,...,$ from $793.1\cdot z$ to $1820.7\cdot z$ seconds) when $m$ increases.\footnote{\scriptsize To determine $\mathcal{S}$'s cost for generating a proof, we considered the case where $\mathcal{S}$ does not store the Merkle tree nodes (to save storage space), instead it generates the tree's paths every time a challenge is given to it. If we let $\mathcal{S}$  store the tree, then it would have a lower computation overhead.}


Next, we analyse $\mathcal{R}$’s cost in phase \ref{RCPoRP::Dispute-Resolution}. First, we evaluate $\mathcal{R}$'s cost when it is invoked by an honest $\mathcal{S}$.  In this case, it invokes the hash function twice and decrypts  $|\bm{v}_{\st\mathcal{S}}|$ queries, where $|\bm{v}_{\st\mathcal{S}}|$ is the total number of verifications that $\mathcal{S}$ complained about and $|\bm{v}_{\st\mathcal{S}}|\leq z$. Now,  we evaluate its cost when it is invoked by an honest $\mathcal{C}$. It invokes the hash function twice to check the correctness of the statement, $\ddot{x}_{\st qp}$, sent by the client. It invokes the hash function  $|\bm{v}_{\st\mathcal{C}}|\cdot (\log_{\st2}(m)+2)$ times and the symmetric key encryption  $|\bm{v}_{\st\mathcal{C}}|\cdot (\log_{\st2}(m)+2)$ times, where $|\bm{v}_{\st\mathcal{C}}|$ is the total number of verifications that $\mathcal{C}$  complained about. Thus, its cost, in phase  \ref{RCPoRP::Dispute-Resolution} is at most $O(z'\cdot \log_{\st2}(m))$, where $z'=Max(|\bm{v}_{\st\mathcal{C}}|, |\bm{v}_{\st\mathcal{S}}|)$ and $z'\leq z$.  Note that due to the use of the proof of misbehaviour in the protocol, $\mathcal{R}$’s cost is about $\frac{1}{\phi}=\frac{1}{460}$ of its computation cost in the absence of such technique where it has to check all $\phi$ proofs for each verification.\footnote{\scriptsize As shown in \cite{DBLP:conf/ccs/AtenieseBCHKPS07}, to ensure 99\% of file blocks is retrievable, it would be sufficient to set the number of challenged blocks to $460$.} Its off-chain run-time is very low and   increases about $1.3\times$ (i.e., from $2\cdot 10^{\st -5}\cdot z'$ to  $4\cdot 10^{\st -5}\cdot z', ...,$ from $9\cdot 10^{\st -5}\cdot z'$ to $10^{\st -4}\cdot z'$ seconds) when $m$ increases.  Phase  \ref{RCPoRP::Dispute-Resolution} also imposes $10^{\st -4}$ ether to $\mathcal{R}$.  
In phase \ref{RCPoRP::CoinTransfer}, SC  invokes the hash function only twice, so its computation complexity is constant. This phase imposes $6\cdot 10^{\st -5}$ ether to the party that calls  $\mathtt{RCPoRP}.\mathtt{pay}$. 

\vspace{-3mm}
\subsection{Communication Cost}\label{subsec:comm_cost}  
\vspace{-2mm}

We first analyse $\mathcal{C}$'s communication cost. In phase \ref{RCPoRP::Client-side-Initiation}, $\mathcal{C}$ sends  ${||u^{\st *}||}+384$ bits. So,  in this phase, its complexity is $O(||u^{\st *}||)$. In phase \ref{RCPoRP::Dispute-Resolution}, it sends  $(\ddot{x}_{\st qp}, m_{\st\mathcal{C}})$, where $\ddot{x}_{\st qp}$ contains (a) padding information  whose size is a few bits and (b) the symmetric-key encryption's key whose size is  $128$ bits. Also,   $m_{\st\mathcal{C}}$  contains at most $z$ invalid paths of the Merkle tree. Thus, in this phase, its communication cost is  $z\cdot \log_{\st 2}(||u^{\st *}||)+128$ bits or $O(z\cdot \log_{\st 2}(||u^{\st *}||))$.  
$\mathcal{S}$'s complexity for $z$ verifications is $O(z\cdot ||\bm{\pi}^{\st *}_{\st j}||)$, as in phase  \ref{RC-PoR-P::Server-side-Proof-Generation}, for each verification, it sends out a proof vector $\bm{\pi}^{\st *}_{\st j}$. 
%
%
$\mathcal{R}$'s communication cost is constant, as it only sends a transaction containing four values in phase \ref{RCPoRP::Dispute-Resolution}. 

\vspace{-3mm}
\subsection{Comparison} 
\vspace{-2mm}

The fair PoR scheme in \cite{s-PoR} assumes that $\mathcal{C}$ is trusted. 
%
 %
 The initiation phase involves
 $O(z)$ modular exponentiations and $O(m+z\phi)$ modular multiplications to generate puzzles and MACs respectively. Given the puzzles,  $\mathcal{S}$ has to \emph{continuously} solve them sequentially until all $z$ verifications end, which requires $\mathcal{S}$ to perform the exponentiations even between two consecutive verifications. This requires $\mathcal{S}$ to perform $O(Tz)$ exponentiations and $z$ modular multiplications, where $T$ is a time parameter. For $z$ verifications, $\mathcal{S}$ performs $O(z\phi)$ multiplications to generate $z$  proofs. A verifier performs $O(z\phi)$ multiplications to verify all proofs. Now we focus on the scheme in \cite{CampanelliGGN17}.  
 As we showed in Section \ref{sec:attack-mitigation}, this scheme is not secure against a malicious $\mathcal{C}$.  
 %
 %
 In the initiation phase, $\mathcal{C}$ generates a signature for each file block which involves $O(m)$ exponentiations and $O(m)$ hash function invocations. For $\mathcal{S}$ to generate $z$ proofs, it (i) performs $O(z\phi)$ exponentiations to combine the signatures,  (ii) invokes the hash function at least $O(1)$ times, and (iii) invokes zk proof system $O(z\phi)$ times.  The scheme imposes the same computation complexity on the verifier as it does on the prover. Campanelli \textit{et al.} \cite{CampanelliGGN17} provide an implementation of zkCSP for publicly  and privately verifiable PoRs. We were informed by Campanelli that the total size of the outsourced file used in their experiment is at most  $256$ bits, which is very small. In contrast, in our experiment, we used a much large file size, i.e., $4$-GB.

 Since both schemes in \cite{s-PoR} and \cite{CampanelliGGN17} use homomorphic tags, proofs for each verification can be combined resulting in constant proof size, i.e., $O(1)$.  These schemes do not address the privacy issue we highlighted in Section \ref{sec::Lack-of-Privacy}. However, RC-PoR-P is secure against a malicious $\mathcal{C}$ and rectifies the privacy issue. Similar to the other two schemes, its initiation complexity is  $O(m)$; but, unlike them, it does not require any modular exponentiations.  Instead, it involves only invocations of the hash function which imposes a much lower overhead. Moreover, unlike the other two schemes that have $O(z\phi)$ complexity in the prove and verify phases, RC-PoR-P's complexity, in theory, is slightly higher, i.e., it is $O(z\phi \log_{\st 2}(m))$. However, the  extra factor: $\log_{\st 2}(m)$ is not very high in practice. For instance, for a $4$-GB file (or $2^{\st 28}$ blocks), it is only $28$.   RC-PoR-P's prove and verify algorithms, similar to the ones  \cite{s-PoR}, involve only symmetric key operations; whereas, the ones in \cite{CampanelliGGN17} need asymmetric key operations. Also, RC-PoR-P's the proof size complexity is larger than the other two schemes; but, each message length in RC-PoR-P is much shorter than the one in \cite{CampanelliGGN17}, i.e., $128$-bit vs $2048$-bit. 

Thus, RC-PoR-P is computationally more efficient than  \cite{CampanelliGGN17} and  \cite{s-PoR} while offering stronger security guarantees (i.e., security against a malicious client and privacy).

\vspace{2mm}

%% file: RC-PoR-P-run-time.tex

 \begin{table*}[ht]
\begin{footnotesize}
\begin{center}
\caption{ \small RC-PoR-P  off-chain run-time (in seconds) and on-chain cost, of $z$ verifications; breakdown by phases. In the table, $z'$ is the maximum  number of complaints the client and server send to the arbiter,  and $m$ is  the number of blocks in a file.}\label{table::RC-PoR-P-runTime} 
\renewcommand{\arraystretch}{1.3}
{
\setlength{\tabcolsep}{1.2pt}
\begin{tabular}{|c|c|c|c|c|c|c|c|c|c|c|c|c|c|} 
   \hline
\cellcolor[gray]{0.9} &
 \multicolumn{7}{c|}{\cellcolor[gray]{0.9}\scriptsize Off-chain cost}&\multicolumn{2}{c|}{\cellcolor[gray]{0.9}\scriptsize On-chain cost}\\
      \cline{2-10}  
      \cline{2-10}  
\cellcolor[gray]{0.9} \multirow{-2}{*} {\scriptsize Phase}&\cellcolor[gray]{0.9}\scriptsize$m:2^{22}$&\cellcolor[gray]{0.9}\scriptsize$m:2^{23}$&\cellcolor[gray]{0.9}\scriptsize$m:2^{24}$&\cellcolor[gray]{0.9}\scriptsize$m:2^{25}$ &\cellcolor[gray]{0.9}\scriptsize$m:2^{26}$&\cellcolor[gray]{0.9}\scriptsize$m:2^{27}$&\cellcolor[gray]{0.9}\scriptsize$m:2^{28}$&\cellcolor[gray]{0.9}\scriptsize Ether&\cellcolor[gray]{0.9}\scriptsize US Dollar\\
\hline

\multirow{3}{*}{\rotatebox[origin=c]{0}{\scriptsize }} \cellcolor[gray]{0.9}\scriptsize Client-side Init.&\scriptsize$23.1$&\scriptsize$45.5$ &\scriptsize$89.7$&\scriptsize$185.8$&\scriptsize$413$&\scriptsize$732.1$&\scriptsize$1596.6$&\scriptsize$123\cdot 10^{-5}$&\scriptsize$3.42$\\
     \cline{1-10}  
     
 \cellcolor[gray]{0.9}\scriptsize Server-side Init.&\scriptsize$8.9$ &\scriptsize$16.5$&\scriptsize$33.2$&\scriptsize$134.6$&\scriptsize$149.4$&\scriptsize$248.8$&\scriptsize$548.8$&\scriptsize$9\cdot 10^{-5}$ &\scriptsize$0.22$  \\
     \cline{1-10} 
      \cellcolor[gray]{0.9}\scriptsize Client-side Query Gen.&- &-&-&-&-&-&-&\scriptsize$6\cdot 10^{-5}\cdot z$&\scriptsize$0.17\cdot z$ \\
     \cline{1-10} 
       \cellcolor[gray]{0.9}\scriptsize Server-side Proof Gen.&\scriptsize$22.4\cdot z$&\scriptsize$30.4\cdot z$ &\scriptsize$57.4\cdot z$&\scriptsize$166.8\cdot z$&\scriptsize$376.1\cdot z$&\scriptsize$793.1\cdot z$&\scriptsize$1820.7\cdot z$&-&-\\
     \cline{1-10} 
  \cellcolor[gray]{0.9}\scriptsize Client-side Proof Ver.&\scriptsize$ 0.09\cdot z$&\scriptsize$0.11\cdot z$ &\scriptsize$0.12\cdot z$&\scriptsize$0.16\cdot z$&\scriptsize$0.18\cdot z$&\scriptsize$0.21\cdot z$&\scriptsize$0.24\cdot z$&-&-\\
  \cline{1-10} 
    \cellcolor[gray]{0.9}\scriptsize Arbiter-side Dispute Res.&\scriptsize$ 2\cdot 10^{-5}\cdot z'$&\scriptsize$ 4\cdot 10^{-5}\cdot z'$ &\scriptsize$8\cdot 10^{-5}\cdot z'$&\scriptsize$8\cdot 10^{-5}\cdot z'$&\scriptsize$9\cdot 10^{-5}\cdot z'$&\scriptsize$9\cdot 10^{-5}\cdot z'$&\scriptsize$10^{-4}\cdot z'$&\scriptsize$  10^{-4}$&\scriptsize$ 0.27$\\
  \cline{1-10} 
    \cellcolor[gray]{0.9}\scriptsize Coin Transfer&-&-&-&-&-&-&-&\scriptsize$6\cdot 10^{-5}$&\scriptsize$ 0.17$\\
 \hline
\end{tabular}
}
\end{center}
\end{footnotesize}
\end{table*}

%% file: table-PoR-cost.tex
 \begin{table*}
\caption{ \small Contingent PoRs  comparison.  In the table, $T$ is a time parameter, and $\phi$ is the number of challenged blocks.} 
\label{table::Cost-comp} 
\begin{footnotesize}
\begin{center}
\renewcommand{\arraystretch}{1.2}
\setlength{\tabcolsep}{1.2pt}
\begin{tabular}{|c|c|c|c|c|c|c|c|c|c|c|c|c|c|c|c|} 

   \hline
\cellcolor[gray]{.9}&\cellcolor[gray]{.9}&
 \multicolumn{4}{c|}{\cellcolor[gray]{.9}\scriptsize \underline{    \  \  \   \  \    \  \  \  \ \  \  \  \  \  \  \ \  \       Computation Complexity   \        \   \  \    \  \  \  \ \  \  \  \  \  \  \ \  \  \   }}&\cellcolor[gray]{.9}\scriptsize Proof& \multicolumn{2}{c|}{\cellcolor[gray]{.9}\scriptsize \underline{Secure Against Malicious}}&\cellcolor[gray]{.9}\scriptsize Offers\\
  \cellcolor[gray]{.9}\multirow{-2}{*}{\scriptsize Protocols} &\cellcolor[gray]{.9}\multirow{-2}{*} {\scriptsize Operation}&\cellcolor[gray]{.9}\scriptsize$\mathtt{Initiate}$&\cellcolor[gray]{.9}\scriptsize$\mathtt{Solve\ Puzzle}$&\cellcolor[gray]{.9}\scriptsize$\mathtt{Prove}$&\cellcolor[gray]{.9}\scriptsize$\mathtt{Verify}$&\cellcolor[gray]{.9}\scriptsize  Size&\cellcolor[gray]{.9}\scriptsize \ \ \ Client \ \ \ &\cellcolor[gray]{.9}\scriptsize  Server&\cellcolor[gray]{.9}\scriptsize Privacy\\
\hline
     \cline{2-6}   
      
     \hline 
       
          \hline 
          
 \cellcolor[gray]{.9}  &\multirow{2}{*}{\rotatebox[origin=c]{0}{\scriptsize }}\cellcolor[gray]{.9}\scriptsize Exp.&\scriptsize$O(z)$&\scriptsize$ O(T z)$&$-$&$-$&&&&\\
     \cline{2-6}

\cellcolor[gray]{.9}   \multirow{-2}{*}{\rotatebox[origin=c]{0}{\scriptsize   \cite{s-PoR}}}  &\cellcolor[gray]{.9}\scriptsize Add. or Mul.&\scriptsize$O(m+z\phi )$&\scriptsize$O(z)$&\scriptsize$O( z\phi)$&\scriptsize$O(z\phi )$&\multirow{-2}{*} {\scriptsize $O(1)$}&\multirow{-2}{*} {\scriptsize {\textcolor{red}{$\times$}}}&\multirow{-2}{*} {\scriptsize {\textcolor{blue}{$\checkmark$}}}&\multirow{-2}{*} {\scriptsize {\textcolor{red}{$\times$}}}\\   

   \cline{2-6}

%
%
      
      \hline
      
       \hline
  \cellcolor[gray]{.9}    &\multirow{2}{*}{\rotatebox[origin=c]{0}{\  \scriptsize }}\cellcolor[gray]{.9}\scriptsize Exp.&\scriptsize$O(m)$&$-$&\scriptsize$O(z\phi)$&\scriptsize$O(z\phi )$&&&&\\
     \cline{2-6}
 \cellcolor[gray]{.9}&\cellcolor[gray]{.9}\scriptsize Add. or Mul.&\scriptsize$-$&\scriptsize$-$&\scriptsize$O(z\phi )$&\scriptsize$O(z\phi )$&&&&\\ 
    \cline{2-6}
      \cellcolor[gray]{.9}  &\multirow{2}{*}{\rotatebox[origin=c]{0}{\scriptsize }}\cellcolor[gray]{.9}\scriptsize Hash &\scriptsize$O(m)$&\scriptsize$ -$&\scriptsize$O(1)$&\scriptsize$O(1)$&&&&\\
      
         \cline{2-6}
     
      \cellcolor[gray]{.9} \multirow{-4}{*}{\rotatebox[origin=c]{0}{\scriptsize  \cite{CampanelliGGN17}}}   &\multirow{2}{*}{\rotatebox[origin=c]{0}{\scriptsize }}\cellcolor[gray]{.9}\scriptsize ZK proof&\scriptsize$-$&\scriptsize$ -$&\scriptsize$O(z\phi )$&\scriptsize$O(z\phi )$&\multirow{-4}{*} {\scriptsize $O(1)$}&\multirow{-4}{*} {\scriptsize {\textcolor{red}{$\times$}}}&\multirow{-4}{*} {\scriptsize {\textcolor{blue}{$\checkmark$}}}&\multirow{-4}{*} {\scriptsize {\textcolor{red}{$\times$}}}\\
 \hline
  
 
 \hline
\cellcolor[gray]{.9}& \multirow{2}{*}{\rotatebox[origin=c]{0}{\cellcolor[gray]{.9}\scriptsize }} \scriptsize Hash&\scriptsize$O(m)$&\scriptsize$-$&\scriptsize$O(z\phi \log_{\scriptscriptstyle 2}(m))$ &\scriptsize$O(z\phi  \log_{\scriptscriptstyle 2}(m))$&&&&\\
     \cline{2-6}  
 \multirow{-2}{*}{\rotatebox[origin=c]{0}{\cellcolor[gray]{.9}\scriptsize  \text{RC-PoR-P} }}&\cellcolor[gray]{.9}\scriptsize Sym. key enc.&\scriptsize$-$ &\scriptsize$-$&\scriptsize$O(z\phi \log_{\scriptscriptstyle 2}(m))$&\scriptsize$O(z\phi  \log_{\scriptscriptstyle 2}(m))$&\multirow{-2}{*} {\scriptsize $O(\phi\log_{\scriptscriptstyle 2}(m))$}&\multirow{-2}{*} {\scriptsize {\textcolor{blue}{$\checkmark$}}}&\multirow{-2}{*} {\scriptsize {\textcolor{blue}{$\checkmark$}}}&\multirow{-2}{*} {\scriptsize {\textcolor{blue}{$\checkmark$}}}\\
     \cline{2-6}    
     \hline 
 

\end{tabular}  
\end{center}
\end{footnotesize}
\end{table*}

%% file: Conclusion.tex
\vspace{-4mm}
\section{Conclusion}
\vspace{-2mm}

%
%

In this work, we formally defined and proposed a generic blockchain-based construction called ``recurring contingent service payment'' (RC-S-P) that supports a fair exchange of digital coins and verifiable service. RC-S-P is the first scheme of its kind that remains secure if either the client or server is malicious. It also preserves the two parties' privacy. We have also proposed an efficient concrete instantiation of RC-S-P when the service is PoR and implemented this concrete instantiation. Our run-time evaluation has demonstrated that the instantiation is computationally more efficient than the state-of-the-art. 

We also showed that the state-of-the-art fair exchange protocols that have been designed to allow a fair exchange of digital coins and a digital (verifiable) service are susceptible to a free-riding attack. We also showed that these protocols do not preserve the parties' privacy.

As illustrated in this work, a free-rider attacker can directly benefit from the lack of privacy in the schemes we studied. 
The lack of privacy is common among other blockchain-based fair exchange protocols too, e.g., those in \cite{EckeyFS20,DziembowskiEF18,TramerZLHJS17}.  
We hope the attack we described will shed light on the seriousness of the lack of privacy in fair exchange protocols.

%% file: Acknowledgments.tex

\section*{Acknowledgements}
Aydin Abadi and Steven J. Murdoch were supported in part by REPHRAIN: The National Research Centre on Privacy, Harm Reduction and Adversarial Influence Online, under UKRI grant: EP/V011189/1. Steven J. Murdoch was also supported by The Royal Society under grant UF160505. 

%% file: Survery-of-Related-work.tex


\section{Survey of Related Work}\label{Survey-of-Related-Work}

 As stated in the introduction, blockchain technology and in particular smart contracts have the potentials to replace the third party in fair exchange protocols.  Ethereum is the most predominant generic smart contract platform. Although the third-party’s role can be directly encoded/programmed in an Ethereum smart contract, it would not be efficient. Moreover, Bitcoin, as the most popular cryptocurrency, supports smart contracts with very limited functionalities. Therefore, the third party’s full role cannot be directly encoded  in a contract on the Bitcoin blockchain.

\subsection{Zero-knowledge Contingent Payment}  For the first time in \cite{contingent-payment}  it was shown how to construct a fair exchange protocol, called ``zero-knowledge contingent payment’’, that utilises Bitcoin’s smart contract capabilities. The protocol allows a fair exchange of  digital goods and payments over  Bitcoin's network. Its main  security requirement  is that a seller is paid  if and only if  a buyer learns a correct secret.  The protocol uses a feature of  Bitcoin's scripting language, called ``hash-lock transaction''. This type of transaction   lets one  create a payment transaction that specifies a hash value $y$ and allows anyone that can provide its preimage $k$, i.e $\mathtt{H}(k) = y$, to claim the amount of coin  specified in the transaction.  The  contingent payment protocol  in \cite{contingent-payment} works as follows. The seller first picks a secret key, $k$, of  a symmetric-key encryption and uses it to encrypt the secret information, $s$. This yields a ciphertext, $c$. It also computes the key's hash, $y = \mathtt{H}(k)$. The seller sends $c, y$, and a (zero-knowledge) proof to the buyer, where the proof asserts that $c$ is the encryption of $s$ under  key $k$ and  $\mathtt{H}(k) = y$. 

After the buyer verifies and accepts the  proof, it sends a transaction to the blockchain that pays the seller a fixed amount of  coin  if the seller  provides, to the blockchain, a value $k$ such that $\mathtt{H}(k) = y$. Next, the seller  sends $k$ to the blockchain and receives the coins. Now, the buyer can read the blockchain and learn $k$ which allows it to decrypt $c$, and extract the secret,  $s$. Later, after the advancement of the ``succinct non-interactive argument of knowledge'' (zk-SNARK) \cite{GennaroGP013}, that results in a more efficient implementation of zero-knowledge proofs, the contingent payment protocol was modified to use zk-SNARK. However, all zk-SNARKs require a trusted third party for a trusted setup, i.e., to generate a ``common reference string'' (CRS), which means there would be a need for the involvement of  an additional third party in those protocols that use them, including the contingent payment protocol. As such involvement is undesirable, the contingent payment protocol, that uses zk-SNARK, lets the buyer play the role of the third party and generate the parameter.

\subsection{Zero-knowledge Contingent Service Payment}\label{sec::lit-rev-zkcsp}
 Later, Campanelli \textit{et al.}   \cite{CampanelliGGN17} identify a serious security issue of the above contingent payment (that uses zk-SNARK and lets a buyer pick a CRS). In particular, the authors show that a malicious buyer (which generates the CRS) can construct the CRS in a way that lets it learn the secret from the seller's proof without paying the seller. Campanelli \textit{et al.} propose a set of fixes; namely, (a) jointly computing the CRS using a secure two-party computation, (b)  allowing the seller to check the well-formedness of the buyer's CRS, or (c) using a new scheme called  ``zero-knowledge Contingent Service Payments'' (zkCSP). The latter solution  is  a more efficient approach than the other two and offers an additional interesting feature; namely, supporting contingent payment for \emph{digital (verifiable) services}. In short, zkCSP works as follows. Let $\mathtt{v}(.)$ be the verification algorithm for a certain service and $s$ be the service's proof, where if the proof is valid it holds that  $\mathtt{v}(s)=1$.  The parties agree on two claw-free hash functions, e.g., $\mathtt{H}_{\st 1}(.)$ and $\mathtt{H}_{\st 2}(.)$. 
 The seller picks a random value, $r$. Then, it computes either $y=\mathtt{H}_{\st 1}(r)$ if it knows $s$ such that $\mathtt{v}(s)=1$, or  $y=\mathtt{H}_{\st 2}(r)$ otherwise. The seller also generates a witness indistinguishable proof of knowledge (WIPoK), $\pi$, using a compound sigma protocol to prove that it knows either the preimage of $y=\mathtt{H}_{\st 1}(r)$ if it knows a valid $s$, i.e., $\mathtt{v}(s)=1$, or  the preimage of $y=\mathtt{H}_{\st 2}(r)$. Note, due to the witness indistinguishability of $\pi$ and the flaw-freeness of the hash functions, the verifier cannot tell which statement the prover is proving. 
 
 The seller sends the proof along with $y$ to the buyer which first ensures $\pi$ is valid. Then, if the check passes,  the buyer sends to the blockchain a hash-lock transaction that would send $n$ coins  to the party that can provide  $r$ to the blockchain such that $y=\mathtt{H}_{\st 1}(r)$. After a seller provides a valid $r$ to the blockchain it gets paid, accordingly the buyer concludes that it has been served honestly by the seller, as the seller  demonstrated the knowledge of the service proof, $s$. Otherwise (if the seller does not provide a valid $r$) it would not get paid and the buyer learns nothing about $s$. To improve the efficiency of the above zkCSP and to make it practical,  the authors suggest using SNARKs in the setting that the buyer generates the CRS but the seller initially performs minimal efficient checks.  Also, as  concrete instantiations of the zkCSP, the authors propose  two schemes in which the  service is  ``proof of retrievability'' (PoR) \cite{DBLP:conf/asiacrypt/ShachamW08}.  One of the schemes relies on a publicly verifiable PoR and the other one relies on a privately verifiable one.  In these  schemes, the buyer uploads its data to a server and  pays if and only if the server provides  valid proof that asserts the buyer's data is retrievable. 

 \subsection{Known Zero-knowledge Contingent (Service) Payment's Flaw in the Literature}
Fuchsbauer  \cite{Fuchsbauer19} identifies a flaw in the above zkCSP. The author shows that the minimal efficient check that the seller performs in  the zkCSP is not sufficient, because it does not prevent the buyer from cheating and learning the secret. He highlights that the use of computationally expensive verification on the CRS is inevitable to address the issue. Very recently, Nguyen \textit{et al.}  \cite{NguyenAA20} show that by relying on a slightly stronger notion of WI (i.e., trapdoor subversion witness indistinguishability), the zkCSP can remain secure and would not be susceptible to the issues Fuchsbauer pointed out. Moreover, they propose an efficient  scheme that relies on an \emph{interactive}  ZK proof system which is based on garbled circuits and oblivious transfer. However, the two issues (i.e., the attack and the lack of privacy), we highlighted in Section \ref{sec::previous-work-limitations}, are not  identified and addressed in  \cite{Fuchsbauer19,NguyenAA20}.

  \subsection{Using Ethereum Smart Contracts in  Contingent Payment}
    Tramer \textit{et al.}  \cite{TramerZLHJS17} propose a fair exchange scheme that uses a combination of  trusted hardware, i.e., Intel SGX, and Ethereum smart contracts. Interestingly, unlike the common assumption that  secure hardware maintains private states, this scheme relies on weaker security assumptions, i.e., it only relies on the integrity of SGX’s computation and the authenticity of a message it sends. At a high level, in this scheme, the buyer and seller agree on a smart contract and then the buyer  deposits a fixed amount of coin in the smart contract.  Then, the seller sends its messages (that contains proofs) to SGX which verifies the messages’ correctness and then sends its verdict to the smart contract. Next, the contract  distributes the deposit according to the SGX's verdicts. The scheme in addition to achieving  fair exchange wants to ensure that after the parties' initial interaction and after the seller makes an offer, the buyer cannot abort without paying the seller. To this end, in the scheme, the  contract needs to validate SGX's signature (or in general attestation). However, as the authors state, in practice the signature scheme used in SGX  (i.e., EPID signature) is not supported  by standard Ethereum  contracts. Therefore, the suggested technique,  to ensure the buyer cannot abort, remains only of theoretical interest. Also,  in the protocol SGX is always involved, regardless of the parties' behaviour.  
    
    Later, Dziembowski \textit{et al.}  \cite{DziembowskiEF18} propose FairSwap, an efficient protocol for a fair exchange of digital goods (i.e.,   files) and  coins. It is mainly based on the Ethereum smart contracts and the notion of proof of misbehaviour \cite{CanettiRR11}. Briefly, a proof of misbehaviour scheme is usually based on a Merkle tree;  in this scheme,   proving that a party has misbehaved is much cheaper than proving it has behaved honestly.  FairSwap offers two main features: (a) imposes a low computation cost to a smart contract, and (b) avoids using zero-knowledge proofs.  At a high level, FairSwap works as follows. First, the seller and buyer agree on a smart contract. Then, the seller picks a key $k$ (for symmetric-key encryption),   encrypts the secret (i.e., file) under $k$, and sends the ciphertext to the buyer. The seller also commits to $k$ and sends the commitment to the smart contract.   Next, the buyer verifies the correctness of the buyer's messages and if approved, it sends a fixed amount of coin to the smart contract. After that, the seller reveals the opening of the commitment, that contains $k$, to the smart contract. This allows the buyer to read from the contract and learn $k$ with which it can decrypt the ciphertext, extract the secret, and then verify the secret's correctness. In the case where the buyer rejects the secret, it can send a short proof (of misbehaviour) to the contract which performs an efficient verification and distributes the deposit according to the verification's result.  
    
    Very recently, Eckey \textit{et al.} \cite{EckeyFS20} propose OPTISWAP that improves FairSwap’s performance. It also   ensures a malicious seller cannot force the buyer to submit a large transaction to the blockchain, which ultimately imposes transaction costs to the buyer, i.e., the grieving attack. Similar to FairSwap, OPTISWAP uses a smart contract and proof of misbehaviour. Nevertheless, to achieve a better efficiency (than FairSwap),  OPTISWAP uses an \emph{interactive} dispute resolution protocol, previously proposed and used in  \cite{Arbitrum}.  The interactive phase is a challenge-response procedure between the two parties and lets an honest buyer efficiently generate proof of misbehaviour. After computing the proof, the buyer sends it to the contract which verifies the proof and distributes the deposit according to the verification result. To prevent the grieving attack, the protocol requires the seller to deposit coins to the contract as well, which allows the contract to compensate an honest buyer which reports the seller's misbehaviour. 
    
    We highlight that the  protocols in \cite{DziembowskiEF18,EckeyFS20,TramerZLHJS17} have been designed and  are suitable for  a fair exchange of digital items, e.g., file, and digital coins. Nevertheless, they are not suitable for verifiable services, e.g., PoR. If they are \emph{directly} used for verifiable services, then they would suffer from   the two  issues we stated in Section \ref{sec::previous-work-limitations}  (i.e., a malicious client can mount a free-riding attack and lack of privacy). For instance, if they are naively used for PoR, then a malicious client (as a buyer) can simply avoid engaging in the payment protocol with the server (as a seller), even though the server has honestly maintained the buyer's data. This means the client can waste the server's resources. This issue would not  be fully addressed by simply forcing the client to deposit coins at the point where it outsources its data. Because, the client can encode its data in a way that makes the server compute an invalid proof, that ultimately allows the client to withdraw its deposit and avoid paying the server. Moreover, the amount of deposit  leaks non-trivial information about the secret (or the file in the PoR context) in  real-time to the public. 
    
    Very recently, outsourced (fair) PoR schemes that allow a client to delegate the verifications to a smart contract have been proposed in \cite{s-PoR,9436004}. The scheme in \cite{s-PoR} uses  message authentication code (MAC) and time-lock puzzle that results in  low cost in the proof generation and verification phases while the one in \cite{9436004} is based on polynomial commitment and involves a high number of modular exponentiations that lead to higher proof generation and verification cost than the former scheme.  The schemes in  \cite{s-PoR,9436004} do not address the above privacy issue either and rely on a stronger security assumption than the rest of the work studied in this section, as these two protocols assume the client is fully honest while the rest assume either party can be corrupt. 
    
    




%% file: notation-Table.tex


\section{Notations}\label{sec:notation-table}

We summarise our notation in Table \ref{table:notation-table}.

\begin{table}[!h]
\begin{center}

\caption{ \small{Notation Table}.}\label{commu-breakdown-party} 
\renewcommand{\arraystretch}{.97}
\scalebox{0.87}{
\begin{tabular}{|c|c|c|c|c|c|c|c|c|c|c|c|c|c|} 

\hline 

\cellcolor{gray!15}\scriptsize \textbf{Setting} &\cellcolor{gray!15} \scriptsize \textbf{Symbol}&\cellcolor{gray!15} \scriptsize \textbf{Description}  \\
    \hline
    
     \hline

\multirow{29}{*}{\rotatebox[origin=c]{90}{\scriptsize \textbf{Generic}}}

 &\cellcolor{white!20}\scriptsize$z$&\cellcolor{white!20}\scriptsize \text{Number of verifications}\\   
 
&\cellcolor{gray!20}\scriptsize$\lambda$ &\cellcolor{gray!20}\scriptsize Security parameter  \\  

&\cellcolor{white!20}\scriptsize$\mathtt{PRF}$ &\cellcolor{white!20}\scriptsize  Pseudorandom function \\ 

&\cellcolor{gray!20}\scriptsize$\zeta$ &\cellcolor{gray!20}\scriptsize $\mathtt{PRF}$'s description\\ 

&\cellcolor{white!20}\scriptsize$Pr$ &\cellcolor{white!20}\scriptsize Probability\\

 &\cellcolor{gray!20}\scriptsize$\mathtt{Com}$ &\cellcolor{gray!20}\scriptsize  Commit algorithm in commitment\\
  
&\cellcolor{white!20}\scriptsize$\mathtt{Ver}$ &\cellcolor{white!20}\scriptsize  Verify algorithm in commitment\\   
               

&\cellcolor{gray!20}\scriptsize$\mu$ &\cellcolor{gray!20}\scriptsize Negligible function\\  
                      
&\cellcolor{white!20}\scriptsize$\mathtt{H}$ &\cellcolor{white!20}\scriptsize Hash function\\

&\cellcolor{gray!20}\scriptsize$\mathtt{MT}$ &\cellcolor{gray!20}\scriptsize Merkle tree\\                    

&\cellcolor{white!20}\scriptsize${sk,pk}$ &\cellcolor{white!20}\scriptsize Secret and public keys\\     
   
&\cellcolor{gray!20}\scriptsize$\mathtt{PoR}$ &\cellcolor{gray!20}\scriptsize Proof of retrievability\\
     
&\cellcolor{white!20}\scriptsize$u$ &\cellcolor{white!20}\scriptsize Service input, e.g., file\\   
        
&\cellcolor{gray!20}\scriptsize$u^{\scriptscriptstyle*}$ &\cellcolor{gray!20}\scriptsize Encoded  input\\   
     
&\cellcolor{white!20}\scriptsize$\sigma$ &\cellcolor{white!20}\scriptsize Metadata\\ 
           
&\cellcolor{gray!20}\scriptsize$\omega_{\scriptscriptstyle\sigma}$ &\cellcolor{gray!20}\scriptsize Proof for metadata's correctness\\ 
             
&\cellcolor{white!20}\scriptsize$e$ &\cellcolor{white!20}\scriptsize $e:=(\sigma,\omega_{\scriptscriptstyle\sigma})$\\

&\cellcolor{gray!20}\scriptsize$pp$ &\cellcolor{gray!20}\scriptsize Public parameter\\

&\cellcolor{white!20}\scriptsize$q,\bm{q}$ &\cellcolor{white!20}\scriptsize Query and query vector\\

&\cellcolor{gray!20}\scriptsize$\bm{\omega}_{\scriptscriptstyle q}$ &\cellcolor{gray!20}\scriptsize Proof for $\bm{q}$'s correctness\\

&\cellcolor{white!20}\scriptsize$c$ &\cellcolor{white!20}\scriptsize $c:=(\bm{q},\bm{\omega}_{\scriptscriptstyle q})$\\

&\cellcolor{gray!20}\scriptsize$\pi, \bm{\pi}$ &\cellcolor{gray!20}\scriptsize Service proof and proof vector\\

&\cellcolor{white!20}\scriptsize$\mathtt{VS}$ &\cellcolor{white!20}\scriptsize Verifiable service\\

&\cellcolor{gray!20}\scriptsize$\mathtt{VSID}$ &\cellcolor{gray!20}\scriptsize Verifiable service with identifiable abort\\

&\cellcolor{white!20}\scriptsize$\mathtt{RCSP}$ &\cellcolor{white!20}\scriptsize Recurring contingent service payment\\

&\cellcolor{gray!20}\scriptsize$\mathtt{SAP}$ &\cellcolor{gray!20}\scriptsize Statement agreement protocol\\

&\cellcolor{white!20}\scriptsize$\mathcal C$ &\cellcolor{white!20}\scriptsize Client\\

&\cellcolor{gray!20}\scriptsize$\mathcal S$ &\cellcolor{gray!20}\scriptsize Server\\

&\cellcolor{white!20}\scriptsize$\mathcal {R}$ &\cellcolor{white!20}\scriptsize Arbiter\\ 

      &\cellcolor{gray!20}\scriptsize SC  &\cellcolor{gray!20}\scriptsize Smart contract\\


    
\hline

\multirow{11}{*}{\rotatebox[origin=c]{90}{\scriptsize  \textbf{Generic}}}

&\cellcolor{gray!20}\scriptsize${F}$ &\cellcolor{gray!20}\scriptsize Function run on $u^{\scriptscriptstyle *}$  by $\mathcal S$\\

&\cellcolor{white!20}\scriptsize${M}$ &\cellcolor{white!20}\scriptsize Metadata generator  function \\

&\cellcolor{gray!20}\scriptsize${Q}$ &\cellcolor{gray!20}\scriptsize Query generator function\\

&\cellcolor{white!20}\scriptsize$\text{aux}$ &\cellcolor{white!20}\scriptsize  Auxiliary information\\

&\cellcolor{gray!20}\scriptsize$m$ &\cellcolor{gray!20}\scriptsize  Number of a file blocks, $m=|u^{\scriptscriptstyle *}|$\\

&\cellcolor{white!20}\scriptsize$||u^{\scriptscriptstyle *}||$ &\cellcolor{white!20}\scriptsize  Bit size of $u^{\scriptscriptstyle *}$\\

&\cellcolor{gray!20}\scriptsize$\delta$ &\cellcolor{gray!20}\scriptsize  Proof of $F$'s evaluation correctness\\

&\cellcolor{white!20}\scriptsize$j$ &\cellcolor{white!20}\scriptsize  Verification index,  $1\leq j\leq z$ \\

&\cellcolor{gray!20}\scriptsize$adr$ &\cellcolor{gray!20}\scriptsize  Address\\

&\cellcolor{white!20}\scriptsize$\phi$ &\cellcolor{white!20}\scriptsize  Number of challenged blocks\\


\hline 

 \multirow{18}{*}{\rotatebox[origin=c]{90}{\scriptsize \textbf{RC-S-P}\ \  \ \ \ or \ \ \ \ \  \textbf{RC-PoR-P}}}

&\cellcolor{gray!20}\scriptsize$r_{\scriptscriptstyle qp},r_{\scriptscriptstyle cp}$ &\cellcolor{gray!20}\scriptsize  Random values\\
                  
&\cellcolor{white!20}\scriptsize$\ddot{x}_{\scriptscriptstyle qp},\ddot{x}_{\scriptscriptstyle cp}$ &\cellcolor{white!20}\scriptsize  $\ddot{x}_{\scriptscriptstyle qp}:=(qp,r_{\scriptscriptstyle qp}),\ddot{x}_{\scriptscriptstyle cp}:=(cp,r_{\scriptscriptstyle cp})$\\

&\cellcolor{gray!20}\scriptsize$coin^{\scriptscriptstyle *}_{\scriptscriptstyle\mathcal{C}}, coin^{\scriptscriptstyle *}_{\scriptscriptstyle\mathcal{S}}$ 

&\cellcolor{gray!20}\scriptsize  Encoded coins deposited by $\mathcal{C}$ and $\mathcal{S}$\\ 
                   
&\cellcolor{white!20}\scriptsize$enc$ &\cellcolor{white!20}\scriptsize Encoding/decoding functions $enc := (E, D)$\\ 
                    
&\cellcolor{gray!20}\scriptsize$m_{\scriptscriptstyle\mathcal{C}}, m_{\scriptscriptstyle\mathcal{S}}$ &\cellcolor{gray!20}\scriptsize Complaints of $\mathcal{C}$ and $\mathcal{S}$\\ 
                           
&\cellcolor{white!20}\scriptsize$pad_{\scriptscriptstyle \pi},pad_{\scriptscriptstyle q}$ &\cellcolor{white!20}\scriptsize Number of elements used to pad $\pi$ and $q$\\ 
                   
&\cellcolor{gray!20}\scriptsize$y_{\scriptscriptstyle\mathcal{C}},y_{\scriptscriptstyle\mathcal{S}}$ &\cellcolor{gray!20}\scriptsize  Number of times  $\mathcal{C}$ and $\mathcal{S}$ misbehave towards each other\\

&\cellcolor{white!20}\scriptsize$y'_{\scriptscriptstyle\mathcal{C}},y'_{\scriptscriptstyle\mathcal{S}}$ &\cellcolor{white!20}\scriptsize  Number of times  $\mathcal{C}$ and $\mathcal{S}$ unnecessarily invoke $\mathcal{A}r$\\

&\cellcolor{gray!20}\scriptsize${cp}$ &\cellcolor{gray!20}\scriptsize  Coin secret parameters\\ 
               
&\cellcolor{white!20}\scriptsize$T_{\scriptscriptstyle cp}$ &\cellcolor{white!20}\scriptsize  Coin encoding token\\ 
                
&\cellcolor{gray!20}\scriptsize${qp}$ &\cellcolor{gray!20}\scriptsize  Query/proof secret parameters\\ 
                 
&\cellcolor{white!20}\scriptsize$T_{\scriptscriptstyle qp}$ &\cellcolor{white!20}\scriptsize  Query/proof encoding token\\ 
                  
&\cellcolor{gray!20}\scriptsize$T$ &\cellcolor{gray!20}\scriptsize  $T:=(T_{\scriptscriptstyle cp},T_{\scriptscriptstyle qp})$\\

&\cellcolor{white!20}\scriptsize$g_{\scriptscriptstyle\mathcal{C}},g_{\scriptscriptstyle\mathcal{S}}$ &\cellcolor{white!20}\scriptsize  Commitments computed by $\mathcal{C}$ and $\mathcal{S}$ \\

&\cellcolor{gray!20}\scriptsize$pl$ &\cellcolor{gray!20}\scriptsize  Price list: $\{(o,l),...,(o'',l'')\}$\\ 
        
   &\cellcolor{white!20}\scriptsize$o$ &\cellcolor{white!20}\scriptsize  Coins $\mathcal{S}$ must get for a valid proof, where $o\in pl$ \\ 
                 
&\cellcolor{gray!20}\scriptsize$l$ &\cellcolor{gray!20}\scriptsize  Coins  $\mathcal{A}r$ must get for resolving a dispute, where $l\in pl$ \\

     &\scriptsize$l_{\scriptscriptstyle max}$ &\scriptsize   $Max(l,...,l'')$\\            
                
         &\cellcolor{gray!20}\scriptsize$o_{\scriptscriptstyle max}$ &\cellcolor{gray!20}\scriptsize   $Max(o,...,o'')$\\  
                     
&\cellcolor{white!20}\scriptsize$p_{\scriptscriptstyle\mathcal{S}}$ &\cellcolor{white!20}\scriptsize  Total coins $\mathcal{S}$ should deposit\\

         \hline

\end{tabular}\label{table:notation-table}

}
\end{center}
\end{table}


%% file: appendix_preliminaries.tex
\section{Preliminaries}\label{app:preliminaries}

\input{appendix_blocks}

\input{PoR-definition}

%% file: appendix_blocks.tex
\subsection{Details on Building Blocks}\label{app:blocks}

\subsubsection{Pseudorandom Function}\label{subsec:PRF}

Informally, a pseudorandom function ($\mathtt{PRF}$) is a deterministic function that takes as input a key and some argument and outputs a value  indistinguishable from that of a truly random function with the same domain and range.  Pseudorandom functions have many applications in cryptography as they provide an efficient and deterministic way to turn  input into a value that looks random. Below, we restate the formal definition of $\mathtt{PRF}$, taken from \cite{DBLP:books/crc/KatzLindell2007}. 
\begin{definition} Let $W:\{0,1\}^{\st\psi}\times \{0,1\}^{\st \eta}\rightarrow \{0,1\}^{\st  \iota}$ be an efficient  keyed function. It is said $W$ is a pseudorandom function if for all probabilistic polynomial-time distinguishers $B$, there is a negligible function, $\mu(.)$, such that:
\begin{equation*}
\bigg | \Pr[B^{\st W_{\hat{k}}(.)}(1^{\st \psi})=1]- \Pr[B^{\st \omega(.)}(1^{\st \psi})=1] \bigg |\leq \mu(\psi)
\end{equation*}
where  the key, $\hat{k}\stackrel{\st\$}\leftarrow\{0,1\}^{\st\psi}$, is chosen uniformly at random and $\omega$ is chosen uniformly at random from the set of functions mapping $\eta$-bit strings to $\iota$-bit strings.
\end{definition}

\subsubsection{Commitment Scheme}\label{subsec:commit}

 A commitment scheme involves two parties,  \emph{sender} and  \emph{receiver}, and includes  two phases: \emph{commit} and  \emph{open}. In the commit phase, the sender  commits to a message: $x$ as $\mathtt{Com}(x,r)=\mathtt{Com}_{\st x}$, that involves a secret value: $r\stackrel{\st\$}\leftarrow \{0,1\}^{\st\lambda}$. In the end of the commit phase,  the commitment $\mathtt{Com}_{\st x}$ is sent to the receiver. In the open phase, the sender sends the opening $\ddot{x}:=(x,r)$ to the receiver who verifies its correctness: $\mathtt{Ver}(\mathtt{Com}_{\st x},\ddot{x})\stackrel{\st ?}=1$ and accepts if the output is $1$.  A commitment scheme must satisfy two properties: (a) \textit{hiding}: it is infeasible for an adversary (i.e., the receiver) to learn any information about the committed  message $x$, until the commitment $\mathtt{Com}_{\st x}$ is opened, and (b) \textit{binding}: it is infeasible for an adversary (i.e., the sender) to open a commitment $\mathtt{Com}_{\st x}$ to different values $\ddot{x}':=(x',r')$ than that was  used in the commit phase, i.e., infeasible to find  $\ddot{x}'$, \textit{s.t.} $\mathtt{Ver}(\mathtt{Com}_{\st x},\ddot{x})=\mathtt{Ver}(\mathtt{Com}_{\st x},\ddot{x}')=1$, where $\ddot{x}\neq \ddot{x}'$.  There exist efficient non-interactive  commitment schemes both in (a) the standard model, e.g., Pedersen scheme \cite{Pedersen91}, and (b)  the random oracle model using the well-known hash-based scheme such that committing  is : $\mathtt{H}(x||r)=\mathtt{Com}_{\st x}$ and $\mathtt{Ver}(\mathtt{Com}_{\st x},\ddot{x})$ requires checking: $\mathtt{H}(x||r)\stackrel{\st ?}=\mathtt{Com}_{\st x}$, where $\mathtt{H}:\{0,1\}^{\st *}\rightarrow \{0,1\}^{\st\lambda}$ is a collision resistant hash function; i.e., the probability to find $x$ and $x'$ such that $\mathtt{H}(x)=\mathtt{H}(x')$ is negligible in the security parameter $\lambda$.

\subsubsection{Publicly Verifiable Non-interactive Zero-knowledge Proof}\label{subsec:PV-NIZK}

In a non-interactive zero-knowledge proof (NIZK), a prover $\mathcal{P}$, given a witness $w$ for some statement $x$ in an NP language $L$, wants to convince a verifier $\mathcal{V}$ of the validity of $x\in L$. The main security property of the scheme is \emph{Zero-knowledge}; meaning,   a potentially malicious verifier  cannot learn anything beyond the validity of the statement.  The procedure is non-interactive, i.e., $\mathcal{P}$ generates a proof $\pi$ and provides $\mathcal{V}$ with $\pi$, who accepts (or rejects) verification. A NIZK is publicly verifiable when any party by obtaining $\pi$ can verify the validity of $x\in L$. Publicly verifiable NIZKs have been constructed under trust assumptions such as the presence of a common reference string, or setup assumptions such as the existence of a random oracle which is used in this work. For a formal definition of NIZKs we refer the reader to~\cite{DBLP:books/cu/Goldreich2001}. 
 
\subsubsection{Symmetric-key Encryption Scheme} \label{subsec:SKE}

A symmetric-key encryption scheme consists of three algorithms $(\mathtt{SKE.keyGen},  \mathtt{Enc}, \mathtt{Dec})$,  defined as follows. (1) $\mathtt{SKE.keyGen}(1^{\st \lambda})\rightarrow k$ is a probabilistic algorithm that outputs a symmetric key $k$. (2) $\mathtt{Enc}(k,m)\rightarrow c$ takes as input $k$ and a message $m$ in some message space and outputs a ciphertext $c$. (3) $\mathtt{Enc}(k,c)\rightarrow m$ takes as input $k$ and a ciphertext $c$ and outputs a message $m$.

The correctness requirement is that for all messages $m$  in the message space, it holds that 
\[\Pr\Big[\  \  \mathtt{Dec}( k, \mathtt{Enc}(k, m))=m: \mathtt{SKE.keyGen}(1^{\st \lambda})\rightarrow k  \Big]=1\;.\]
 The symmetric-key encryption scheme satisfies \emph{indistinguishability against chosen-plaintext attacks (IND-CPA)}, if any probabilistic polynomial time (PPT) adversary $\mathcal{A}$ has no more than $\frac{1}{2}+\mathsf{negl}(\lambda)$ probability in winning the following game: the challenger generates a symmetric key $\mathtt{SKE.keyGen}(1^{\st \lambda})\rightarrow k$ . The adversary $\mathcal{A}$ is given access to an encryption oracle $\mathtt{Enc}(k,\cdot)$ and eventually sends to the challenger a pair of messages $m_0,m_1$ of equal length. In turn, the challenger chooses a random bit $b$ and provides $\mathcal{A}$ with a ciphertext $\mathtt{Enc}(k,m_b)\rightarrow c_b$. Upon receiving $c_b$,  $\mathcal{A}$ continues to have access to $\mathtt{Enc}(k,\cdot)$ and wins if its guess $b'$ is equal to $b$.

\subsubsection{Digital Signature Scheme}\label{subsec:DS}

A digital signature is a scheme for verifying the authenticity of digital messages. It involves three algorithms, $(\mathtt{Sig.keyGen},  \mathtt{Sig.sign}, $ $\mathtt{Sig.ver})$,  defined as follows. (1) $\mathtt{Sig.keyGen}(1^{\st \lambda})\rightarrow (sk,pk)$ is probabilistic algorithm run by  a  signer that outputs a key pair $(sk,pk)$, consisting of secret key $sk$, and public key $pk$. (2) $\mathtt{Sig.sign}(sk, pk, u)\rightarrow sig$ is an algorithm run by the signer. It takes as input  key pair $(sk,pk)$ and a message $u$. It outputs a signature $sig$. (3) $\mathtt{Sig.ver}( pk, u, sig)\rightarrow h\in\{0,1\}$ is an algorithm run by a verifier. It takes as input  public key $pk$,  message $u$, and signature $sig$. It checks the signature's validity.   If the verification passes, then it outputs $1$; otherwise, it outputs $0$.  

A digital signature scheme should meet two properties. (1) \textit{Correctness:} for every input $u$ it holds that:
\begin{equation*}
\begin{split}
\Pr\Big[  \mathtt{Sig.ver}&( pk, u, \mathtt{Sig.sign}(sk, pk, u))=1:\\
&\mathtt{Sig.keyGen}(1^{\st \lambda})\rightarrow (sk, pk)  \Big]=1
\end{split}
\end{equation*}
(2) \textit{Existential unforgeability under chosen message attacks (EUF-CMA)}: a probabilistic polynomial time PPT adversary that obtains $pk$ and has access to a signing oracle for messages of its choice, cannot create a valid pair $(u^{\st *},sig^{\st *})$ for a new message $u^{\st *}$, except with a negligible probability, $\sigma$. For a formal definition of digital signatures, we refer readers to~\cite{DBLP:books/crc/KatzLindell2007}.

\subsubsection{Merkle Tree}\label{sec::merkle-tree}

In the setting where  a Merkle tree is used to remotely check a file, the file  is  split into blocks and the tree is built on top of the file blocks. Usually, for the sake of simplicity, it is assumed the number of blocks, $m$, is a power of $2$. The height of the tree, constructed on $m$ blocks, is $\log_{\st 2}(m)$. The 
Merkle tree scheme includes three algorithms $(\mathtt{MT.genTree}, \mathtt{MT.prove}, \mathtt{MT.verify})$ as follows: 


\begin{itemize}
\item[$\bullet$] The algorithm that constructs a Merkle tree, $\mathtt{MT.genTree}$, is run by $\mathcal{V}$. It takes file blocks, $u:=u_{\st 1},...,u_{\st m}$, as input. Then, it groups the blocks  in pairs. Next,   a collision-resistant hash function, $\mathtt{H}(.)$, is used to hash each pair. After that, the hash values are grouped in pairs and each pair is further hashed, and this process is repeated until only a single hash value, called ``root'', remains. This yields a  tree with the leaves corresponding to the blocks of the input file and the root corresponding to the last remaining hash value.  $\mathcal{V}$ locally stores the root, and sends the file and tree to $\mathcal{P}$.

\item[$\bullet$] The proving algorithm, $\mathtt{MT.prove}$, is run by $\mathcal{P}$. It takes a block index, $i$, and a tree as inputs. It outputs  
a vector proof, of  $\log_{\st 2}(m)$ elements. The proof asserts the membership of $i$-th block in the tree, and consists of  all the sibling nodes on a path from the $i$-th block to the root of the Merkle tree (including $i$-th block). The proof is given to $\mathcal{V}$.

\item[$\bullet$] The verification algorithm, $\mathtt{MT.verify}$, is run by $\mathcal{V}$. It takes as input $i$-th block, a proof and tree's root. It checks if the $i$-th block corresponds to the root. If the verification passes, it outputs $1$; otherwise, it outputs $0$.

\end{itemize}

The Merkle tree-based scheme has two properties: \emph{correctness} and \emph{security}. Informally, the correctness requires that if both parties run the algorithms correctly, then a proof  is always accepted by  $\mathcal{V}$. The security requires that a computationally bounded malicious $\mathcal{P}$ cannot convince  $\mathcal{V}$ into accepting an incorrect proof, e.g., proof for non-member block. The security  relies on the assumption that it is computationally infeasible to find the hash function's collision.

%% file: PoR-definition.tex

\subsection{Definition of PoR}\label{sec::PoR-definition}

A PoR scheme considers the case where an honest client wants to outsource the storage of its file to a potentially malicious server, i.e., an active adversary. It is a challenge-response interactive protocol, where the server proves to the client that its file is intact and retrievable. Below, we restate PoR's formal definition  initially proposed  in \cite{DBLP:journals/iacr/JuelsK07,DBLP:conf/asiacrypt/ShachamW08}. A PoR scheme comprises of five algorithms: 
\begin{itemize}[leftmargin=5mm]
\item[$\bullet$] $\mathtt{PoR.keyGen}(1^{\st\lambda})\rightarrow k:=(sk,pk)$.  A probabilistic algorithm, run by a client, $\mathcal{C}$. It takes as input the security parameter  $1^{\st\lambda}$. It outputs  private-public verification key, $k:=(sk,pk)$. 
\item[$\bullet$] $\mathtt{PoR.setup}(1^{\st\lambda}, u,k)\rightarrow (u^{\st *}, \sigma, {pp})$. A probabilistic algorithm, run by $\mathcal{C}$. It takes as input  $1^{\st\lambda}$,  a file $u$, and key $k$. It encodes $u$ yielding  $u^{\st *}$ and generates   metadata, $\sigma$.  It outputs  $u^{\st *}$,  $\sigma$, and public parameters $pp$. 
\item[$\bullet$] $\mathtt{PoR.genQuery}(1^{\st\lambda},k,{ pp})\rightarrow \bm{q}$. A probabilistic algorithm, run by $\mathcal{C}$. It takes as input  $1^{\st\lambda}$,  key $k$, and {  public parameters $pp$}. It outputs a  query vector $\bm{q}$, possibly picked uniformly at random.
%
%
\item[$\bullet$] $\mathtt{PoR.prove}(u^{\st *}, \sigma,\bm{q},pk,{ pp})\rightarrow \pi$. It is run by the server, $\mathcal{S}$.  It takes as input the encoded file $u^{\st *}$, metadata $\sigma$, query $\bm{q}$,  public key $pk$, and {  public parameters $pp$}. It  outputs a proof, $\pi$. 
\item[$\bullet$] $\mathtt{PoR.verify}(\pi,\bm{q},k,{ pp})\rightarrow d\in\{0,1\}$. It is run by $\mathcal{C}$. It takes as input $\pi$,   $\bm{q}$,   $k$, and  $pp$. It outputs  $0$ if it rejects the  proof,  or $1$ if it accepts the proof. 
\end{itemize}

A PoR scheme has two  properties: \emph{correctness} and \emph{soundness}. Correctness requires that the verification algorithm accepts  proofs generated by an honest verifier; formally, PoR requires that for any key $k$, any file $u\in\{0,1\}^{\scriptscriptstyle *}$, and any  pair $(u^{\scriptscriptstyle *},\sigma)$ output by $\mathtt{PoR.setup}(1^{\scriptscriptstyle\lambda}, u,k)$, and any query $\bm{q}$, the verifier  accepts when it interacts with an honest prover. Soundness requires that if a prover convinces the verifier (with high probability) then the file is stored by the prover. This is formalized via the notion of an extractor algorithm, that is able to extract the file in interaction with the adversary using a polynomial number of  rounds. Before we define  soundness, we restate the  experiment, defined in \cite{DBLP:conf/asiacrypt/ShachamW08}, that takes place between an environment $\mathcal{E}$ and  adversary $\mathcal{A}$. In this experiment, $\mathcal{A}$ plays the role of a corrupt party and $\mathcal{E}$ simulates an honest party's role. 
\begin{enumerate}
\item $\mathcal{E}$ executes $\mathtt{PoR.keyGen}(1^{\st\lambda})$ algorithm and provides public key, $pk$, to $\mathcal{A}$.   
\item $\mathcal{A}$ can pick  arbitrary file $u$, and  uses it to make queries to  $\mathcal{E}$ who runs  $\mathtt{PoR.setup}(1^{\st\lambda}, u,k)\rightarrow (u^{\scriptscriptstyle *}, \sigma, { pp})$   and returns the output to $\mathcal{A}$. Also, upon receiving the output of $\mathtt{PoR.setup}(1^{\st\lambda}, u,k)$, $\mathcal{A}$ can ask $\mathcal{E}$  to run   $\mathtt{PoR.genQuery}(1^{\st\lambda}$ $, k,{ pp})\rightarrow \bm{q}$ and give the output to it. $\mathcal{A}$ can locally run $\mathtt{PoR.prove}(u^{\scriptscriptstyle *}, \sigma, \bm{q}, $ $pk,{ pp})\rightarrow \pi$ to get its outputs as well. 
\item $\mathcal{A}$ can request from $\mathcal{E}$ the execution of $\mathtt{PoR.verify}(\pi,\bm{q},k, $ ${ pp})$ for any $u$ used to query $\mathtt{PoR.setup}(\cdot)$. Accordingly, $\mathcal{E}$ informs  $\mathcal{A}$ about the verification output. The adversary can send to $\mathcal{E}$ a polynomial number of queries. Finally, $\mathcal{A}$ outputs metadata $\sigma$ returned from a setup query and the description of a prover, $\mathcal{\hat A}$, for any file it has already chosen above. 
\end{enumerate}
It is said that a cheating prover, $\mathcal{\hat A}_{\epsilon}$, is \emph{$\epsilon$-admissible} if it convincingly answers $\epsilon$ fraction of verification challenges (for a certain file). Informally, a PoR scheme supports extractability, if there is an extractor algorithm $\mathtt{Ext}(k,\sigma,\mathcal{\hat A}_{\epsilon})$, that takes as input the key $k$, metadata $\sigma$, and the description  of the  machine implementing the prover's role $\mathcal{\hat A}_{\epsilon}$ and outputs the file, $u$. The extractor has the ability to reset the adversary to the beginning of the challenge phase and repeat this step polynomially many times for the purpose of extraction, i.e., the extractor can rewind $\mathcal{\hat A}_{\epsilon}$.

\begin{definition}[$\epsilon$-soundness]\label{extractable} A PoR scheme is $\epsilon$-sound if   there exists an extraction algorithm $\mathtt{Ext}(\cdot)$ such that, for every adversary $\mathcal{A}$ who plays  experiment $ \mathsf{Exp}^{\mathcal{\scriptscriptstyle A}}_{\scriptscriptstyle \mathsf{PoR}}$ and outputs an $\epsilon$-admissible cheating prover $\mathcal{\hat A}_{\epsilon}$ for a file $u$,  the extraction algorithm  recovers $u$ from $\mathcal{\hat A}_{\epsilon}$, given honest party's private key, public parameters, metadata and the description of $\mathcal{\hat A}_{\epsilon}$,  except with $\mathsf{negl}(\lambda)$ probability. Formally:

\[ \Pr\left[{\small{
  \begin{array}{l}
 \hs  \mathtt{PoR.keyGen}(1^{\scriptscriptstyle\lambda})\rightarrow k:=(sk,pk)\\
\hs  \mathcal{A}(1^{\lambda},pk)\rightarrow u\\
\hs  \mathtt{PoR.setup}(1^{\lambda}, u,k)\rightarrow (u^{\scriptscriptstyle *},\sigma, {pp})\\
\hs  \mathcal{A}(u^{\scriptscriptstyle *},\sigma, {pp})\rightarrow \text{state}\\
 \hs  \mathtt{PoR.genQuery}(1^{\lambda}, k,{pp})\rightarrow \bm{q}\\
 \hs  \Big(\big(\mathcal{A}(\bm{q},\text{state})\rightarrow \pi)\rightleftharpoons (\mathtt{PoR.verify}(\pi,\\\bm{q},k,{pp})\big)\Big) \rightarrow \mathcal{\hat A}_{\epsilon}\\
  \hline
\hs \mathtt{Ext}(k, { pp},\sigma, \mathcal{\hat A}_{\epsilon})\neq u
\end{array}   }}  \right]=\mathsf{negl}(\lambda)\]
\end{definition}

In contrast to the PoR definition in \cite{DBLP:journals/iacr/JuelsK07,DBLP:conf/asiacrypt/ShachamW08} where $\mathtt{PoR.genQuery}(\cdot)$ is implicit, in the above definition we have explicitly defined\\  $\mathtt{PoR.genQuery}(\cdot)$, as it    plays an important role in this paper. Also, there are PoR protocols, e.g., in \cite{MillerPermacoin}, that do not involve $\mathtt{PoR.keyGen}(\cdot)$. Instead, a set of public parameters/keys (e.g., file size or a root of Merkle tree) are output by $\mathtt{PoR.setup}(\cdot)$. To make the PoR definition generic to capture both cases, we have explicitly included the public parameters $pp$ in the algorithms' definitions too.

%% file: proof-status-issue-further-discussion.tex

\section{Further Discussion on Proof Status Leakage}\label{appendix::proof-status-further-discussion}

As we already highlighted, the leakage of proof status (which reveals whether the server is suffering from hardware/software failure) might be problematic in certain circumstances. In Section \ref{sec:attack-mitigation}, we have described a case where such leakage could benefit adversaries. 

However, this is not the only case.  An adversary may directly target business and/or individual clients by mounting social engineering attacks on them. It can exploit such leakage to increase the chance of success in its attack. A survey conducted by Kaspersky lab suggests that 33\% of attacks that affect business clients of cloud servers are of type social engineering \cite{KasperskySurvey} which is a high rate.  Moreover, social engineering attacks on individuals are still prevalent.  In the UK, scammers have been impersonating the Post Office and sending ``extremely convincing'' text messages to the citizens and asking them to use the link provided in the message to schedule redelivery and pay for the redelivery using the given online form. The attackers would steal victims' payment details once they use the form and insert their card details. This message may seem relevant and convincing to a certain percentage of recipients as they are indeed waiting for parcel delivery. A survey conducted by a UK-based consumer protection organisation suggests that 61\% of surveyed people had received such a text message \cite{Scam-Delivery}.

The same scam can be simply adjusted to target users of the cloud servers which are suffering a failure.  Given real-time evidence of servers' failure (that is also known to the clients), an adversary can provide more convincing evidence to their victims as a part of social engineering.  This  ultimately increases the adversary's chance of success. The adversary can target a large set of people  a subset of which is likely to be the cloud's clients  (akin to the above parcel delivery scam), or it can target specific cloud clients by using the techniques used to link the blockchain's addresses to certain parties \cite{AndroulakiKRSC13,BarberBSU12,ReidH11}.

%
%
%
%
%

%% file: VS-definition.tex

\section{Verifiable Service (VS) Definition}\label{subsec:VS}
At a high level, a verifiable service scheme is a two-party protocol in which a client chooses a function, $F$, and provides (an encoding of) $F$, its input $u$, and a query $\bm{q}$ to a server.  The server is expected to evaluate $F$ on $u$ and $\bm{q}$ (and some public parameters) and respond with the output. Then, the client  verifies that the output is indeed the output of the function computed on the provided input. In verifiable services, either the computation (on the input) or both the computation and storage of the input are delegated to the server. A verifiable service is defined as follows.

\begin{definition}[VS Scheme]\label{service-def}
A verifiable service scheme VS $:=(\mathtt{VS.keyGen}, \mathtt{VS.setup},\mathtt{VS.genQuery},$ $\mathtt{VS.prove}, $ $ \mathtt{VS.verify})$ with function $F$, metadata generator function $M$, and query generator function $Q$ consists of five algorithms defined as follows.

\begin{itemize}

\item[$\bullet$] $\mathtt{VS.keyGen}(1^{\lambda})\rightarrow k:=(sk,pk)$.  A probabilistic algorithm run by the client. It takes as input the security parameter $1^{\lambda}$ and outputs a secret/public verification key pair $k$. The server is given $pk$.

\item[$\bullet$] $\mathtt{VS.setup}(1^{\lambda}, u,k)\rightarrow (u^{\st *},\sigma,{pp})$. It is run by the client. It takes as input the security parameter $1^{\lambda}$,  the service input $u$,  and  key pair $k$. If an encoding is needed, then it encodes $u$, that results in $u^{\st *}$; otherwise, $u^*=u$. It outputs encoded input $u^{\st *}$, metadata $\sigma=M(u^{\st *},k,{pp})$, and (possibly input dependent) public parameters {$pp$}. Right after that, the server is  given $u^{\st *}$,  $\sigma$, and $pp$.



\item[$\bullet$] $\mathtt{VS.genQuery}(1^{\lambda}, \text{aux},k,{pp})\rightarrow \bm{q}$. A probabilistic algorithm run by the client. It takes as input the security parameter $1^{\lambda}$, auxiliary information $\text{aux}$,   the key pair $k$,  and {public parameters $pp$}. It outputs  a query vector $\bm{q}=Q( \text{aux},k,{pp})$.  Depending on service types, $\bm{q}$ may be empty or contain only random strings. The output  is given to the server.

\item[$\bullet$] $\mathtt{VS.prove}(u^{\st *},\sigma, \bm{q}, pk,{pp})\rightarrow \bm\pi$. It is run by the server. It  takes as input the service encoded input $u^{\st *}$, metadata $\sigma$,   queries $\bm{q}$, public key $pk$, and { public parameters  $pp$}. It outputs a proof, $\bm\pi = [F(u^{\st *}, \bm{q},{pp}),\delta]$, containing the function evaluation for service input $u$, {public parameters $pp$}, and query $\bm{q}$, i.e., $h=F(u^{\st *},\bm{q},{pp})$, and a proof $\delta$ asserting the evaluation is performed correctly, where generating $\delta$ may involve $\sigma$. The output is given to the client.

\item[$\bullet$] $\mathtt{VS.verify}(\bm\pi, \bm{q}, k, {pp})\rightarrow d\in\{0,1\}$. It is run by the client. It takes as input the proof $\bm\pi$,   query vector $\bm{q}$,   key  $k$, and {public parameters $pp$}.   In the case where $\mathtt{VS.verify}(\cdot)$ is publicly verifiable then $k:=(\bot,pk)$,  and when it is privately verifiable $k:=(sk,pk)$. The algorithm  outputs $d=1$, if the proof is accepted; otherwise, it outputs $d=0$.

\end{itemize}
\end{definition}

A verifiable service scheme has two main properties, \emph{correctness} and \emph{soundness}. Correctness requires that  the verification algorithm always accepts a proof generated by an honest prover.  It is formally stated below.

\begin{definition}[VS Correctness] A verifiable service scheme VS with functions $F,M,Q$ is  correct for  an auxiliary information $aux$, if for any service input $u$ it holds that:

\[
\Pr\left[{\small{
\begin{array}{l}
\mathtt{VS.keyGen}(1^{\lambda})\rightarrow k:=(sk,pk)\\
\mathtt{VS.setup}(1^{\lambda}, u,k)\rightarrow (u^{\st *},\sigma, {pp})\\
\mathtt{VS.genQuery}(1^{\lambda}, \text{aux},k,{pp})\rightarrow \bm{q}\\
\mathtt{VS.prove}(u^{\st *},\sigma, \bm{q},pk, {pp})\rightarrow \bm\pi\\
\hline
\mathtt{VS.verify}(\bm\pi, \bm{q}, k, {pp})\rightarrow 1.\\
\end{array}
}}\right]=1
\]

\end{definition}

Intuitively, a verifiable service is sound  if a malicious server cannot convince the verification algorithm to accept an incorrect output of $F$ except with negligible probability. Soundness is formally stated as follows.



\begin{definition}[VS Soundness] A verifiable service VS with functions $F,M,Q$ is sound for an auxiliary information \text{aux}, if for any probabilistic polynomial time adversary $\mathcal{A}$, it holds that:

\[\Pr\left[{\small{
    \begin{array} {l}
 \mathtt{VS.keyGen}(1^{\st\lambda})\rightarrow k:=(sk,pk)\\
 \mathcal{A}(1^{\lambda},pk, F,M,Q)\rightarrow u\\
\mathtt{VS.setup}(1^{\lambda}, u,k)\rightarrow (u^{\st *},\sigma, {pp})\\
 \mathtt{VS.genQuery}(1^{\lambda}, \text{aux},k,{pp})\rightarrow \bm{q}\\
 \mathcal{A}(\bm{q},u^{\st *},\sigma,{pp})\rightarrow \bm\pi=[h,\delta]\\
 \mathtt{VS.verify}(\bm\pi,\bm{q},k,{pp})\rightarrow d\\
 \hline
 F(u^{\st *},\bm{q},{pp})\neq h \wedge d=1
\end{array} }}    \right]=\mathsf{negl}(\lambda)\]

\end{definition}

The above generic definition captures the core requirements of a wide range of verifiable services such as verifiable outsourced storage, i.e., Proofs of Retrievability \cite{DBLP:journals/iacr/JuelsK07,DBLP:conf/asiacrypt/ShachamW08} or Provable Data Possession \cite{DBLP:conf/ccs/AtenieseBCHKPS07,ShenT11}, verifiable computation \cite{GennaroGP10,LaiDPW14}, verifiable searchable encryption \cite{9075374,LiuLLJL18}, and verifiable information retrieval \cite{ZhangS14,WangZ18}, to name a few. Other additional  security properties (e.g., privacy) mandated by certain services can be added to the above definition. Alternatively, the definition can be upgraded to capture the  additional requirements.  The verifiable service with identifiable abort (VSID) and recurring contingent service payment (RC-S-P) definitions presented in this paper are two examples.


\begin{remark}
It is not hard to see that the original PoR definition (presented in Section \ref{PoR-def}) is a VS's special case. In particular, PoR's $\epsilon$-soundness captures VS's soundness;  in $\epsilon$-soundness, the extractor algorithm interacts (many times) with the cheating prover  which must not be able to persuade the extractor to accept an invalid proof with a high probability and should provide  accepting proofs for non-negligible $\epsilon$ fraction of verification challenges.  The former property is exactly what VS soundness states. 
\end{remark}



%% file: appendix_VSID.tex
\section{Verifiable Service with Identifiable Abort (VSID)}\label{app:VSID}

A protocol that realises only VS's definition (cf. Appendix~\ref{subsec:VS}) would be merely secure against a malicious server and  assumes the client is honest.  Although this  assumption would suffice in certain settings and has been used before (e.g., in \cite{ShenT11,9075374}), it is rather strong and not suitable in the real world, especially when there are monetary incentives (e.g., service payment)  that encourage a client to misbehave. Therefore, in the following we enhance the VS notion to allow  (a) either party to be malicious and (b) a trusted third party, \emph{arbiter}, to identify a corrupt party. We call an upgraded verifiable service scheme with these features \emph{verifiable service with identifiable abort} (VSID), inspired by the notion of secure multi-party computation with identifiable abort \cite{DBLP:conf/crypto/IshaiOZ14}. 

\input{VSID-definition.tex}

\input{VSID-Protocol}

%% file: VSID-definition.tex

\subsection{VSID Definition}\label{subsec:VSID}

The definition of a VSID scheme is provided below.

\begin{definition}[VSID Scheme]\label{service-def} A verifiable service with identifiable abort  \\VSID $:=(\mathtt{VSID.keyGen}, $ $\mathtt{VSID.setup}, $ $\mathtt{VSID.serve},$ $\mathtt{VSID.genQuery}$,\\ $\mathtt{VSID.checkQuery},$ $\mathtt{VSID.prove},$ $\mathtt{VSID.verify}, \mathtt{VSID.identify})$\\ with function $F$, metadata generator function $M$, and query generator function $Q$ involves four entities; namely, client, server, arbiter, and bulletin board. It  consists of eight algorithms defined below.

\begin{itemize}
\item[$\bullet$] $\mathtt{VSID.keyGen}(1^{\lambda})\rightarrow k:=(sk, pk)$.  A probabilistic algorithm run by the client $\mathcal{C}$. It takes as input the security parameter $1^{\st\lambda}$ and outputs a secret/public verification key pair $k$. It sends $pk$ to the bulletin board.


\item[$\bullet$] $\mathtt{VSID.setup}(1^{\lambda}, u,k)\rightarrow (u^{\st *}, {pp},e)$. It is run by the client. It takes as input the security parameter $1^{\st\lambda}$,  the service  input $u$, and the key pair $k$. If an encoding is needed, then it encodes $u$, that results $u^{\st *}$; otherwise, $u^{\st *}=u$. It outputs  $u^{\st *}$, {(possibly file dependent) public parameters $pp$} and $e:=(\sigma,w_{\sigma})$, where $\sigma=M(u^{\st *},k, {pp})$ is  metadata and $w_{\sigma}$ is a proof asserting the metadata is well-structured.  It sends the output (i.e., $u^{\st *}, {pp},e$)  to the bulletin board.

\item[$\bullet$] $\mathtt{VSID.serve}(u^{\st *},e,pk, {pp})\rightarrow a\in\{0,1\}$. It is run by the server $\mathcal{S}$. It takes as input the encoded service input $u^{\st *}$, the pair $e:=(\sigma,w_{\sigma})$, public key $pk$, and {public parameters $pp$}. It outputs $a=1$, if the proof $w_{\sigma}$ is accepted, i.e., if the metadata is well-formed. Otherwise, it outputs $a=0$. The output is sent to the bulletin board.

\item[$\bullet$] $\mathtt{VSID.genQuery}(1^{\lambda},  \text{aux}, k,{pp})\rightarrow c:=(\bm{q},\bm{w}_{\st q})$. A probabilistic algorithm run by the client. It takes as input the security parameter $1^{\lambda}$, auxiliary information $ \text{aux}$,  the key pair $k$,   and public parameters $pp$. It outputs a pair $c$ containing a query vector, $\bm{q}=Q( \text{aux},k,{pp})$,  and proofs, $\bm{w}_{\st q}$, proving the queries are well-structured. Depending on service types, $c$ might be  empty or  contain only random strings. It sends $c$ to the bulletin board.

\item[$\bullet$] $\mathtt{VSID.checkQuery}(c, pk,{pp})\rightarrow b\in\{0,1\}$. It is run by the server. It takes as input a pair $c:=(\bm{q}, \bm{w}_{\st q})$ including queries and their proofs, as well as public key $pk$, and {public parameters $pp$}. It outputs $b = 1$ if the proofs $\bm{w}_{\st\sigma}$ are accepted, i.e., the queries are well-structured. Otherwise, it outputs $b = 0$.

\item[$\bullet$] $\mathtt{VSID.prove}(u^{\st *},\sigma,c,pk, { pp})\rightarrow \bm\pi$. It is run by the server. It takes as input the  encoded service input $u^{\st *}$, metadata $\sigma$, a pair $c:=(\bm{q},\bm{w}_{\st q})$,  public key $pk$, and { public parameters $pp$}. It outputs a proof, $\bm\pi=[F(u^{\st *},\bm{q},{ pp}), \delta)]$ containing the function evaluation, i.e., $h=F(u^{\st *},\bm{q},{ pp})$, and a proof $\delta$ asserting the evaluation is performed correctly, where computing $h$ may involve $pk$ and computing $\delta$ may involve $\sigma$. It sends $\bm\pi$ to the board.

\item[$\bullet$] $\mathtt{VSID.verify}(\bm\pi, \bm{q}, k, { pp})\rightarrow d\in\{0,1\}$. It is run by the client. It takes as input the proof $\bm\pi$, queries $\bm{q}$,  key pair $k$, and  { public parameters pp}. If the proof is accepted, it outputs $d=1$; otherwise, it outputs $d=0$.

\item[$\bullet$]  $\mathtt{VSID.identify}(\bm\pi,c,k,e,u^{\st *}, { pp})\rightarrow I\in \{ \mathcal{C},\mathcal{S},\bot\}$. It is run by a third party arbiter. It takes as input the proof $\bm\pi$, query pair $c:=(\bm{q},\bm{w}_{\st q})$,  key pair $k$, metadata pair  $e:=(\sigma,w_{\sigma})$,  $u^{\st *}$, and { public parameters pp}. If   proof $w_{\st\sigma}$ or $\bm{w}_{\st q}$  is rejected, then it outputs $I=\mathcal C$; otherwise, if proof $\bm\pi$ is rejected it outputs $I=\mathcal S$.  Otherwise, if  $w_{\st\sigma}, \bm{w}_{\st q}$, and $\bm\pi$ are accepted, it outputs   $I=\bot$.

\end{itemize}
\end{definition}

A VSID scheme  has four main properties; namely, it is (a) correct, (b) sound, (c)  inputs of clients are well-formed, and (d) a corrupt party can be identified by an arbiter, i.e., detectable abort. In the following, we formally define each of them.  Correctness requires that  the verification algorithm always accepts a proof generated by an honest prover and both parties are identified as honest.  It is formally stated as follows.

\begin{definition}[VSID Correctness] \label{deff::VSID-Correctness}  
A verifiable service with identifiable abort scheme with functions $F,M,Q$ is correct for an auxiliary information $\text{aux}$, if for any service input $u$ it holds that: 

\[
\Pr\left[{\small{
\begin{array}{l}
\mathtt{VSID.keyGen}(1^{\lambda})\rightarrow k:=(sk,pk)\\
\mathtt{VSID.setup}(1^{\lambda}, u,k)\rightarrow (u^{\st *},{pp},e)\\
\mathtt{VSID.serve}(u^{\st *},e,pk,{pp})\rightarrow a\\
\mathtt{VSID.genQuery}(1^{\lambda}, \text{aux},k,{pp}) \rightarrow c\\
\mathtt{VSID.checkQuery}(c, pk,{pp})\rightarrow b\\
\mathtt{VSID.prove}(u^{\st *},\sigma,c,pk,{pp})\rightarrow \bm\pi\\
\mathtt{VSID.verify}(\bm\pi,\bm{q},k,{pp})\rightarrow d\\
\hline
\mathtt{VSID.identify}(\bm\pi,c,k,e,u^{\st *},{pp})\rightarrow I=\bot \  \  \wedge \\ a=1 \ \ \wedge \  b=1 \ \ \wedge \  d=1
\end{array}
}}\right]=1
\]
\end{definition}

Intuitively, a VSID is sound  if a malicious server cannot convince the client to accept an incorrect output of $F$ except with negligible probability.  It is  formally stated as follows.

\begin{definition}[VSID Soundness]\label{deff::VSID-Soundness}  A VSID  with functions $F,M,Q$ is sound for an auxiliary information \text{aux}, if for any probabilistic polynomial time adversary $\mathcal{A}$, it holds that the following probability is $\mathsf{negl}(\lambda)$:
\[\Pr\left[{\small{
    \begin{array}{l}
    \mathtt{VSID.keyGen}(1^{\lambda})\rightarrow k:=(sk,pk)\\
    \mathcal{A}(1^{\lambda},pk, F,M,Q)\rightarrow u\\
    \mathtt{VSID.setup}(1^{\lambda}, u, k)\rightarrow (u^{\st *},{pp},e)\\
    \mathtt{VSID.genQuery}(1^{\lambda},  \text{aux},k,{pp})\rightarrow c:=(\bm{q}, \bm{w}_{q})\\
     \mathcal{A}(c,e, u^{\st *},{pp})\rightarrow \bm\pi=[h,\delta]\\
     \mathtt{VSID.verify}(\bm\pi,\bm{q},k,{pp})\rightarrow d\\
     \hline
     F(u^{\st *}, \bm{q},{pp})\neq h \wedge d=1
\end{array}   } }\right]\]
\end{definition}

A VSID has   well-formed inputs, if a malicious client cannot persuade a server to serve it on  ill-structured inputs (i.e., to accept  incorrect outputs of $M$ or $Q$). Below, we state the property formally.

\begin{definition}[VSID Inputs Well-formedness]\label{deff::VSID-Inputs-Well-formedess}  A  VSID with functions $F,M,Q$  has  well-formed inputs for  an auxiliary information \text{aux},  if for  any probabilistic polynomial time adversary $\mathcal{A}$, it holds that the following probability is $\mathsf{negl}(\lambda)$:

 
\[ \Pr\left[{\small{
    \begin{array}{l}    
    \mathcal{A}(1^{\lambda},F,M,Q)\rightarrow \Big(u^{\st *},k:=(sk,pk),\\e:=(\sigma,w_{\sigma}),{pp}\Big)\\
    \mathtt{VSID.serve}(u^{\st *},e,pk,{pp})\rightarrow a\\
       \mathcal{A}(1^{\lambda},  \text{aux},k,{pp})\rightarrow c:=(\bm{q},\bm{w}_{\st q})\\
    \mathtt{VSID.checkQuery}(c, pk,{pp})\rightarrow b\\
    \hline
      (M(u^{\st *},k,{pp})\neq \sigma \wedge a=1) \vee\\(Q(\text{aux},k,{pp})\neq \bm{q} \wedge  b=1)\\
\end{array}}  }  \right]\]

\end{definition}

The   above property ensures an honest server can detect  a malicious client if the client provides ill-structured inputs. It is further required that a malicious party  be identified by an honest third party, arbiter. This ensures that in the case of dispute (or false accusation) a malicious party can be pinpointed. A VSID supports  detectable abort if a corrupt party can escape from being identified, by the arbiter,  with only negligible probability.  Formally:

\begin{definition}[VSID Detectable Abort]\label{def::VSID-Identifiable-Abort}  A  VSID  with functions $F,M,Q$ supports  detectable abort for an auxiliary information \text{aux}, if  the following hold: 

\begin{enumerate}
\item For any PPT adversary $\mathcal{A}_{\st 1}$, the following probability is $\mathsf{negl}(\lambda)$:

\[\Pr\left[{\small{
    \begin{array}{l}
    \mathtt{VSID.keyGen}(1^{\st\lambda})\rightarrow k:=(sk,pk)\\
    \mathcal{A}_{\st 1}(1^{\lambda},pk, F,M,Q)\rightarrow u\\
    \mathtt{VSID.setup}(1^{\lambda}, u, k)\rightarrow (u^{\st *},pp,e)\\
    \mathtt{VSID.genQuery}(1^{\lambda},  \text{aux},k,{pp})\rightarrow c:=(\bm{q}, \bm{w}_{q})\\
     \mathcal{A}_{\st 1}(c,e, u^{\st *},{pp})\rightarrow \bm\pi=[h,\delta]\\
     \mathtt{VSID.verify}(\bm\pi,\bm{q},k,{pp})\rightarrow d\\
     \mathtt{VSID.identify}(\bm\pi,c,k,e,u^{\st *},{pp})\rightarrow I\\
     \hline
  d=0 \wedge I\neq\mathcal{S}\\  
\end{array} } }  \right]\]

\item For any PPT adversary $\mathcal{A}_{\st 2}$, the following probability is $\mathsf{negl}(\lambda)$:

\[ \Pr\left[{\small{
    \begin{array}{l}
    \mathcal{A}_{\st 2}(1^{\lambda},F,M,Q)\rightarrow \Big(u^{\st *},k:=(sk,pk),\\e:=(\sigma,w_{\sigma}),{pp}\Big)\\
    \mathtt{VSID.serve}(u^{\st *},e,pk,{pp})\rightarrow a\\
       \mathcal{A}_{\st 2}(\text{aux},k)\rightarrow c:=(\bm{q},\bm{w}_{\st q})\\
    \mathtt{VSID.checkQuery}(c, pk,{pp})\rightarrow b\\
    \mathtt{VSID.prove}(u^{\st *},\sigma,c,pk,{pp})\rightarrow \bm\pi\\
    \mathtt{VSID.identify}(\bm\pi,c,k,e,u^{\st *},{pp})\rightarrow I\\
    \hline
     (a=0 \vee b=0)\wedge I\neq\mathcal{C}\\
\end{array}  } } \right]\]

\end{enumerate}
\end{definition}

\subsubsection{Lighter VSID Scheme ($\text{VSID}_{\st\text{light}}$)}\label{remark::light-VSID}
In the VSID  definition, algorithm  $\mathtt{VSID.identify}(\cdot)$    allows an arbiter  to identify a misbehaving client even in the setup phase. Nevertheless, it is often sufficient to let the arbiter pinpoint a corrupt party \emph{after} the client and server agree to deal with each other, i.e., after the setup when the server runs  $\mathtt{VSID.serve}(\cdot)$ and outputs $1$. A VSID protocol that meets the latter (lighter) requirements,  denoted by $\text{VSID}_{\st \text{light}}$, would impose lower costs  especially when $u$ and elements of $e$ are of large size.  Because    the arbiter  is not required to identify a misbehaving client in setup; therefore, it does not need to have access to the entire file $u^{\st *}$ and metadata $e$. This means (a) the server or client does not need to send $u^{\st *}$ and $e$ to the arbiter that leads to lower communication cost, and (b) the arbiter skips  checking the correctness of metadata in $\mathtt{VSID.identify}(\cdot)$, which ultimately saves it computation cost too.  In $\text{VSID}_{\st\text{light}}$, algorithm $\mathtt{VSID.identify}(\cdot)$ needs to take only $(\pi, c, k, e',{pp})$ as input, where $e'\subset e$.  So, this requires two changes to the VSID definition, (a) the arbiter algorithm would be   $\mathtt{VSID.identify}(\pi,c,k,e',{pp})\rightarrow I$, and (b) in case 2, in Definition \ref{def::VSID-Identifiable-Abort} we would have $b=0\wedge I\neq \mathcal C$, so event $a=0$ is excluded. In this paper, any time we refer to $\text{VSID}_{\st\text{light}}$, we assume the above minor adjustments are applied to the VSID definition. 


%% file: VSID-protocol.tex

\subsection{VSID Protocol}\label{subsec:VSID_protocol}

In this section, we present the VSID protocol. We show how it  can be built upon a protocol that satisfies the VS definition.  As stated previously, a VS scheme inherently protects an honest client from a malicious server. Therefore, at a high-level, VSID needs to have two added features; namely, it protects an honest server from a malicious client and allows an arbiter to detect a corrupt party. VSID can be built upon VS using  the following standard techniques; Briefly, (a) all parties sign their outgoing messages, (b) they post the signed messages on a bulletin board, and (c) the client, using a publicly verifiable NIZK scheme, proves to the server that its inputs have been correctly constructed.   In particular, like VS, the client first generates its secret and public parameters. Then, in the setup, it processes its input, $u$, to generate encoded input and metadata using the metadata generation function, $M$. Also, the client   utilizes a publicly verifiable NIZK scheme to prove to the server that the metadata has been constructed correctly.  The client posts the encoded input, metadata and the proofs along with their signatures to a bulletin board. Next, the server verifies the signatures and proofs. It  agrees to serve the client, if they are accepted.  Like VS, when the client wants the server to run  function $F$ on its input, it uses  function $Q$ to generate a query. However, it uses the zero-knowledge scheme to prove to the server that the query has been constructed correctly. The client posts the query,  proofs, and their signatures to the board. After that, the server verifies the  signatures and proofs. The server-side proves  and client-side verifies algorithms remain unchanged with a difference that the server posts its proofs (i.e., the output of the prove algorithm) and their signatures to the board and the client first verifies the signatures before checking the proofs.  In the case of any dispute/abort, either party invokes the arbiter which, given the signed posted messages,  checks the signatures and proofs in turn to identify a corrupt party. Below, we present the VSID protocol in which we assume all parties sign their outgoing messages and their counter-party first verifies the signature on the messages, before they feed them to their local algorithms. 

\

%
%
\begin{enumerate}[leftmargin=4mm]
\item\textbf{Key Generation}\label{VSID::keygen}. $\mathtt{VSID.keyGen}(1^{\lambda})$ 
\begin{enumerate}
\item Calls  $\mathtt{VS.keyGen}(1^{\lambda})$ to generate a pair of secret and public keys, $k:(sk,pk)$.
\item Commits to the secret key and appends the commitment: $\mathtt{Com}_{\st sk}$ to $pk$.
\item Posts $pk$ to a bulletin board. 
\end{enumerate}

\item\textbf{Client-side Setup}\label{VSID::Client-side-Setup}. $\mathtt{VSID.setup}(1^{\lambda}, u,k)$
\begin{enumerate}
\item Calls  $\mathtt{VS.setup}(1^{\lambda}, u,k)\rightarrow (\sigma,u^{\st *})$, to generate a   metadata: $\sigma=M(u^{\st *},k,{pp})$, encoded file service input and { (input dependent) parameters $pp$}. 

\item Generates non-interactive publicly verifiable zero-knowledge proofs asserting  $\sigma$ has been generated correctly, i.e., $\sigma$ is the output of  $M$ that is evaluated on $u^{\st *}$, $pk$,  $sk$, and {$pp$} without revealing $sk$. Let $w_{\st\sigma}$ contain the proofs.
\item\label{vsid::post-metadata} Posts $ e:=(\sigma,w_{\st\sigma})$, {$pp$},  and $u^{\st *}$ to the bulletin board. 
\end{enumerate}

\item\textbf{Server-side Setup}. $\mathtt{VSID.serve}(u^{\st *},e,pk,{pp})$

Ensures the metadata $\sigma$ has been constructed correctly, by verifying the proofs in $w_{\sigma}$ (where $\sigma,w_{\st\sigma}\in e$). If the proofs are accepted, then it outputs $a=1$  and proceeds to the next step; otherwise, it outputs $a=0$ and halts. 

\item\textbf{Client-side Query Generation}\label{VSID::Client-side-QueryGeneration}. \\$\mathtt{VSID.genQuery}(1^{\lambda},  \text{aux},k,$ ${pp} )$. 
\begin{enumerate}
\item Calls $\mathtt{VS.genQuery}(1^{\lambda}, \text{aux},k,{pp} )\rightarrow \bm{q}$, to generate a query vector, $\bm{q}=Q(\text{aux},k,{pp})$. If $\text{aux}$ is a private input, then it also commits to it,  that yields  $\mathtt{Com}_{\st \text{aux}}$
\item Generates non-interactive publicly verifiable zero-knowledge proofs proving $\bm{q}$ has been generated correctly, i.e., $\bm{q}$ is the
output of $Q$ which is evaluated on $\text{aux}$, $pk$, $sk$, and $pp$ without revealing $sk$ (and $\text{aux}$, if it is a private input). Let $\bm{w}_{\st q}$ contain the proofs and  $\text{aux}$ (or  $\mathtt{Com}_{\st \text{aux}}$ if $\text{aux}$ is a private input).
\item Posts $c:(\bm{q},\bm{w}_{\st q})$ to the board. 
\end{enumerate}

\item\textbf{Server-side Query Verification}. \\$\mathtt{VSID.checkQuery}(c, pk, $ $ {pp})$

Checks if   the query: $\bm{q}\in c$ has been constructed correctly by verifying the proofs  $\bm{w}_{\st q}\in c$. If the check passes, then it outputs $b=1$; otherwise, it outputs $b=0$.

\item\textbf{Server-side Service Proof Generation}. \\$\mathtt{VSID.prove}(u^{\st *},\sigma,$ $c,$ $ pk,{pp} )$ 
This phase starts only if the query was accepted, i.e., $b=1$ .
\begin{enumerate}
\item Calls $\mathtt{VS.prove}(u^{\st *},$ $\sigma, \bm{q},pk,{pp})\rightarrow \bm\pi$, to generate $\bm\pi=[F(u^{\st *}, $ $\bm{q},{pp}),\delta]$. Recall that $\bm{q}\in c$.
\item Posts $\bm\pi$ to the board. 
\end{enumerate}

\item\textbf{Client-side Proof Verification}. \\$\mathtt{VSID.verify}(\bm\pi,\bm{q},k,{pp})$

Calls $\mathtt{VS.verify}(\bm\pi,\bm{q},k,{pp})\rightarrow d$, to verify  proof $\bm\pi$. It accepts the proof if $d=1$; otherwise, it rejects it.

\item\textbf{Arbiter-side Identification}. \\$\mathtt{VSID.identify}(\bm\pi,c,k,e,u^{\st *}, {pp})$ 
\begin{enumerate}

\item\label{Arb-VSID.serve} Calls $\mathtt{VSID.serve}(u^{\st *},e,pk,{pp})\rightarrow a$. If $a=1$, then it proceeds to the next step. Otherwise, it outputs $I=\mathcal C$ and halts. 

\item Calls $\mathtt{VSID.checkQuery}(c, pk,{pp})\rightarrow b$. If $b=1$, then it proceeds to the next step. Otherwise, it outputs $I=\mathcal C$ and halts. 

\item  If $\bm\pi$ is  privately verifiable, then the arbiter  first checks if $sk\in k$ (provided by the client along with other opening information) matches $\mathtt{Com}_{\st sk}\in pk$. If they do not match, then the arbiter outputs $I=\mathcal C$. Otherwise,  it calls $\mathtt{VS.verify}(\bm\pi, \bm{q},k,{pp})\rightarrow d$. If $d=1$, then it outputs $I=\bot$; otherwise, it outputs $I=\mathcal S$.
\end{enumerate}
\end{enumerate}

\begin{theorem}\label{thm:VSID}
The VSID protocol with functions $F,M,Q$ satisfies the correctness, soundness, inputs well-formedness, and detectable abort properties for auxiliary information $\text{aux}$, (cf. Definitions \ref{deff::VSID-Correctness}-\ref{def::VSID-Identifiable-Abort}), if the underlying VS protocol with functions $F,M,Q$ is correct and sound for $\text{aux}$ and the underlying commitment, publicly verifiable non-interactive zero-knowledge, and signature schemes are correct/complete  and secure. 
\end{theorem}

\input{VSID-Proof}

\begin{remark}\label{remark::light-VSID}
As we mentioned before, it is often sufficient  to let the arbiter pinpoint a corrupt party \emph{after} the client and server agree to deal with each other. We denoted a VSID protocol that meets the latter (lighter) requirement, by $\text{VSID}_{\st \text{light}}$. This  version would impose lower costs, when $u$ and elements of $e$ are of large size. In  $\text{VSID}_{\st \text{light}}$ protocol, the client and server  run phases 1-3 of the VSID protocol as before, with a difference that the client does not post $e$ and $u^{\st*}$ to the board; instead, it sends them directly to the server. In $\text{VSID}_{\st\text{light}}$  the arbiter algorithm, i.e., $\mathtt{VSID.identify}(\cdot)$, needs to take only $(\bm\pi,c,k,e',{pp})$ as input, where $e'$ contains the opening of $\mathtt{Com}_{\st {sk}}$ if $\mathtt{VSID.verify}(\cdot)$ is privately verifiable or $e'=\bot$ if  it is publicly verifiable. In this light version, the arbiter   skips step \ref{Arb-VSID.serve}. Thus, $\text{VSID}_{\st\text{light}}$ saves (a)  communication cost, as   $u^{\st*}$ and $e$ are never sent to the board and  arbiter, and (b) computation  cost as the arbiter does not need to run $\mathtt{VSID.serve}(\cdot)$ anymore. 
\end{remark}

%% file: VSID-Proof.tex

\begin{proof}[Proof (sketch)]
 Correctness is implied by the correctness/completeness of the underlying primitives. The soundness  of VSID  stems  from the hiding property of the commitment, zero-knowledge property of the publicly verifiable NIZK proofs, and soundness of  the  verifiable service (VS) schemes. In particular, in VSID, the verifier (i.e., in this case, the client) makes block-box calls to the algorithms of VS to ensure soundness. However, the prover (i.e.,  the server) is given  additional messages, i.e.,   $\mathtt{Com}_{\scriptscriptstyle sk}$,  $\mathtt{Com}_{\scriptscriptstyle \text{aux}}$, $w_{\scriptscriptstyle\sigma}$ and $w_{\scriptscriptstyle q}$. The hiding property of the commitment scheme and zero-knowledge property of the zero-knowledge system ensure, given the messages, the prover learns nothing about the verification key and auxiliary information, except with negligible probability. Moreover, the soundness of VS scheme ensures a corrupt prover cannot convince an honest verifier, except with a negligible probability.  Inputs well-formedness property boils down to the security of the commitment and publicly verifiable NIZK proofs   schemes that are used in steps \ref{VSID::keygen}, \ref{VSID::Client-side-Setup} and \ref{VSID::Client-side-QueryGeneration} in VSID protocol. Specifically,  the binding property of the commitment and the soundness of the publicly verifiable NIZK proofs schemes  ensure that a corrupt prover (i.e., in this case the client) cannot convince a verifier (i.e., the server) to accept metadata proofs, $w_{\scriptscriptstyle\sigma}$ and $\mathtt{Com}_{\scriptscriptstyle sk}\in pk$, while  $M(u^{\scriptscriptstyle *},k,{pp})\neq \sigma $ or to accept query proofs, $\bm{w}_{\scriptscriptstyle q}$ and $\mathtt{Com}_{\scriptscriptstyle \text{aux}}$, while $Q(\text{aux},k,{pp})\neq \bm{q}$, except with negligible probability.  
  
  Moreover, the detectable abort property holds as long as both previous properties (i.e., soundness and inputs well-formedness) hold, the commitment is secure, the zero-knowledge proofs are publicly verifiable and the signature scheme is secure. The reason is that the algorithm $\mathtt{VSID.identify}(\cdot)$, which ensures detectable abort, is a wrapper function that is invoked by the arbiter, and  sequentially makes subroutine calls to algorithms  $\mathtt{VSID.serve}(\cdot)$,  $\mathtt{VSID.checkQuery}(\cdot)$ and $\mathtt{VS.verify}(\cdot)$,   where the first two ensure input well-formedness, and the last one ensures soundness. Also, due to the security of the commitment (i.e., binding), the malicious client cannot provide the arbiter with  another secret verification key than what  was initially committed. Moreover, due to the public verifiability of the zero-knowledge proofs, the arbiter can verify all proofs input to $\mathtt{VSID.serve}(\cdot)$ and $\mathtt{VSID.checkQuery}(\cdot)$. The signature's security ensures if a proof is not signed correctly, then it can also be rejected by the arbiter; on the other hand, if a proof is signed correctly, then  it cannot be repudiated by the signer later on (due to signature's unforgeability); this guarantees that the signer is held accountable for a rejected proof it provides.
  \end{proof}

%% file: Recurring-Contingent-Service-Payments.tex

\section{Recurring Contingent Service Payment (RC-S-P) Protocol}\label{app::RC-S-P}

In this section, we present our RC-S-P protocol.


\input{RCSP-protocol}

%% file: RCSP-protocol.tex

\subsection{Recurring Contingent Service Payment (RC-S-P) Protocol}\label{sec::RC-S-P-protocol}

 In this section, we present the ``recurring contingent service payment'' (RC-S-P) protocol for a generic service. It utilises a novel combination of $\text{VSID}_{\st\text{light}}$, SAP, the private time bubble notion, and symmetric-key encryption schemes along with the coin masking and padding techniques. At a high level, the protocol works as follows.  The  client and server use SAP to provably agree on two private statements; the first statement includes  payment details, while another one specifies a secret key, $k$, and the pads' length. They also  agree on  public parameters such as (a) the private time bubble's length, that is the total number of billing cycles, $z$, plus a waiting period, $\texttt{J}$, and (b) a smart contract which  specifies $z$ and  the total amount of masked coins each party should deposit. The client   deploys the contract. Each party deposits its masked coins in the contract. If either party does not deposit enough coins on time,   later each party has a chance to withdraw its coins and terminate the contract. To start using/providing the service, they invoke $\text{VSID}_{\st\text{light}}$ protocol. In particular, they engage in the  $\mathtt{VSID.keyGen}(\cdot)$, $\mathtt{VSID.setup}(\cdot)$, and $\mathtt{VSID.serve}(\cdot)$ algorithms. If the server decides not to serve, e.g., it detects the client's misbehaviour,  it sends $0$  within a fixed time; in this case, the parties can withdraw their deposit and terminate the contract. Otherwise, the server sends $1$ to the contract.

At the end of each billing cycle, the client generates an encrypted query, by calling $\mathtt{VSID.genQuery}(\cdot)$ and encrypting its output using the key, $k$. It pads the encrypted query and  sends the result  to the contract. The encryption and pads ensure  nothing about the client's input (e.g., outsourced file) is revealed to the public within the private time bubble.  In the same cycle, the server retrieves the query, removes the pads and decrypts the result. Then, it locally checks its validity, by calling $\mathtt{VSID.checkQuery}(\cdot)$. If the query is rejected, the server locally stores the index of the billing cycle and then  generates a dummy proof.  Otherwise, if the server accepts the query, it generates a proof of service by calling $\mathtt{VSID.prove}(\cdot)$. In either case, the server encrypts the proof, pads it and sends the result to the contract. Note that sending (padded encrypted) dummy proofs ensures that the public, during the private time bubble, does not learn if the client generates invalid queries.  After the server sends the messages to the contract, the client removes the pads, decrypts the proof and locally verifies it, by calling $\mathtt{VSID.verify}(\cdot)$. If the verification is passed, then the client knows the server has delivered the service honestly. But, if the proof is rejected,  it waits until the private time bubble passes and dispute resolution time arrives. During the dispute resolution period, in the case the client or server  rejects any proofs, it  invokes the arbiter,  refers it to the invalid encrypted proofs in the contract, and  sends to it the decryption key and the pads' detail. The arbiter checks the validity of the key and pads, by using SAP. If they are accepted,  then the arbiter locally removes the pads from the encrypted proofs, decrypts the related proofs, and runs $\mathtt{VSID.identify}(\cdot)$ to check the validity of the party's claim. The arbiter sends to the contract a report of its findings that includes  the total number of times  the server and client provided invalid proofs. In the next phase, to distribute the coins, either client or server  sends: (a) ``pay'' message, (b) the agreed statement that specifies the payment details, and (c) the statement's proof to the contract which verifies the statement and if approved it distributes the coins according to the statement's detail, and the arbiter's report. 

Now we outline why RC-S-P addresses the issues, raised in Section \ref{sec::previous-work-limitations}.    In the setup, if the client provides ill-formed inputs (so later it can  accuse the server) then the server can detect and avoid serving it.  After the setup, if the client avoids sending any input,  then the server still gets paid for the service it provided. Also, in the case of a dispute between the parties, their claim is checked, and  the corrupt party is identified. The corrupt party has to pay the arbiter and if that is the client, then it has to pay the server as well.  These  features not only do guarantee the server's resource is not wasted, but also ensures fairness (i.e.,  if a potentially malicious server is paid, then it must have provided the service and if a potentially malicious client does not pay, then it will learn nothing). Furthermore, as during the private time bubble  (a) no plaintext proof is given to the contract, and (b) no dispute resolution and coin transfer take place on contract, the public cannot figure out the outcome of each verification. This  preserves the server's privacy. Also, because the  deposited coins are masked and the agreed statement is kept private,   nothing about the detail of the service  is leaked to the public before the bubble bursts. This preserves the client's privacy. Also, as either party can prove to the contract the validity of the agreed statement, and ask the contract to distribute the coins, the coins will be not be locked forever.

\vspace{-2mm}
\subsubsection{Protocol description}\label{subsec:RCSP_description}
The RC-S-P protocol is parameterized by the functions $F,M,Q$ of the underlying VSID and encoding/decoding functions $(E,D)$  that refer to ``encrypt then pad''/``remove pad then decrypt'' procedures, respectively.  It is assumed that (a) each party $\mathcal{P} \in\{\mathcal{C,S,R}\}$ already has a blockchain public address, $adr_{\st\mathcal{P}}$, which is known to all parties, (b) it uses that (authorised) address to send transactions to the smart contract,  (c) the contract   before recording a transaction, ensures the transaction is  originated from an authorised address, and (d) there is a public price list $pl$ known to everyone.     The protocol is presented below.

\begin{enumerate}
\item \textbf{Key Generation}. $\mathtt{RCSP}.\mathtt{keyGen}(1^{\lambda})$
\begin{enumerate}

\item  $\mathcal C$ runs  $\mathtt{VSID.keyGen}(1^{\lambda})\rightarrow k:=(sk,pk)$. It picks a random secret key  $\bar{k}$ for a symmetric-key encryption. Also,  it sets two parameters: $pad_{\st \pi}$ and $pad_{\st q}$, where  $pad_{\st \pi}$ and $pad_{\st q}$  refer to the number of  dummy values that will be used to pad encrypted proofs and encrypted queries respectively\footnote{\scriptsize The values of  $pad_{\st \pi}$ and $pad_{\st q}$ is determined as follows, $pad_{\st \pi}=\pi_{\st max}-\pi_{\st act}$ and $pad_{\st q}=q_{\st max}-q_{\st act}$, where $\pi_{\st max}$ and $\pi_{\st act}$ refer to the maximum and actual the service's proof size  while $q_{\st max}$ and $q_{\st act}$ refer to the maximum and actual the service's query size, respectively.},  determined by the security parameter and description of $F$.  Let $sk':=(pad_{\st \pi},pad_{\st q}, \bar{k})$. The keys' size is part of the security parameter.  Let  $\bm{k}=[k,k']$, where $k':=(sk',pk')$   and $pk':=(adr_{\st\mathcal{C}},adr_{\st\mathcal{S}})$. 
\end{enumerate}


\item \textbf{Client-side Initiation}. $\mathtt{RCSP}.\mathtt{cInit}(1^{\lambda}, u,\bm{k},z,pl)$
\begin{enumerate}

\item Calls  $\mathtt{VSID.setup}(1^{\lambda}, u,k)\rightarrow (u^{\st *}, {pp},e)$, to encode  service input, and generate metadata. It  sets $qp=sk'$ and appends $pp$ to $qp$.

\item  Calls $\mathtt{SAP.init}(1^{\lambda}, adr_{\st\mathcal{C}}, adr_{\st\mathcal{S}},qp )\rightarrow(r_{\st qp}, g_{\st qp}, \\ adr_{\st\text{SAP}_{\st 1}})$, to initiate an agreement (with $\mathcal{S}$) on $qp$.  Let $T_{\st qp}:=(\ddot{x}_{\st qp}, g_{\st qp})$ be proof/query encoding token,  where  $\ddot{x}_{\st qp}:=(qp,r_{\st qp})$ is the opening and $g_{\st qp}$ is the commitment stored on the contract as a result of running   SAP.

\item   Sets coin parameters as   follows,   $o$: the amount of coins for each accepting proof, and $l$: the amount of coins  to cover the cost of each potential dispute resolution, given  price list $pl$.  

\item  Sets  $cp:=(o,o_{\st max},l,l_{\st max},z)$, where $o_{\st max}$ is the maximum amount of coins  for an accepting service proof,  $l_{\st max}$  is the maximum amount of coins to resolve a potential dispute, and $z$ is the number of  service proofs/verifications. Then, $\mathcal C$ calls $\mathtt{SAP.init}(1^{\lambda}, adr_{\st\mathcal{C}}, adr_{\st\mathcal{S}},cp )\rightarrow(r_{\st cp},g_{\st cp},adr_{\st\text{SAP}_{\st 2}})$, to initiate an agreement (with $\mathcal{S}$) on $cp$. Let $T_{\st cp}:=(\ddot{x}_{\st cp},g_{\st cp})$ be coin encoding token,  where  $\ddot{x}_{\st cp}:=(cp,r_{\st cp})$ is the opening and $g_{\st cp}$ is the commitment stored on the contract as a result of executing   SAP.  Let  $T:=\{T_{\st qp},T_{\st cp}\}$.

\item Set parameters  $coin^{\st *}_{\st\mathcal C}=z\cdot (o_{\st max}+l_{\st max})$ and $p_{\st\mathcal{S}}=z\cdot l_{\st max}$, where $coin^{\st *}_{\st\mathcal C}$ and $p_{\st\mathcal{S}}$ are   the total number of masked coins $\mathcal C$ and $\mathcal S$ should deposit respectively. It also designs  a smart contract, SC, that explicitly specifies  parameters $z$,  $coin^{\st *}_{\st\mathcal C}$, $p_{\st\mathcal{S}}$,  $adr_{\st\text{SAP}_{\st 1}}$,  $adr_{\st\text{SAP}_{\st 2}},  pk$, and $pk'$.  It sets a set of time points/windows, $\texttt{Time}:\{ \texttt{T}_{\st 0},..., \texttt{T}_{\st 2},\texttt{G}_{\st 1,1},...,\texttt{G}_{\st z,2},\texttt{J}, $ $\texttt{K}_{\st 1},...,\texttt{K}_{\st 3}, \texttt{L}\}$, that are explicitly specified in the contract which will accept a certain party's message only in a specified  time point/window. The time allocation will become clear in the next phases. 


\item Sets also four counters $[y_{\st\mathcal C},y'_{\st\mathcal C},y_{\st\mathcal S}, y'_{\st\mathcal S}]$ in SC, where their initial value is $0$. It signs and deploys SC to the blockchain.  Let $adr_{\st \text{SC}}$ be the address of the deployed SC, and  $\bm{y}: [y_{\st\mathcal C}, $ $y'_{\st\mathcal C},y_{\st\mathcal S}, y'_{\st\mathcal S}, adr_{\st \text{SC}}]$.


\item Deposits $coin^{\st *}_{\st\mathcal C}$ coins in the contract. It sends $u^{\st *}, z, e, \ddot{x}_{\st qp}$, and $\ddot{x}_{\st cp}$ (along with $adr_{\st \text{SC}}$) to $\mathcal S$. Let $\texttt{T}_{\st 0}$ be the time that the above process finishes. 


\end{enumerate}

\item \textbf{Server-side Initiation}\label{RCSP::Server-side-Initiation}. \\$\mathtt{RCSP}.\mathtt{sInit}(u^{\st *}, e, pk, z, T, p_{\st\mathcal{S}} ,\bm{y})$

\begin{enumerate}

\item Checks the parameters  in $T$ (e.g., $qp$ and $cp$) and  in SC (e.g., $p_{\st\mathcal{S}}, \bm{y}$) and ensures a sufficient amount of coins has been deposited by $\mathcal C$.  

\item Calls $\mathtt{SAP.agree}(qp,r_{\st qp},g_{\st qp},adr_{\st\mathcal{C}},adr_{\st\text{SAP}_{1}})\rightarrow (g'_{\st qp},b_{\st 1})$ and $\mathtt{SAP.agree}(cp,r_{\st cp},g_{\st cp},adr_{\st\mathcal{C}},adr_{\st\text{SAP}_{2}})\rightarrow (g'_{\st cp},b_{\st 2})$, to verify the correctness of tokens in $T$ and to agree on the tokens' parameters, where $qp,r_{\st qp}\in \ddot{x}_{\st qp}, \text{and }  cp,r_{\st cp} \in  \ddot{x}_{\st cp}$. Recall that if both $\mathcal{C}$ and $\mathcal{S}$ are honest, then $g_{\st qp}=g'_{\st qp}$ and $g_{\st cp}=g'_{\st cp}$.

\item If any above check is rejected, then it sets $a=0$. Otherwise, it  calls $\mathtt{VSID.serve}(u^{\st *},e,pk, {pp})\rightarrow a$. 

\item Sends $a$ and $coin^{\st *}_{\st\mathcal S}=p_{\st\mathcal{S}}$ coins to SC at time $\texttt{T}_{\st 1}$, where  $coin^{\st *}_{\st\mathcal S}=\bot$ if $a=0$
\end{enumerate}
Note that, $\mathcal S$  and  $\mathcal C$ can withdraw their coins at time $\texttt{T}_{\st 2}$, if  $\mathcal S$ sends $a=0$, fewer coins than $p_{\st\mathcal{S}}$, or nothing to the SC. To withdraw, $\mathcal S$  or  $\mathcal C$  simply sends a ``pay'' message to $\mathtt{RCSP}.\mathtt{pay}(\cdot)$ algorithm (only) at time $\texttt{T}_{\st 2}$.

\textbf{\textit{Billing-cycles Onset}}. $\mathcal{C}$ and $\mathcal{S}$ engage in the following three phases, i.e., phases \ref{Billing-cycles-genQuery}-\ref{Client-sideProofVerification},  at the end of every $j$-th billing cycle, where $1\leq j\leq z$. Each $j$-th cycle includes two  time points, $\texttt{G}_{\st j,1}$ and $\texttt{G}_{\st j,2}$, where $\texttt{G}_{\st j,2}>\texttt{G}_{\st j,1}$, and $\texttt{G}_{\st 1,1}>\texttt{T}_{\st 2}$.


\item\textbf{Client-side Query Generation}.\label{Billing-cycles-genQuery} \\$\mathtt{RCSP}.\mathtt{genQuery}(1^\lambda, \text{aux},$ $k, $ $T_{\st qp})$

\begin{enumerate}
\item Calls $\mathtt{VSID.genQuery}(1^{\lambda},  \text{aux},k,{pp})\rightarrow c_{\st j}:=(\bm{q}_{\st j},\bm{w}_{\st q_{_{\scaleto{j}{\size}}}})$, to generate a query-proof pair. 
\item Encodes  $c_{\st j}$, by first encrypting it, $\mathtt{Enc}(\bar{k},c_{\st j})=c'_{\st j}$, where $\bar{k}\in T_{\st qp}$; and  then, padding (each element of) the result with ${pad}_{\st q}\in T_{\st qp}$ random values that are picked uniformly at random from the encryption's output range, $U$. Let $c^{\st *}_{\st j}$ be the result. 
\item Sends the padded encrypted query-proof pair, $c^{\st *}_{\st j}$, to SC at time $\texttt{G}_{\st j,1}$.
\end{enumerate}

\item\textbf{Server-side Proof Generation}. \\$\mathtt{RCSP}.\mathtt{prove}(u^{\st *},  \sigma,  c^{\st *}_{\st j}, pk, T_{\st qp})$

\begin{enumerate}

\item Constructs an empty  vector, $\bm{m}_{\st\mathcal S}=\bot$, if $j=1$.  
\item Removes the pads from $c^{\st *}_{\st j}$, using  parameters of $T_{\st qp}$. Let  $c'_{\st j}$ be the result. Next, it  decrypts  the result, $\mathtt{Dec(}\bar{k},c'_{\st j})=c_{\st j}$. Then,  it   runs $\mathtt{VSID.checkQuery}(c_{\st j}, pk, {pp})\rightarrow b_{\st j}$, to check the correctness of the queries.

\begin{itemize}
\item[$\bullet$] If $\mathcal S$ accepts the query, i.e., $b_{\st j}=1$, then  calls\\ $\mathtt{VSID.prove}(u^{\st *},\sigma,c_{\st j},pk,{pp})\rightarrow \bm\pi_{\st j}$, to generate the service proof. In this case, $\mathcal S$ encrypts it, $\mathtt{Enc}(\bar{k},\bm\pi_{\st j})=\bm\pi'_{\st j}$. Next, it   pads (every element of) the encrypted proof with ${pad}_{\st \pi}\in T_{\st qp}$ random values  picked uniformly at random from $U$. Let $\bm\pi^{\st *}_{\st j}$ be the result. It sends the padded encrypted proof  to SC at time $\texttt{G}_{\st j,2}$. 

\item[$\bullet$] Otherwise (if $\mathcal S$ rejects the query), it appends  $j$  to $\bm{m}_{\st\mathcal{S}}$,  constructs a dummy proof $\bm\pi'_{\st j}$, picked uniformly at random from $U$,  pads the result as above, and sends the padded dummy proof, $\bm\pi^{\st *}_{\st j}$, to SC at time $\texttt{G}_{\st j,2}$. 

\end{itemize}
When $j=z$  and $\bm{m}_{\st\mathcal{S}}\neq\bot$, $\mathcal S$ sets    $m_{\st\mathcal S}:=(\bm{m}_{\st\mathcal{S}},adr_{\st \text{SC}})$.

\end{enumerate}

\item \textbf{Client-side Proof Verification}.\label{Client-sideProofVerification} \\$\mathtt{RCSP}.\mathtt{verify}(\bm\pi^{\st *}_{\st j},c^{\st *}_{\st j}, k,T_{\st qp})$
\begin{enumerate}

\item Constructs an empty vector, $\bm{m}_{\st\mathcal C}=\bot$, if $j=1$. 

\item Removes the pads from $\bm\pi^{\st *}_{\st j}$, utilising  parameters of $T_{\st qp}$. Let $\bm\pi'_{\st j}$ be the result. It decrypts the service proof: $\mathtt{Dec}(\bar{k},\bm\pi'_{\st j})=\bm\pi''_{\st j}$ and then calls $\mathtt{VSID.verify}(\bm\pi''_{\st j}, \bm{q}_{\st j}, k, {pp})\rightarrow d_{\st j}$, to verify the proof, where $\bm{q}_{\st j}\in c_{\st j}$ (and $c_{\st j}$ is the result of removing pads from  $c^{\st *}_{\st j}$ and then decrypting the result). Note that if $\bm\pi'_{\st j}=\mathtt{Enc}(\bar{k}, \bm\pi_{\st j})$, then $\bm\pi''_{\st j}=\bm\pi_{\st j}$. 
\begin{itemize}
\item[$\bullet$] If $\bm\pi''_{\st j}$ passes the verification (i.e., $d_{\st j}=1$), then $\mathcal C$ concludes that the service  for this verification has been delivered successfully. 

\item[$\bullet$]  Otherwise (when $\bm\pi''_{\st j}$ is rejected), $\mathcal C$ appends $j$  to $\bm{m}_{\st\mathcal{C}}$.

\end{itemize}
\end{enumerate}
When $j=z$  and $\bm{m}_{\st\mathcal{C}}\neq\bot$, $\mathcal C$ sets    ${m}_{\st\mathcal C}:=(\bm{m}_{\st\mathcal{C}},adr_{\st \text{SC}},e')$, where  $e'$  contains the opening of $\mathtt{Com}_{\st sk}$ or  $\bot$, as stated in Remark \ref{remark::light-VSID}.

\item \textbf{Dispute Resolution}.\label{RCSP::Dispute-Resolution} \\$\mathtt{RCSP}.\mathtt{resolve}(\bm m_{\st \mathcal{C}}, \bm m_{\st \mathcal{S}}, z, {\bm{\pi}}^{\st *}, {\bm{c}}^{\st *}, pk, T_{\st qp})$

 The phase takes place only in  case of dispute, e.g., when $\mathcal C$ and/or $\mathcal S$ reject any proofs in the previous phases. 

\begin{enumerate}

\item The arbiter sets  counters: $y_{\st\mathcal C}, y'_{\st\mathcal C},y_{\st\mathcal S}$ and  $y'_{\st\mathcal S}$, that are initially set to $0$,  before time $\texttt{K}_{\st 1}$,  where $\texttt{K}_{\st 1}>\texttt{G}_{\st z,2}+\texttt{J}$. 
\item $\mathcal C$  sends $m_{\st\mathcal C}$ and $\ddot{x}_{\st qp}$ to the arbiter at time $\texttt{K}_{\st 1}$. Or,  $\mathcal S$ sends $m_{\st\mathcal S}$ and $\ddot{x}_{\st qp}$ to the arbiter at time $\texttt{K}_{\st 1}$. 

\item\label{arbiter-checks-statement} At time $\texttt{K}_{\st 2}$, the arbiter checks the validity of  statement $\ddot{x}_{\st qp}$ sent by each party  $\mathcal P\in \{\mathcal {C,S}\}$. To  do so, it sends  each $\ddot{x}_{\st qp}$ to SAP contract which  returns  either $1$ or $0$. The arbiter  constructs an empty vector, $\bm{v}$.  If  party $\mathcal P$'s statement is accepted, then 
 it appends every  element of $\bm{m}_{\st\mathcal{P}}$ to $\bm{v}$. It ensures $\bm{v}$ contains only distinct elements which are in the range $[1,z]$. Otherwise (if the party's statement is rejected) it discards the party's request, $\bm{m}_{\st\mathcal{P}}$. It proceeds to the next step if $\bm{v}$ is not empty, otherwise it halts.

\item\label{the-arbiter-loop} The arbiter  for every element $i\in\bm{v}$:

\begin{enumerate}

\item removes the pads from the related encrypted    query-proof pair  and from encrypted service proof. Let $c'_{\st i}$ and $\bm\pi'_{\st i}$ be the result. 
\item decrypts the encrypted query-proof pair and encrypted service proof as follows, $ \mathtt{Dec}(\bar{k},c'_{\st i})=c_{\st i}$ and 
$ \mathtt{Dec}(\bar{k},\bm\pi'_{\st i})=\bm\pi''_{\st i}$.
\item\label{arbiter-identify} calls   $\mathtt{VSID.identify}(\bm\pi''_{\st i},c_{\st i}, k, e', {pp})\rightarrow I_{\st i}$
\begin{enumerate}

\item[$\bullet$] if $I_{\st i}=\mathcal C$ and $y'_{\st\mathcal C}$ was not incremented for $i$-th verification, it increments $y_{\st\mathcal C}$ by $1$. 
\item[$\bullet$]  if $I_{\st i}=\mathcal S$ and $y'_{\st\mathcal S}$ was not incremented for $i$-th verification, it increments $y_{\st\mathcal S}$ by $1$. 
\item[$\bullet$]  if $I_{\st i}=\bot$, then it increments  $y'_{\st\mathcal C}$ or $y'_{\st\mathcal S}$ by $1$, if $i$ is in  the complaint of $\mathcal {C}$ or $\mathcal {S}$ respectively and $y_{\st\mathcal C}$ or $y_{\st\mathcal S}$ was not incremented  in $i$-th verification. 

\end{enumerate}
\end{enumerate}

 
Let $\texttt{K}_{\st 3}$ be the time that the arbiter finishes the above checks.

\item The arbiter at time $\texttt{K}_{\st 3}$ sends $[y_{\st\mathcal C},y_{\st\mathcal S},  y'_{\st\mathcal C},y'_{\st\mathcal S}]$  to SC that accordingly overwrites the  elements it holds (i.e.,  elements of $\bm{y}$) by  the related vectors elements the arbiter sent.  
\end{enumerate}

\item \textbf{Coin Transfer}\label{RCSP::CoinTransfer}. $\mathtt{RCSP}.\mathtt{pay}(\bm{y},T_{\st cp},a,p_{\st\mathcal S}, coin^{\st *}_{\st\mathcal C},coin^{\st *}_{\st\mathcal S})$
\begin{enumerate}
 
 \item If SC  receives  ``pay'' message  at time $\texttt{T}_{2}$, where $a=0$ or $coins^{*}_{\st\mathcal{S}}<p_{\st\mathcal{S}}$, then it sends $coin^{\st *}_{\st\mathcal C}$ coins to $\mathcal C$ and $coin^{\st *}_{\st\mathcal S}$ coins to $\mathcal S$. In other words, the parties can withdraw their coins if they do not reach to an agreement in the end of   phase \ref{RCSP::Server-side-Initiation}, i.e., server-side initiation.  Otherwise (i.e., they reach to an agreement), they take the following steps. 

\item Either $\mathcal C$ or $\mathcal S$ sends ``pay'' message and the statement, $\ddot{x}_{\st cp}\in T_{\st cp}$,  to SC at time $\texttt{L}>\texttt{K}_{\st 3}$.
\item SC checks the validity of the statement by sending $\ddot{x}_{\st cp}$ to the SAP contract which  returns  either $1$ or $0$. SC only proceeds to the next step if the output is $1$. 
\item SC distributes the coins to the parties as follows:

\begin{itemize}
\item[$\bullet$]   $coin_{\st\mathcal C}=coin^{\st *}_{\st\mathcal C}-o\cdot (z-y_{\st\mathcal S})-l\cdot(y_{\st\mathcal C}+y'_{\st\mathcal C})$ coins  to $\mathcal C$.
\item[$\bullet$] $coin_{\st\mathcal S}=coin^{\st *}_{\st\mathcal S}+o\cdot(z-y_{\st\mathcal S})-l\cdot (y_{\st\mathcal S}+y'_{\st\mathcal S})$ coins to $\mathcal S$.
\item[$\bullet$] $coin_{\st\mathcal{R}}=l\cdot(y_{\st\mathcal S}+y_{\st\mathcal C}+y'_{\st\mathcal S}+y'_{\st\mathcal C})$ coins to the arbiter.  

\end{itemize}

\end{enumerate}
\end{enumerate}


\subsubsection{Discussion on the RC-S-P protocol description}\label{subsec:RCSP_discussion}
We conclude Subsection~\ref{sec::RC-S-P-protocol} with the following remarks:
\begin{itemize}[leftmargin=5mm]
\item The length of a private time bubble can be agreed between  the server and client to be of any size that suits them and can exceed the point where the $z$-th verifications is completed.

\item For the sake of simplicity, in the RC-S-P protocol,  we let each $y\in\{y_{\st\mathcal C}, y'_{\st\mathcal C},y_{\st\mathcal S},   y'_{\st\mathcal S}\}$ be a counter; instead of a binary vector, $\bm{y}\in \{\bm{y}_{\st\mathcal C}, \bm{y}'_{\st\mathcal C}, \bm{y}_{\st\mathcal S},    \bm{y}'_{\st\mathcal S}\}$, defined in the RC-S-P definition. However, it is not hard to see that the sum of all elements $\bm{y}$ of equal $y$, i.e., $y=\sum\limits_{\st j=1}^{\st z}\bm{y}_{\st j}$. The same holds for the  amounts of coin each party receives, $coin\in\{coin_{\st\mathcal C}, coin_{\st\mathcal S}, coin_{\st\mathcal{R}}\}$, in the protocol and the coin vector  used in the definition, $\bm{coin}\in\{\bm{coin}_{\st\mathcal C}, \bm{coin}_{\st\mathcal S}, \bm{coin}_{\st\mathcal{R}}\}$. 

\item In the protocol, the pads are added \emph{after} the actual values are encrypted. This is done to save computation cost. Otherwise (if the pads are added prior to the encryption),  then the pads would have to be encrypted  too, which imposes additional computation cost. 

\item As stated in Section \ref{sec::RC-S-P-Definition},    $\mathtt{RCSP}.\mathtt{genQuery}(\cdot),\mathtt{RCSP}.\mathtt{prove}(\cdot), \\ \mathtt{RCSP}.\mathtt{verify}(\cdot)$ and $\mathtt{RCSP}.\mathtt{resolve}(\cdot)$ implicitly  take $a, coin^{\st*}_{\st\mathcal{S}}, $ $p_{\st\mathcal{S}}$ as another inputs and  execute only if $a=1$ and $coin^{\st*}_{\st\mathcal{S}}=p_{\st\mathcal{S}}$. For the sake of simplicity, we  avoided explicitly stating it in the protocol.   Also, keeping track of  $(y'_{\st\mathcal C},y'_{\st\mathcal S})$ enables the arbiter to make malicious parties, that \emph{unnecessarily} invoke  it for  accepting proofs in step \ref{arbiter-identify}, pay  for the verifications it performs. 

\item The total coin amounts the client receives is as follows; its initial deposit, i.e., $coin^{\st *}_{\st\mathcal C}$, minus the total coin amounts that the server should be paid for those verifications that it has acted honestly towards the client, i.e., $o\cdot(z-y_{\st\mathcal S})$, minus the total coin amounts the client has to pay to the arbiter when it misbehaved towards the server and the arbiter, i.e., $l\cdot(y_{\st\mathcal C}+y'_{\st\mathcal C})$. The total coin amounts  the server receives is as follows. Its initial deposit, i.e., $coin^{\st *}_{\st\mathcal S}$, plus the total coin amounts that it should get paid for those verifications that it  acted honestly towards the client, i.e., $o\cdot(z-y_{\st\mathcal S})$, minus  the total coin amounts   it has to pay to the arbiter when it misbehaved towards the client and the arbiter, i.e., $l\cdot(y_{\st\mathcal S}+y'_{\st\mathcal S})$. Moreover the arbiter receives in total $l\cdot(y_{\st\mathcal S}+y_{\st\mathcal C}+y'_{\st\mathcal S}+y'_{\st\mathcal C})$ coins to cover its cost of resolving disputes, i.e., $l\cdot(y_{\st\mathcal S}+y_{\st\mathcal C})$, plus the cost imposed to it when it is  unnecessarily invoked, i.e.,  $l\cdot(y'_{\st\mathcal S}+y'_{\st\mathcal C})$. If all parties behave honestly,   then the server receives all its deposit back plus the coin amounts  they initially agreed to pay the server if it  delivers accepting proofs for all $z$ cycles, i.e., in total it receives $coin^{\st *}_{\st\mathcal S}+o\cdot z$ coins. Also, in this case an honest client receives all coins minus the coin amounts paid to the server for delivering accepting proofs for $z$ cycles, i.e., in total it receives $coin^{\st *}_{\st\mathcal C}-o\cdot z$ coins. However, the arbiter receives no coins, as it is never invoked.

\item The VSID scheme  does not (need to) preserve the privacy of the proofs. However, in RC-S-P  protocol  each proof's privacy must be preserved, for a certain time; otherwise, the proof itself can leak its status, e.g., when it can be publicly verified. This is the reason why in the RC-S-P protocol,  \emph{encrypted} proofs are sent to the contract.   Moreover, for the sake of simplicity, in the above protocol, we assumed that each arbiter's invocation has a fixed  cost regardless of the number of steps it takes. To define a fine-grained costing, one can simply allocate to each step the arbiter takes a certain rate and also separate counter  for the client and server.  

\item In the case where $\mathtt{VSID.verify}(\cdot)$ is privately verifiable and    the server invokes the arbiter, the client needs to provide inputs to the arbiter too. Otherwise (when it is publicly verifiable and  the server invokes the arbiter), the client's involvement is not required in the dispute resolution phase.  In contrast, if the client invokes the arbiter, the server's involvement is not required in that phase, regardless of the type of verifiability $\mathtt{VSID.verify}(\cdot)$ supports. Furthermore, with a minor adjustment to the RC-S-P protocol, we can let the client and server be compensated (by a misbehaving party) for the transaction they send to the contract.  To do so, briefly, we can let the parties, in initiation phases, agree on and include in $cp$  parameters, $l'$ and $l''$, that  cover the client's and server's cost of sending a transaction, respectively. The parameters are  encoded the same way as  $l$ is encoded. In this setting, in the coin transfer phase, the client and server receive  $coin^{\st *}_{\st\mathcal C}-o\cdot(z-y_{\st\mathcal S})-l\cdot(y_{\st\mathcal C}+y'_{\st\mathcal C})+l'\cdot y_{\st\mathcal S}-l''\cdot y_{\st\mathcal C}$ and $coin^{\st *}_{\st\mathcal S}+o\cdot(z-y_{\st\mathcal S})-l\cdot(y_{\st\mathcal S}+y'_{\st\mathcal S})-l'\cdot y_{\st\mathcal S}+l''\cdot y_{\st\mathcal C}$, coins respectively.  The amount of coins the arbiter receives remains unchanged.  

\item  The server or client, even during the private time bubble, can spend (or more accurately promise to a third party) the amount of coins kept in the contract and will ultimately be transferred to it. With  slight adjustments to the RC-S-P,  they can do so in a privacy-preserving manner. We briefly explain how it can be done.  For the sake of simplicity,  we assume the server  will receive $coin_{\st\mathcal S}$ coins after the bubble bursts and wants to promise $\hat{coin}_{\st\mathcal S}$ coins (where $\hat{coin}_{\st\mathcal S} \leq coin_{\st\mathcal S}$) to   the third party  $\mathcal{D}$ within the bubble. First, the server proves to $\mathcal{D}$ that it will receive $coin_{\st\mathcal S}$ coins after the bubble bursts. To do that, it sends the RC-S-P transcripts (that includes all proofs) to $\mathcal{D}$ which can verify the server’s claim, as all proofs are publicly verifiable. Next, if $\mathcal{D}$ is convinced, the server and $\mathcal{D}$ invoke a new instance of the SAP  and insert the value $\hat{coin}_{\st\mathcal S}$  into the SAP’s private statement. This results in a smart contract, $\text{SC}_{\st\text{SAP}_{\st 3}}$. Next, if both parties agree on the parameters of $\text{SC}_{\st\text{SAP}_{\st 3}}$, then the server sends the address of $\text{SC}_{\st\text{SAP}_{\st 3}}$ to the main contract of RC-S-P, i.e., SC. When the bubble bursts, SC transfers the client’s share of coins to the client as before. But, SC distributes the server’s coins if the server or $\mathcal{D}$ sends to it a valid proof for the above private statement (in addition to the proofs required in the Phase \ref{RCSP::CoinTransfer} of the original RC-S-P).  Upon receiving that proof, SC invokes $\text{SC}_{\st\text{SAP}_{\st 3}}$ to check the validity of the proof. If the proof is accepted, then SC  sends $\hat{coin}_{\st\mathcal S}$ to $\mathcal{D}$ and $coin_{\st\mathcal S}-\hat{coin}_{\st\mathcal S}$ to the server. It is evident that this approach leaks no information about the coins amount (including $\hat{coin}_{\st\mathcal S}$) during the bubble to the public, due to the security of the SAP. The above idea can be further  extended to support multiple parties. For instance, if the server wants to promise $coin_{\st\mathcal S}-\hat{coin}_{\st\mathcal S}$ coins  to  $\mathcal{D}'$ (after its promise to  $\mathcal{D}$), it needs to send to $\mathcal{D}'$  all the proofs, including  the one related to the above private statement. 
 
\item As stated previously, the proofs are sent to the contract to avoid running into the deniability issue, i.e., a malicious client  wrongly claims the server never sent a proof for a certain verification or a malicious server wrongly claims it sent its  proof to the client. However, in the case where the proof size is  large  and posting it to the smart contract would impose a high cost, the parties can use the following technique to directly communicate with each other to send and receive the proof. The server sends a  signed proof  directly to the client which needs to send back to the server  a signed acknowledgment stating  that it received the proof, within a fixed time period.  If the server does not receive a valid acknowledgment on time, it sends the signed proof to the arbiter. Moreover, if  the client does not receive the proof on time, it needs to let the arbiter know about it. In this case, if  the arbiter has already received the proof, it  sends the proof  to the client which allows the client to perform the rest of the computation. On the other hand, if the arbiter does not have the proof, it asks the server to send to it the client's acknowledgment. If the server provides a valid acknowledgment, then the arbiter considers the client as  a misbehaving party; otherwise (if the server could not provide the acknowledgment), it considers the server as a misbehaving one. However, if both the server and client behave honestly in sending and receiving the proof, then they do not need to  invoke the arbiter for this matter and the proof is never stored on the blockchain. 

\end{itemize}



 \input{RCSP-Proof}

%% file: RCSP-Proof.tex

\subsection{Security Analysis of RC-S-P Protocol}
In this section, we analyse the security of RC-S-P protocol, presented in Section \ref{sec::RC-S-P-protocol}. First, we  present the protocol’s primary security theorem.

\begin{theorem}\label{thm:generic}
The RC-S-P protocol with functions $F,M,E,D,Q$ presented in Section \ref{sec::RC-S-P-protocol} is secure for auxiliary information $\text{aux}$, (cf. Definition \ref{def::RC-S-P-Security}), if the underlying VSID protocol with functions $F,M,Q$ satisfies correctness, soundness, inputs well-formedness, and detectable abort  for $\text{aux}_j$, the SAP is secure, the signature scheme is secure, and the symmetric-key encryption scheme is IND-CPA secure. 
\end{theorem}

 To prove Theorem~\ref{thm:generic}, we show that RC-S-P meets all security properties defined in Section \ref{sec::RC-S-P-Definition}.  We start by proving that RC-S-P satisfies security against a malicious server.

 \begin{lemma}\label{lem::RC-S-P-SecurityAgainstMaliciousServer}
 If the SAP and signature scheme are secure and the VSID protocol satisfies correctness, soundness, and detectable abort for auxiliary information $\text{aux}$, then the RC-S-P protocol presented in Section \ref{sec::RC-S-P-protocol} is secure against malicious server for $\text{aux}$.(cf.  Definition \ref{deff::RC-S-P-SecurityAgainstMaliciousServer}). 
 \end{lemma}

 \begin{proof}
 We first consider event  
  $$\Big(F(u^{\st *}, \bm{q}_{\st j},{pp})= h_{\st j}\Big)\ \wedge$$ $$ \Big((coin_{\st\mathcal{C},j}\neq  \frac{coin_{\st\mathcal C}^{\st*}}{z}-o)  \vee (coin_{\st\mathcal{R},j}\neq l\ \wedge\  y'_{\st \mathcal {S},j}=1)\Big)$$
   that captures the case where the server provides an accepting service proof but makes an honest client withdraw an incorrect amount of coins, i.e., $coin_{\st\mathcal{C},j}\neq  \frac{coin_{\st\mathcal C}^{\st*}}{z}-o$, or it makes the arbiter withdraw an incorrect amount of coins, i.e., $coin_{\st\mathcal{R},j}\neq l$, if it unnecessarily invokes the arbiter.  As the service proof is valid, an honest client accepts it and does not raise any dispute. However, the server  would be able to make the client  withdraw incorrect amounts of coins, if it manages to either
\begin{enumerate}
\item convince the arbiter that the client has misbehaved, by making the arbiter output  $y_{\st\mathcal {C},j}=1$ through the dispute resolution phase, or
\item     submit to the contract, in the coin transfer phase,  an accepting  statement $\ddot{x}'_{\st cp}$ other than what was agreed in the initiation phase, i.e., $\ddot{x}'_{\st cp}\neq\ddot{x}_{\st cp}$, so it can change the payments' parameters (e.g., $l$ or $o$) or send a message on the client's behalf to invoke the arbiter unnecessarily.
\end{enumerate}   
    Nevertheless, the server cannot falsely accuse the client of misbehaviour. This is because,  due to the security of SAP (i.e., the underlying commitment's binding property), it cannot  convince the arbiter to accept different decryption key or pads other than what was agreed with the client in the initiation phase. Specifically, it cannot persuade the arbiter to accept  $\ddot{x}'_{\st qp}$, where $\ddot{x}'_{\st qp}\neq \ddot{x}_{\st qp}$, except with a negligible probability. This ensures that the honest client's message is accessed by  the arbiter with a high probability, as the arbiter can extract the client's message using valid pad information and decryption key.  On the other hand, if the adversary provides a valid statement, i.e., $\ddot{x}_{\st qp}$, then due to the correctness of VSID, algorithm $\mathtt{VSID.identify}(\cdot)$ outputs $I_{\st j}=\bot$. Therefore, due to the security of SAP (i.e., the binding property) and correctness of VSID,   $y_{\st\mathcal {C}}$ and $y_{\st\mathcal {S}}$ are not incremented by $1$  in the $j$-th verification, i.e., $y_{\st\mathcal{C},j}=y_{\st\mathcal{S},j}=0$.   Also,  due to the security of SAP (i.e., the binding property), the server cannot change the payment parameters by persuading the contract to accept any statement $\ddot{x}'_{\st cp}$ other than what was agreed initially between the client and server, except with a negligible probability when it finds the hash function's collision (in the SAP scheme). Moreover, since the proof is valid the client never raises a dispute, also due to the digital signature's  unforgeability, the server  cannot send a message on behalf of the client (to unnecessarily invoke the arbiter), and make the arbiter output $y'_{\st\mathcal {C},j}=1$ for the $j$-th verification, except with a negligible probability. So with a high probability $y'_{\st\mathcal {C},j}=0$. Recall, in the protocol, the total coins the client should receive after $z$ verifications is $coin^{\st *}_{\st\mathcal C}-o\cdot(z-y_{\st\mathcal S})-l\cdot (y_{\st\mathcal C}+y'_{\st\mathcal C})$. Since we focus on the $j$-th verification, the amount of  coins that should be credited to the client for that verification is
  \begin{equation}\label{equ::what-client-recives-in-j}
   coin_{\st\mathcal{C},j}=\frac{coin^{\st *}_{\st\mathcal C}}{z}-o\cdot(1-y_{\st\mathcal {S},j})-l\cdot(y_{\st\mathcal {C},j}+y'_{\st\mathcal {C},j})
     \end{equation}

As shown above $y_{\st\mathcal{C},j}=y'_{\st\mathcal{C},j}=y_{\st\mathcal{S},j}=0$.  So, according to Equation \ref{equ::what-client-recives-in-j}, the client is credited $\frac{coin_{\st\mathcal C}^{\st*}}{z}-o$ coins for $j$-th verification, with a high probability.  On the other hand, as stated above, if the adversary invokes the arbiter, the arbiter with a high probability outputs $I_{\st j}=\bot$ which results in $y'_{\st\mathcal{S},j}=1$. Recall, in the RC-S-P protocol, the total coins the arbiter should receive for $z$ verifications is $l\cdot(y_{\st\mathcal S}+y_{\st\mathcal C}+y'_{\st\mathcal S}+y'_{\st\mathcal C})$, so for the $j$-th the credited coins should be: 
     \begin{equation}\label{equ::arbiter}
    coin_{\st\mathcal{R},j}= l\cdot(y_{\st\mathcal {S},j}+y_{\st\mathcal {C},j}+y'_{\st\mathcal {S},j}+y'_{\st\mathcal {C},j})
     \end{equation}

As already shown, in the case where arbiter is unnecessarily invoked by the server, it holds that $y'_{\st\mathcal{S},j}=1$; So,  according to Equation \ref{equ::arbiter}, $l$ coins is credited to the arbiter for the $j$-th verification.  For the server to make the arbiter withdraw other than that amount (for the $j$-th verification), in the coin transfer phase, it has to send to the contract  an accepting  statement $\ddot{x}'_{\st cp}$ other than what was agreed in the initiation phase, i.e., $\ddot{x}'_{\st cp}\neq\ddot{x}_{\st cp}$, so it can change the payments' parameters, e.g., $l$ or $o$. But, as argued above, it cannot succeed with a probability significantly greater than negligible, due to the binding property of the SAP's commitment. We now move on to the following event 
     
     $$\Big(F(u^{\st *}, \bm{q}_{\st j}, {pp})\neq h_{\st j}\Big) \ \wedge$$ $$ \Big(d_{\st j}=1	\vee y_{\st \mathcal {S},j}=0 \ \vee coin_{\st\mathcal{C},j}\neq \frac{coin_{\st\mathcal C}^{\st*}}{z} \ \vee  coin_{\st\mathcal{R},j}\neq l\Big)$$
     
     This event captures the case where the server provides an invalid service proof but either persuades the client to accept the proof, or  persuades the arbiter to accept the proof (e.g., when the client raises a dispute) or makes the client or arbiter withdraw an incorrect amount of coins, i.e.,    $coin_{\st\mathcal{C},j}\neq \frac{coin_{\st\mathcal C}^{\st*}}{z}$ or $coin_{\st\mathcal{R},j}\neq l$ respectively.   Nevertheless, due to the soundness of VSID, the probability that a corrupt server can convince an honest client to accept invalid proof (i.e., outputs $d_{\st j}=1$) is negligible. So, the client detects it with a high probability and raises a dispute.  On the other hand, the server may try to convince the arbiter, and make it output $y_{\st \mathcal {S},j}=0$, e.g., by sending a complaint. For $y_{\st \mathcal {S},j}=0$ to happen, the server has to either provide a different accepting statement $\ddot{x}'_{\st qp}$, than what was initially agreed with the client (i.e., $\ddot{x}'_{\st qp}\neq \ddot{x}_{\st qp}$) and passes the verification, which  requires finding the hash function's collision (in the SAP scheme), and its probability of success is negligible. Or it makes the arbiter  accept an invalid proof, but due to the detectable abort property of VSID, its probability of success is also negligible. Also, as we discussed above, the probability that the adversary makes the arbiter to recognise the client as misbehaving, and output $y_{\st \mathcal {C},j}=1$ is  negligible too. Therefore, the arbiter outputs $y_{\st \mathcal {S},j}=1$ and $y_{\st \mathcal {C},j}=0$  with a high probability, in both events when it is invoked by the client or  server. Also,  in this case, $y'_{\st \mathcal {C},j}=y'_{\st \mathcal {S},j}=0$ as the arbiter has already identified a misbehaving party. So, according to Equation \ref{equ::what-client-recives-in-j}, the client is credited $\frac{coin_{\st\mathcal C}^{\st*}}{z}$ coins for that verification, with a high probability.      Moreover, according to Equation \ref{equ::arbiter}, the arbiter is credited $l$ coins for that verification, with a high probability.  The adversary may  try to  make them withdraw an incorrect amount of coins, e.g., in the case where it does not succeed in convincing the client or arbiter. To this end,  in the coin transfer phase, it has to send a  different accepting statement  than what was initially agreed with the client. But, it would succeed only with a negligible probability, due to the security of SAP, i.e., its binding property. 
  \end{proof}

\begin{lemma}\label{lem::RC-S-P-Security-Against Malicious-Client}
 If the SAP and signature scheme are secure and the VSID scheme satisfies correctness, inputs well-formedness, and detectable abort for auxiliary information $\text{aux}$, then the RC-S-P protocol presented in Section \ref{sec::RC-S-P-protocol} is secure against malicious client for $\text{aux}$ (cf. Definition \ref{deff::RC-S-P-Security-Against Malicious-Client}). 
\end{lemma}

\begin{proof}First, we consider event 
   $$\Big(M(u^{\st *},k,{pp})= \sigma \  \wedge \ Q(\text{aux},k, {pp})= \bm{q}_{\st j}\Big)\ \wedge$$ $$
  \Big((coin_{\st\mathcal{S},j}\neq  \frac{coin_{\st\mathcal S}^{\st*}}{z}+o)\ \vee \ (coin_{\st\mathcal{R},j}\neq l\ \wedge\  y'_{\st \mathcal {C},j}=1)\Big)$$

This event captures the case where the client provides  accepting metadata and query  but makes the server withdraw an incorrect amount of coins, i.e., $coin_{\st\mathcal{S},j}\neq  \frac{coin_{\st\mathcal S}^{\st*}}{z}+o$, or makes the arbiter withdraw an incorrect amount of coins, i.e., $coin_{\st\mathcal{R},j}\neq l$, if it unnecessarily invokes the arbiter.  Since the metadata and query's proofs are valid, an honest server accepts them and does not raise any dispute, so we have $y_{\st\mathcal{C},j}=0$.  The client could  make the server  withdraw incorrect amount of coins, if it manages to either convince the arbiter, in phase \ref{RCSP::Dispute-Resolution}, that the server has misbehaved, i.e., makes the arbiter output $y_{\st\mathcal{S},j}=1$, or submit to the contract  an accepting  statement $\ddot{x}'_{\st cp}$ other than what was agreed at the initiation phase, i.e., $\ddot{x}_{\st cp}$, in phase \ref{RCSP::CoinTransfer}, or send a message on the server's behalf to invoke the arbiter unnecessarily. However, it cannot falsely accuse the server of misbehaviour, as due to the security of SAP (i.e., the binding property) it cannot  convince the arbiter to accept different decryption key and pads' detail, by providing a different accepting statement $\ddot{x}'_{\st qp}$  (where $\ddot{x}'_{\st qp}\neq \ddot{x}_{\st qp}$), than what was initially agreed with the server, except with negligible probability. This ensures the arbiter is given the honest server's messages, with a high probability. So, with a high probability $y_{\st\mathcal{S},j}=0$. On the other hand, if the adversary provides a valid statement, i.e., $\ddot{x}_{\st qp}$, then due to the correctness of VSID, algorithm $\mathtt{VSID.identify}(\cdot)$ outputs $I_{\st j}=\bot$. So, due to the security of SAP and correctness of VSID,  we would have $y_{\st\mathcal{C},j}=y_{\st\mathcal{S},j}=0$ with a high probability. Moreover, due to the security of SAP, the client cannot convince the contract to accept any statement $\ddot{x}'_{\st cp}$ other than what was initially agreed  between the client and server (i.e., $\ddot{x}'_{\st cp}\neq \ddot{x}_{\st cp}$), except with  negligible probability. Also, it holds that  $y'_{\st\mathcal{S},j}=0$ because an honest server never invokes the arbiter when the client's messages are well-structured and due to  the signature's unforgeability, the client cannot send a signed message on the server's behalf to unnecessarily invoke the arbiter.  According to  RC-S-P protocol, the total coins the server should receive after $z$ verifications is $coin^{\st *}_{\st\mathcal S}+o\cdot(z-y_{\st\mathcal S})-l\cdot(y_{\st\mathcal S}+y'_{\st\mathcal S})$. Since we focus on the $j$-th verification, the amount of  coins that should be credited to the server for the $j$-th verification is
  \begin{equation}\label{equ::what-server-recives-in-j}
   coin_{\st\mathcal{S},j}=\frac{coin^{\st *}_{\st\mathcal S}}{z}+o\cdot(1-y_{\st\mathcal {S},j})-l\cdot(y_{\st\mathcal {S},j}+y'_{\st\mathcal {S},j})
     \end{equation}

  As shown above, the following holds $y_{\st\mathcal{S},j}=y'_{\st\mathcal{S},j}=0$, which   means, according to  Equation \ref{equ::what-server-recives-in-j}, the server is credited $\frac{coin_{\st\mathcal S}^{\st*}}{z}+o$ coins for the $j$-th verification, with a  high probability.  Furthermore, if the adversary invokes the arbiter, the arbiter with a high probability outputs $I_{\st j}=\bot$ which yields  $y'_{\st\mathcal{C},j}=1$. Also, as stated above,   $y'_{\st\mathcal{S},j}=0$.   Hence, according to Equation \ref{equ::arbiter}, the arbiter for the $j$-th verification is credited $l$ coins, if it is unnecessarily invoked.  As previously stated,  due to the security of SAP, the client cannot make the arbiter withdraw incorrect amounts of coin by changing the payment parameters  and persuading the contract to accept any statement $\ddot{x}'_{\st cp}$ other than what was agreed initially between the client and server, except with negligible probability. We now turn our attention to 
  $$\Big(M(u^{\st *},k,{pp})\neq \sigma\ \wedge \ a=1\Big)$$
     that captures the case where the server accepts an ill-formed metadata. However, due to inputs well-formedness of VSID, the probability that event happens is negligible. So, with a high probability $a=0$.  Note, in the  case where $a=0$, the server does not raise any dispute, instead it avoids serving the client. Next, we move on to 
     $$\Big(Q(\text{aux},k,{pp})\neq \bm{q}_{\st j}\Big)\ \wedge$$ $$  \Big(b_{\st j}=1\ \vee \ y_{\st\mathcal{C},j}=0 \ \vee \ coin_{\st\mathcal{S},j}\neq \frac{coin_{\st\mathcal S}^{\st*}}{z}+o \ \vee \ coin_{\st\mathcal{R},j}\neq l\Big)$$
     
 This event considers the case where the client provides an invalid query, but  either convinces the server or arbiter to accept it, or  makes the server or arbiter  withdraw an incorrect amount of coins, i.e., $coin_{\st\mathcal{S},j}\neq \frac{coin_{\st\mathcal S}^{\st*}}{z}+o$ or  $coin_{\st\mathcal{R},j}\neq l$ respectively. Nevertheless, due to inputs well-formedness of VSID, the probability that the server outputs $b_{\st j}=1$ in this case is negligible. When the server rejects the query and raises a dispute, the client may try to convince the arbiter and make it output $y_{\st \mathcal {C},j}=0$, e.g., by sending a complaint. However, for the adversary to win, either 
\begin{enumerate}
\item it has to provide a different accepting statement $\ddot{x}'_{\st qp}$, than what was initially agreed with the server (i.e., $\ddot{x}'_{\st qp}\neq \ddot{x}_{\st qp}$) and passes the verification. Due to the security of SAP, its probability of success is negligible. Or,
\item  it has to make the arbiter  accept an invalid query, i.e., makes the arbiter output $y_{\st\mathcal{C},j}=0$. Due to the detectable abort property of VSID, its probability of success is negligible too.
\end{enumerate} 
 Therefore, with a high probability, we have $y_{\st\mathcal{C},j}=1$. Also, as discussed above (due to the security of SAP), the client cannot make the arbiter recognise the  honest server as a misbehaving party with a probability significantly greater than negligible. That means with a high probability $y_{\st\mathcal{S},j}=0$. Furthermore, as we already discussed, since the arbiter 
has identified a misbehaving party, the following holds $y'_{\st\mathcal{C},j}=y'_{\st\mathcal{S},j}=0$.  Hence, according to Equation \ref{equ::what-server-recives-in-j} the server is credited  $\frac{coin_{\st\mathcal S}^{\st*}}{z}+o$ coins for this verification. Also, the arbiter is credited $l$ coins, according to Equation \ref{equ::arbiter}.  Note that the adversary may still try to  make them withdraw an incorrect amount of coins (e.g., if the adversary does not succeed in convincing the server or arbiter). To this end, at the coin transfer phase, it has to send a  different accepting statement  than what was initially agreed with the server. However, due to the security of SAP (i.e., binding property), its  success probability is negligible. 
 \end{proof}
  
  
  Prior to proving  RC-S-P's privacy, we provide a lemma that will be used in the privacy's proof. Informally, the lemma states that encoded coins leaks no information about the actual amount of coins $(o,l)$, agreed between the client and server.

\begin{lemma}\label{lemma::encoded-coins} Let $\beta\stackrel{\st\$}\leftarrow\{0,1\}$, price list be $\{(o_{\st 0},l_{\st 0}),(o_{\st 1},l_{\st 1})\}$, and encoded coin amounts be $coin^{\st *}_{\st\mathcal{C}}= z\cdot(Max(o_{\st\beta}, $ $o_{\st1-\beta})$ $+Max(l_{\st\beta},l_{\st1-\beta}))$ and $coin^{\st *}_{\st\mathcal{S}}=z\cdot(Max(l_{\st\beta}, l_{\st 1-\beta}))$. Then, given the price list, $z$, $coin^{\st *}_{\st\mathcal{C}}$, and $coin^{\st *}_{\st\mathcal{S}}$, an adversary $\mathcal{A}$ cannot tell the value of $\beta$ with a probability significantly greater than $\frac{1}{2}$ (where the probability is taken over the choice of $\beta$ and the randomness of $\mathcal{A}$).
\end{lemma}

\begin{proof}
As it is evident, the list and $z$ contains no information about $\beta$. Also, since  $z$ is a public value, it holds that   $coin'^{\st *}_{\st\mathcal C}=\frac{coin^{\st *}_{\st\mathcal C}}{z} = Max(o_{\st\beta},o_{\st1-\beta})+Max(l_{\st\beta}, l_{\st1-\beta})$. It is not hard to see $coin'^{\st *}_{\st\mathcal C}$ is a function of maximum value of $(o_{\st 0},o_{\st 1})$, and maximum value of $(l_{\st 0},l_{\st 1})$. It is also  independent of $\beta$. Therefore (given the list, $z$ and $coin'^{\st *}_{\st\mathcal C}$) the adversary learns nothing about $\beta$, unless it guesses the value,  with  success probability $\frac{1}{2}$. The same also holds for $coin^{\st *}_{\st\mathcal S}$. 
 \end{proof}

\begin{lemma}\label{lem::RC-S-P-Privacy} If SAP is secure and the symmetric-key encryption scheme is IND-CPA secure, then the RC-S-P protocol presented in Section \ref{sec::RC-S-P-protocol} preserves privacy for auxiliary information $\text{aux}$, (cf. Definition \ref{deff::RC-S-P-Privacy}). 
\end{lemma}

\begin{proof} We start with  case $1$, i.e., the privacy of service input. Due to the privacy property of SAP, that stems from the hiding property of the commitment scheme, given the commitments $g_{\st qp}$ and $g_{\st cp}$, (that are stored in the blockchain as a result of running SAP) the adversary learns no information about the committed values (e.g., $o, l, pad_{\st\pi}, pad_{\st q},$ and $\bar{k}$), except with a negligible probability.   Also, given  price list $pl$, encoded coins $coin^{\st *}_{\st\mathcal{C}}=z\cdot (o_{\st max}+l_{\st max})$ and  $coin^{\st *}_{\st\mathcal{S}}=z\cdot l_{\st max}$, the adversary learns nothing about the actual price that was agreed between the server and client,  $(o,l)$, for each verification, due to Lemma \ref{lemma::encoded-coins}. Next we analyse the privacy of padded encrypted query vector $\bm{c}^{\st *}$. For  the sake of simplicity, we focus on   $\bm{q}^{\st *}_{\st j}\in c^{\st *}_{\st j}\in\bm{c}^{\st *}$, that is a padded encrypted query vector for $j$-th verification. Let $\bm{q}_{\st j,0}$ and $\bm{q}_{\st j,1}$  be query vectors, for $j$-th verification, related to the service inputs $u_{\st 0}$ and $u_{\st 1}$ that    are picked by the adversary according to  Definition  \ref{deff::RC-S-P-Privacy} which lets  the environment pick $\beta\stackrel{\st\$}\leftarrow \{0,1\}$. Also, let $\{\bm{q}_{\st j,0},...,\bm{q}_{\st j,\bar{m}}\}$ be a  list of all queries of different sizes. In the experiment, if $\bm{q}_{\st j,\beta}$  is only   encrypted (but not padded), then given the ciphertext, due to semantical security of the encryption, an adversary cannot tell if the ciphertext corresponds to $\bm{q}_{\st j,0}$ or $\bm{q}_{\st j,1}$  (accordingly to $u_{\st 0}$ or $u_{\st 1}$) with probability significantly greater than $\frac{1}{2}+\mathsf{negl}(\lambda)$,  under the assumption  that the size of $\bm{q}_{\st j,\beta}$ is equal to the size of largest query size \footnote{\scriptsize The assumption that all queries have the same size is subsumed under the above assumption.}, i.e.,  $Max(|\bm{q}_{\st j,0}|,...,|\bm{q}_{\st j,\bar{m}}|)=|\bm{q}_{\st j,\beta}|$. The  above assumption is  relaxed with the use of a pad; as each encrypted query is padded  to the queries' maximum size, i.e., $Max(|\bm{q}_{\st j,0}|,...,|\bm{q}_{\st j,\bar{m}}|)$, the adversary cannot tell with a probability greater than $\frac{1}{2}+\mathsf{negl}(\lambda)$ if the  padded encrypted proof corresponds to  $\bm{q}_{\st j,0}$ or $\bm{q}_{\st j,1}$, as the padded encrypted query \emph{always has the same size} and the pad values are picked from the same range as the encryption's ciphertext are defined. The same argument holds for $\bm{w}^{\st *}_{\st q_{_{\scaleto{j}{\size}}}}\in c^{\st *}_{\st j}\in\bm{c}^{\st *}$.  Next we analyse the privacy of padded encrypted proof vector $\bm{\pi}^{\st *}$. The argument is similar to the one presented above, however, we provide it for  the sake of completeness.  We focus on an element of the vector,  $\pi_{\st j}^{\st *}\in\bm{\pi}^{\st *}$, that is a padded encrypted proof for $j$-th verification. Let $\pi_{\st j,0}$ and $\pi_{\st j,1}$  be proofs, for $j$-th verification, related to the service inputs $u_{\st 0}$ and $u_{\st 1}$, where the inputs   are picked by the adversary, w.r.t. Definition  \ref{deff::RC-S-P-Privacy} in which  the environment picks $\beta\stackrel{\st\$}\leftarrow \{0,1\}$.  Let $\{\pi_{\st j,0},...,\pi_{\st j,\bar{m}}\}$ be proof list including all proofs of different sizes.  If we assume $\pi_{\st j,\beta}$  is only   encrypted, then given the ciphertext, due to semantical security of the encryption, an adversary cannot tell if the ciphertext corresponds to $\pi_{\st j,0}$ or $\pi_{\st j,1}$  (accordingly to $u_{\st 0}$ or $u_{\st 1}$) with a probability significantly greater than $\frac{1}{2}+\mathsf{negl}(\lambda)$,  if $Max(|\pi_{\st j,0}|,...,|\pi_{\st j,\bar{m}}|)=|\pi_{\st j,\beta}|$. However, the  assumption is  relaxed with the use of a pad. In particular, since each encrypted proof is padded  to the proofs' maximum size, the adversary cannot tell with a probability greater than $\frac{1}{2}+\mathsf{negl}(\lambda)$ if the  padded encrypted proof corresponds to  $\pi_{\st j,0}$ or $\pi_{\st j,1}$. Also, since the value of $a$ is independent of $u_{\st 0}$  or $u_{\st 1}$, and only depends on whether the metadata is well-formed, it leaks nothing about the service input $u_{\st\beta}$, $\beta$, the query-proof pair and service proof. Thus (given  $\bm{c}^{\st *},coin^{\st *}_{\st \mathcal S},coin^{\st *}_{\st\mathcal C}, g_{\st cp},  g_{\st qp},$  $\bm{\pi}^{\st *},pl$, and $a$) the probability that the adversary can tell the value of $\beta$ is at most $\frac{1}{2}+\mathsf{negl}(\lambda)$, due to IND-CPA security of the symmetric-key encryption scheme.




Now we move on to  case 2, i.e., the privacy of proof's status. Recall that in the experiment, an \emph{invalid} query-proof pair is generated with probability $Pr_{\st 0,j}$ and a \emph{valid} query-proof pair is generated with probability $Pr_{\st 1,j}$.  As stated above, each encoded query-proof pair $c^{\st *}_{\st j}\in\bm{c}^{\st *}$ has a fixed size and contains random elements of $U$, i.e., they are uniformly random elements in the symmetric-key encryption scheme's output range. Also, it is assumed that for each $j$-th verification, an encoded query-proof is always provided to the contract. Therefore, each encoded pair leaks nothing, not even the query's status to the adversary (due to the use of padding and IND-CPA security of the encryption). So, given only a vector of $c^{\st *}_{\st j}$ (i.e., $\bm{c}^{\st *}$) it can learn a query-proof's status with probability at most $Pr'+\mu(\lambda)$, where $Pr':=Max\{Pr_{\st 0,1},Pr_{\st 1,1},..., Pr_{\st 0,z},Pr_{\st 1,z}\}$.  On the other hand, for each $j$-th verification, an encoded service proof $\pi_{\st j}^{\st *}\in\bm{\pi}^{\st *}$ is always provided to the contract, regardless of the query's status. As stated above, each $\pi_{\st j}^{\st *}$ has a fixed size and contains random element of $U$ too.   As we showed above, $g_{\st cp},  g_{\st qp},pl$, and $a$ leak no information about the service input, except with a negligible probability, $\mu(\lambda)$. They are also independent of the query-proof pair and service proof, so they leak no information about the pair and service proof too. So, given   $\bm{c}^{\st *},coin^{\st *}_{\st \mathcal S},coin^{\st *}_{\st\mathcal C}, g_{\st cp},  g_{\st qp},$  $\bm{\pi}^{\st *},pl$, and $a$, an adversary has to learn a proof's status from the aforementioned values or by correctly guessing a query's status. In other words, its probability of learning a proof' status is at most $Pr'+\mu(\lambda)$.
 \end{proof}

%
%
%

%% file: trade-offs.tex

\section{Trade-offs}\label{sec::Trade-offs}
As we highlighted in Section \ref{sec::Overview-of-Our-Solution}, to design an efficient RC-S-P we have made certain trad-offs compared to the previous work in  \cite{CampanelliGGN17}. In this section, we explain the main trad-offs that we have made. 

\subsection{Involving a Third-party Arbiter}

Similar to various works that aim for efficiency (e.g., in  \cite{AsokanSW97,eurocrypt/AsokanSW98,BaoDM98,DongCCR13}) in the RC-S-P we involved an arbiter third party that remains offline for the most part of the protocol execution and is invoked only in the case where parties raise a dispute. This is a trade-off because the original (but insecure) zkCSP in \cite{CampanelliGGN17} did not involve any additional (third) party.

Instead, we could allow Ethereum smart contracts to play the arbiter's role if we would use only publicly verifiable proofs and delegate the arbiter-side computations to the smart contracts. However, this would impose a much higher financial cost on the parties than what RC-S-P does. Because performing a certain computation by smart contract would cost much more than a third party performing the same computation locally.

\subsection{Requiring Parties to Deposit Extra Coins}

The amount of coins that the RC-S-P requires parties to deposit is higher than what zkCSP in \cite{CampanelliGGN17} requires; for two reasons.  First, in the RC-S-P, parties hide the amount of the deposit by masking it; to do that, they increase the amount to the maximum amount of coins in the server’s price list. This trade-off has been made to achieve privacy while keeping the computation cost low. 

Second, in the RC-S-P, every party needs to deposit a certain amount to cover the cost of the arbiter's computation; this deposit will be transferred to the arbiter in the case where the arbiter is invoked and the party was identified as misbehaving; however, when the party is honest, it can retrieve its deposit when the protocol terminates.  This trade-off has been made to efficiently detect a misbehaving party. 

\subsection{Storing more Messages on Blockchain}

In the RC-S-P most of the messages exchanged between the client and server are stored in the blockchain, unlike the original zkCSK which requires a few messages to be stored on the blockchain. Therefore, the RC-S-P would impose a higher communication cost than the zkCSK does.  
This trade-off has been made to satisfy the RC-S-P security requirement. Specifically, in the RC-S-P, both parties can be malicious, whereas the zkCSK (wrongly) assumes that only the server is malicious.  in particular, to ensure a malicious party cannot deny receiving a message.  

\subsection{Use of Ethereum Smart Contracts}

Unlike the original zkCSP that can be implemented in Bitcoin (which supports very limited functionalities from smart contracts), the RC-S-P relies on Ethereum smart contracts that support wider functionalities. However, the RC-S-P  relies on very minimal capabilities of Ethereum smart contracts, such as running very simple functions and distributing deposits according to the output of the functions. This trade-off has been made to satisfy the RC-S-P security requirements, e.g., privacy, ensuring a malicious client cannot waste the server's resources,  and ensuring a fair exchange.

%% file: SAP-discussion.tex


\section{Further Discussion of the SAP}\label{sec:Discussion-on-the-SAP}

In this section, first we outline why the SAP satisfies all four security properties set out in Section \ref{SAP} and then discuss  why naive solutions are not suitable replacements of the SAP. After that,  we highlight that the SAP's verification phase can be locally performed with low costs.

\subsection{SAP's Security Analysis} Intuitively,  the SAP meets Property 1  due to the binding property of the  commitment scheme. Property 2 is satisfied due to the security of the blockchain and smart contract; namely, due to blockchain's liveness property an honestly generated transaction, containing the opening,  eventually gets into  chains of honest miners, and due to  the security and correctness of smart contracts a valid opening is always accepted by the contract. Property 3 is met due to the hiding property of the commitment, while  Property 4 is satisfied due to the  signature scheme's security.


\subsection{Unsuitability of Naive Solutions} As a replacement of the SAP, one may  let each party  sign the statement and send it to the other party, so later  each party can send both signatures to the contract which verifies them. However, this would not work,  as the party who first receives the other party's signature  may refuse  to send its own signature, that prevents the other party from proving that it has  agreed on the statement with its counter-party, i.e., cannot satisfy  Property 2. Alternatively, one may want to use a protocol for a fair exchange of digital signature (or fair contract signing) such as  those in \cite{BonehN00,DBLP:conf/fc/GarayJ02}. In this case, after both parties have the other party's signature, they can sign the statement themselves and send the two signatures to the contract which first checks the validity of both  signatures. Although this satisfies the four security requirements, it yields two main \emph{efficiency} and \emph{practical} issues; namely, it (a)  imposes very high computation costs, as  protocols for a fair exchange of signatures involve generic zero-knowledge proofs and require a high number of modular exponentiations, and (b)  is impractical because protocols for the fair exchange of signatures  support only certain signature schemes (e.g., RSA, Rabin, or Schnorr) that are not directly supported by the most predominant  smart contract framework,  Ethereum, that only supports  Elliptic Curve Digital Signature Algorithm (EDCSA).

\subsection{Off-chain Verification in the SAP}
The SAP's verification algorithm can  be executed \emph{off-chain}. In particular, given  statement $\ddot{x}$, anyone can read $(g_{\st\mathcal C},g_{\st\mathcal S},adr_{\st\mathcal{C}}, adr_{\st\mathcal{S}})$ from the SAP smart contract and locally run $\mathtt{SAP.verify}(\ddot{x}, g_{\st\mathcal C}, g_{\st\mathcal S},adr_{\st\mathcal{C}}, $ $adr_{\st\mathcal{S}})$ to check the statement's correctness.  This relieves  the verifier from the  transaction  and smart contract's execution costs.

%% file: correctness.tex

\section{RC-PoR-P's Correctness}\label{sec::RC-PoR-P-Correctness}

In this section, we briefly discuss why the correctness of the RC-PoR-P protocol holds, w.r.t. Definition \ref{def::RC-S-P-Correctness}. Recall, correctness requires that $\mathcal{S}$ accepts an honest $\mathcal{C}$'s encoded data and query while honest $\mathcal{C}$ accepts $\mathcal{S}$'s valid service proof. Also,  honest $\mathcal{C}$ gets back all its deposited coins minus the service payment, the honest $\mathcal{S}$ gets back all its deposited coins plus the service payment and the arbiter receives nothing. In short, this protocol's correctness holds dues to the correctness of PoR, symmetric key encryption, SAP, and smart contract. Below, we elaborate on that:

\begin{enumerate}
\item\label{smart-contract-correctness} due to the correctness of smart contracts,  any message sent by a party (i.e., $\mathcal{C}$ or $\mathcal{S}$) to a contract is kept intact by the contract.
\item\label{sym-key-correctness} due to the correctness of symmetric key encryption, the encrypted messages sent by a party to another one can be correctly decrypted by its counterparty who knows the correct key.  
\item\label{contract-and-sym-key-correctness} due to points \ref{smart-contract-correctness} and \ref{sym-key-correctness} above, an encrypted message sent to a contract by a party can be correctly decrypted by its counterparty who knows the correct key. 
\item\label{PoR-correctness} due to the correctness of the underlying PoR scheme, proofs generated by honest  $\mathcal{S}$ are always accepted by honest $\mathcal{C}$. Also, for the same reason, a query generated by honest $\mathcal{C}$ is always accepted by honest $\mathcal{S}$. 
\item  due to points \ref{smart-contract-correctness}-\ref{PoR-correctness}, the counters remain $0$, i.e., $y_{\st\mathcal S}=y_{\st\mathcal C}=y'_{\st\mathcal S}=y'_{\st\mathcal C}=0$. 
\item due to the correctness of SAP and point \ref{smart-contract-correctness}, a private statement's proof (e.g., $\ddot{x}_{\st cp}\in T_{\st cp}$) sent by an honest $\mathcal{C}$ or $\mathcal{S}$ will always be accepted by the smart contract. 
\end{enumerate}

Therefore,  $\mathcal C$ receives $coin_{\st\mathcal C}=coin^{\st *}_{\st\mathcal C}-o\cdot z$ coins, $\mathcal S$ receives $coin_{\st\mathcal S}=coin^{\st *}_{\st\mathcal S}+o\cdot z$, and arbiter receives $0$ coins.

%% file: appendix_RC-PoR-P.tex
\section{Proof of Theorem~\ref{thm:RC-PoR-P}}\label{app:RC-PoR-P_proof}

This section contains the security analysis of the RC-PoR-P consrtuction presented in Section~\ref{sec:RC-PoR-P_protocol}. First, we prove the security of the PoR scheme in Subsection~\ref{protocol::PoR} in the following lemma.

\input{PoR-proof}
\input{RCPoRP-Proof}

%% file: PoR-proof.tex


\begin{lemma}\label{lemma::PoR-protocol}
Let $\epsilon$ be non-negligible in the security paramenter $\lambda$. Then, the PoR scheme presented in Subsection \ref{protocol::PoR} is $\epsilon$-sound w.r.t. Definition~\ref{extractable}, if the underlying Merkle tree and pseudorandom function $\mathtt{PRF}$ are secure.
\end{lemma}

 \begin{proof}[Proof (sketch)]
 As stated above, the proposed PoR differs from the standard Merkle tree-based PoR by a couple of perspectives. However, the changes do not affect the security and soundness of the proposed PoR.  Its security proof is similar to the existing Merkle tree-based PoR schemes, e.g., \cite{DBLP:conf/ccs/HaleviHPS11,DBLP:journals/iacr/JuelsK07,MillerPermacoin}. Alternatively, our protocol can be proven based on the security analysis of the PoR schemes that use  MACs or BLS signatures, e.g., \cite{DBLP:conf/asiacrypt/ShachamW08}. In this case, the extractor design (in the Merkle tree-based PoR)  would be simpler because it does not need to extract blocks from a linear combination of MACs or signatures, as the  blocks are included in  proofs, i.e., they are part of the Merkle tree proofs.  Intuitively, in either case, the extractor interacts with any adversarial prover that passes a non-negligible $\epsilon$ fraction of audits. It initialises an empty array. Then it  challenges a subset of file blocks and asks the prover (server) to generate a proof.  If the received proof passes the verification, then it adds the related block (in the proof) to the array. It then rewinds the prover and challenges a fresh set of blocks, and repeats the process  many times. Since the prover has a good chance of passing the audit, it is easy to show that the extractor can eventually extract a large fraction of the entire file, as it is shown in \cite{DBLP:conf/asiacrypt/ShachamW08}. In particular, the following hold:
\begin{enumerate}
\item Due to the security of the Merkle tree, which relies on the collision resistance of the utlized hash function, whenever the extractor is convinced of a prover's proof of membership, it is ascertained that the retrieved values are the valid and correct file blocks.
\item Due to the security of the pseudorandom function, the challenges $ q_{\st i}=\big(\mathtt{PRF}(\hat{k},i)\bmod m\big)+1$, for $i=1,\ldots,\phi$ are not predictable by $\mathcal{S}$ before the time that the server receives the PRF key $\hat{k}$ by $\mathcal{C}$. 
\end{enumerate}
 After collecting a sufficient number of blocks, the extractor can use the error-correcting code to decode and recover the entire file blocks, given the retrieved ones. 
\end{proof}

%% file: RCPoRP-Proof.tex



By applying Lemma~\ref{lemma::PoR-protocol}, we prove the main  theorem of Section~\ref{sec:RC-PoR-P_protocol}. Note that, given the value of a counter, e.g., $y_{\st \mathcal{S}}$, we can easily parse it as a binary vector, e.g. $y_{\st \mathcal{S},1},..., y_{\st \mathcal{S},z}$. This is because each counter is maintained and incremented by the smart contract, hence the counter's history/log can be retrieved from the blockchain.   \\

\noindent\textbf{Theorem~\ref{thm:RC-PoR-P}.}
\emph{The RC-PoR-P scheme with functions $F_{\st \mathsf{PoR}},M_{\st \mathsf{PoR}},E_{\st \mathsf{PoR}},$ $D_{\st \mathsf{PoR}},Q_{\st \mathsf{PoR}}$ described in Subsections~\ref{protocol::PoR} and ~\ref{sec::RC-PoR-P-protocol} is secure (cf. Definition \ref{def::RC-S-P-Security}), if the underlying Merkle tree, pseudorandom function, commitment scheme, digital signature scheme are secure, and the underlying symmetric-key encryption scheme is IND-CPA secure.}

\begin{proof}
We show that  the RC-PoR-P scheme meets all security properties defined in Section \ref{sec::RC-S-P-Definition} by proving a series of claims. First, we recall  that  $coin_{\st\mathcal{P},j}$ denotes the coins that are credited to the party $\mathcal{P}\in\{\mathcal{C},\mathcal{S},\mathcal{R}\}$ for the $j$-th verification and $h_{\st j}$ is a value included in the decoded proof $\mathbf{\pi}_{\st j}$ that should match $F_{\st \mathsf{PoR}}(u^{\st *}, \hat{k}_{\st j},{pp})$. In addition, $y_{\st \mathcal{C},j}=1$ (resp. $y_{\st \mathcal{S},j}=1$) if $\mathcal{C}$ (resp. $\mathcal{S}$) misbehaved in the $j$-th billing cycle, and $y'_{\st \mathcal{C},j}=1$ (resp. $y'_{\st \mathcal{S},j}=1$) if $\mathcal{C}$ (resp. $\mathcal{S}$) has provided a complaint that does not allow $\mathcal{R}$ to identify a misbehaved party in the $j$-th verification .

\begin{claim}\label{claim::RC-PoR-P-SecurityAgainstMaliciousServer}
 The RC-PoR-P scheme with functions $F_{\st \mathsf{PoR}},M_{\st \mathsf{PoR}},E_{\st \mathsf{PoR}},$ $D_{\st \mathsf{PoR}},Q_{\st \mathsf{PoR}}$ is secure against a malicious server (cf. Definition \ref{deff::RC-S-P-SecurityAgainstMaliciousServer}), if the SAP and signature scheme are secure,  and the PoR scheme satisfies correctness and soundness. 
\end{claim}
\noindent\textbf{Proof of Claim~\ref{claim::RC-PoR-P-SecurityAgainstMaliciousServer}.}
First, we  consider the event    $\Big(F_{\st \mathsf{PoR}}(u^{\st *}, \hat{k}_{\st j},{pp})= h_{\st j}\ \wedge \Big((coin_{\st\mathcal{C},j}\neq  \frac{coin_{\st\mathcal C}^{\st*}}{z}-o)  \vee (coin_{\st\mathcal{R},j}\neq l\ \wedge\  y'_{\st \mathcal {S},j}=1)\Big)\Big)$ that captures the case where the server provides an accepting  proof, but one of the following happen:
\begin{enumerate}
\item an honest client withdraws an incorrect amount of coins, i.e., $coin_{\st\mathcal{C},j}\neq  \frac{coin_{\st\mathcal C}^{\st*}}{z}-o$, or
\item the arbiter withdraws an incorrect amount of coins, i.e., $coin_{\st\mathcal{R},j}\neq l$, if the server unnecessarily invokes the arbiter, i.e., $y'_{\st \mathcal {S},j}=1$
\end{enumerate}
 
 Because the  proof is valid, an honest client accepts it and does not raise a dispute. However, the server  could make the client  withdraw incorrect amount of coins, if it manages to (i) convince the arbiter that the client has misbehaved, by making the arbiter output  $y_{\st\mathcal {C},j}=1$ through the dispute resolution phase, or (ii)  submit an accepting  statement $\ddot{x}'_{\st cp}$ to SC which is other than what was agreed in the initiation phase, i.e., $\ddot{x}'_{\st cp}\neq\ddot{x}_{\st cp}$, so it can change the payments' parameters, or (iii) send a message on the client's behalf to unnecessarily invoke the arbiter. We argue that in any of the above three cases, the server cannot falsely accuse the client of misbehaviour. 

First, due to the binding property of the SAP commitment scheme, $\mathcal{S}$ cannot convince the arbiter to accept a different decryption key (that will be used to decrypt queries) other than what was agreed with the client in the SAP initiation phase. In particular,  it cannot persuade the arbiter to accept  $\ddot{x}'_{\st qp}$, where $\ddot{x}'_{\st qp}\neq \ddot{x}_{\st qp}$, except with $\mathsf{negl}(\lambda)$ probability. This ensures that the honest client's queries are accessed by the arbiter with a high probability.   Furthermore, if the adversary provides a valid statement, i.e., $\ddot{x}_{\st qp}$, then due to the correctness of the PoR and query-checking process (specified in step \ref{RC-PoR-P::checkquery}), no one is identified as a misbehaving party in the dispute resolution phase, i.e., so we would have  $I_{\st j}=\bot$. Therefore, due to the binding property of SAP and correctness of PoR and query-checking process,   the following holds  $y_{\st\mathcal{C},j}=y_{\st\mathcal{S},j}=0$. 

Moreover, due to the binding property of the SAP commitment scheme, the server cannot change the payment parameters by convincing the contract to accept any statement $\ddot{x}'_{\st cp}$ other than what was agreed initially between the client and server, except with $\mathsf{negl}(\lambda)$ probability. Also, due to the EUF-CMA security of the underlying digital signature scheme, the adversary  cannot send a message on behalf of the client to unnecessarily invoke the arbiter and make  it output $y'_{\st\mathcal {C},j}=1$, except with $\mathsf{negl}(\lambda)$ probability; so with high probability, it holds that $y'_{\st\mathcal {C},j}=0$.  Recall that by the description of the coin transfer phase, in RC-PoR-P or RC-S-P  protocol, according to Equation \eqref{equ::what-client-recives-in-j}, the amount of  coins that should be credited to the client for the $j$-th verification is $coin_{\st\mathcal{C},j}=\frac{coin^{\st *}_{\st\mathcal C}}{z}-o\cdot(1-y_{\st\mathcal {S},j})-l\cdot(y_{\st\mathcal {C},j}+y'_{\st\mathcal {C},j})$. Since it holds that $y_{\st\mathcal{C},j}=y_{\st\mathcal{S},j}=y'_{\st\mathcal {C},j}=0$,  the client is credited $\frac{coin_{\st\mathcal C}^{\st*}}{z}-o$ coins for the $j$-th verification, with high probability.   

As stated above, if the adversary invokes the arbiter, the arbiter with a high probability outputs $I_{\st j}=\bot$ that yields  $y'_{\st\mathcal{S},j}=1$. In  RC-PoR-P or RC-S-P protocol, according to Equation \ref{equ::arbiter}, the amount of coins the arbiter should be credited for $j$-th verification is $coin_{\st\mathcal{R},j}= l\cdot(y_{\st\mathcal {S},j}+y_{\st\mathcal {C},j}+y'_{\st\mathcal {S},j}+y'_{\st\mathcal {C},j})$. As shown above $y_{\st\mathcal{C},j}=y_{\st\mathcal{S},j}=y'_{\st\mathcal{C},j}=0$ and $y'_{\st\mathcal{S},j}=1$, which   means $l$ coins  is credited to the arbiter for the $j$-th verification if it is unnecessarily invoked by the adversary.  In this case, for the server to make the arbiter withdraw other than this amount, it has to send to SC (in the coin transfer phase) an accepting  statement $\ddot{x}'_{\st cp}$ other than what was agreed in  the initiation phase, i.e., $\ddot{x}'_{\st cp}\neq\ddot{x}_{\st cp}$, so it can change the payments' parameters. However, as stated above, it cannot succeed with a probability significantly greater than $\mathsf{negl}(\lambda)$ due to the binding property of the underlying commitment scheme.

 We now study the event  $\Big(\Big(F_{\st \mathsf{PoR}}(u^{\st *}, \hat{k}_{\st j}, {pp})\neq h_{\st j}\Big) \ \wedge\ \Big(d_{\st j}=1	\vee y_{\st \mathcal {S},j}=0 \ \vee coin_{\st\mathcal{C},j}\neq \frac{coin_{\st\mathcal C}^{\st*}}{z} \ \vee  coin_{\st\mathcal{R},j}\neq l\Big)\Big)$, which captures the case where the server provides an invalid  proof and causes at least one of the following to happen:
 \begin{enumerate}
 \item convinces the client/arbiter to accept the proof, or
 \item persuades the arbiter to consider it as honest, i.e., keep  $y_{\st \mathcal {S},j}=0$, or
 \item makes the client or arbiter withdraw incorrect amount of coins, i.e.,    $coin_{\st\mathcal{C},j}\neq \frac{coin_{\st\mathcal C}^{\st*}}{z}$ or $coin_{\st\mathcal{R},j}\neq l$ respectively.  
 \end{enumerate}
 
By the security of the underlying pseudorandom function and the Merkle tree, Lemma~\ref{lemma::PoR-protocol} holds, i.e., the tree-based PoR building block is sound. This implies that (i) the probability that the adversary can convince an honest client/arbiter to accept invalid proof is $\mathsf{negl}(\lambda)$ and (ii) the file is extractable within a polynomial number of interactions with an $\epsilon$-admissible adversary, where $\epsilon$ is some non-negligible function. Therefore, the client outputs $d_{\st j}=0$  with a high probability and raises a dispute. 

Furthermore, the server may try to convince the arbiter that the client has misbehaved, and output $y_{\st \mathcal {C},j}=1$; if it succeeds, then the arbiter would consider the server as honest.  Therefore, it would keep $y_{\st \mathcal {S},j}=0$ (according to the protocol's description).  
In this case, according to RC-PoR-P protocol, if the honest client sends a complaint,  its complaint (for the $j$-th verification) would not be processed by the arbiter, because the arbiter has already detected a misbehaving party, i.e., the client in this case.
This  allows   $y_{\st \mathcal {S},j}$ to remain $0$. Nonetheless, as we argued in the study of the previous event, due to the binding property of the SAP commitment scheme and the EUF-CMA security of the digital signature scheme, the probability that the adversary makes the arbiter  recognise the client as misbehaving is $\mathsf{negl}(\lambda)$. 

By the above two paragraphs, with high probability the honest client will raise a dispute and the malicious server cannot convince the arbiter that the client has misbehaved. This means that, with high probability, $y_{\st \mathcal {S},j}=1$ and $y_{\st \mathcal {C},j}=0$, after the arbiter  is invoked by the client or  server.

In addition, it holds that $y'_{\st \mathcal {C},j}=0$ with a high probability, as the honest client will not unnecessarily invoke the arbiter and the server cannot send a message to the arbiter (to unnecessarily invoke it) on behalf of the client (to make the arbiter set $y'_{\st \mathcal {C},j}=1$), due to the digital signature’s EUF-CMA security. 
 Also, it holds that $y'_{\st \mathcal {S},j}=0$, because if the malicious server unnecessarily invokes the arbiter after it is detected, then the arbiter discards its complaint without carrying out any investigation/computation as a malicious party (i.e., the server in this case) has already been identified. 
 Moreover, due to the binding property of the SAP commitment scheme, the probability that the adversary succeeds in changing the payment parameters to make the client or arbiter withdraw an incorrect amount of coins is $\mathsf{negl}(\lambda)$ too.   So, according to Equations \eqref{equ::what-client-recives-in-j} and \eqref{equ::arbiter}  the client and arbiter are credited $\frac{coin_{\st\mathcal C}^{\st*}}{z}$ and $l$ coins for the $j$-th verification respectively. Also, due to the liveness of the blockchain and the security and correctness of the smart contract (cf. Appendix~\ref{sec:Discussion-on-the-SAP}), the second security property of SAP holds, i.e., after the parties agree on the statement, an honest party can almost always prove to the verifier that it has the agreement. Thus, the adversary cannot block an honest client's  messages, ``pay'' and $\ddot{x}_{\st cp}$, to the contract in the coin transfer phase.\hfill$\dashv$

\begin{claim}\label{claim::RC-PoR-P-SecurityAgainstMaliciousClient}
  The RC-PoR-P scheme with functions $F_{\st \mathsf{PoR}},M_{\st \mathsf{PoR}},E_{\st \mathsf{PoR}},$ $D_{\st \mathsf{PoR}},Q_{\st \mathsf{PoR}}$ is secure against a malicious client (cf. Definition \ref{deff::RC-S-P-Security-Against Malicious-Client}), if SAP and signature scheme are secure and  the Merkle tree scheme is secure.
\end{claim}
\noindent\textbf{Proof of Claim~\ref{claim::RC-PoR-P-SecurityAgainstMaliciousClient}.}
We first consider  the event $\Big(\Big(M_{\st \mathsf{PoR}}(u^{\st *},k,{pp})= \sigma\   \wedge\  Q_{\st \mathsf{PoR}}(\text{aux},k,$ $ {pp})= \hat{k}_{\st j}\Big)\ \wedge\ \Big((coin_{\st\mathcal{S},j}\neq  \frac{coin_{\st\mathcal S}^{\st*}}{z}+o)\ \vee\  (coin_{\st\mathcal{R},j}\neq l\ \wedge\  y'_{\st \mathcal {C},j}=1)\Big)\Big)$. It captures the case where the client provides  accepting metadata (i.e., a Merkle tree and its root) and query  but one the following happen:
\begin{enumerate}
\item the server withdraws  incorrect amount of coins, i.e., $coin_{\st\mathcal{S},j}\neq  \frac{coin_{\st\mathcal S}^{\st*}}{z}+o$, or
\item the arbiter withdraw incorrect amount of coins, i.e. $coin_{\st\mathcal{R},j}\neq l$, when the client unnecessarily invokes the arbiter.
\end{enumerate}

 Since the metadata and queries  are valid and correctly structured, an honest server accepts them and does not raise a dispute, so $y_{\st\mathcal{C},j}=0$.  However, the client could  make the server  withdraw an incorrect amount of coins if it manages to (i) persuade the arbiter to recognise the server as misbehaving, i.e., makes the arbiter output $y_{\st\mathcal{S},j}=1$, or (ii) submit to the contract  an accepting  statement $\ddot{x}'_{\st cp}$ other than what was agreed at the initiation phase, i.e., $\ddot{x}_{\st cp}$, or (iii) send a message on the server's behalf to unnecessarily invoke the arbiter.  Nevertheless, it cannot falsely accuse the server of misbehaviour, as,  due to the binding property of SAP commitment scheme, it cannot  convince the arbiter to accept different decryption key and pads' details by providing a different accepting statement $\ddot{x}'_{\st qp}$  (where $\ddot{x}'_{\st qp}\neq \ddot{x}_{\st qp}$), than what was initially agreed with the server, except with $\mathsf{negl}(\lambda)$ probability. This ensures  that the arbiter is given the honest server's messages, with high probability. Therefore, with high probability, (i) and (ii) cannot happen, implying that  $y_{\st\mathcal{S},j}=0$. In addition, (iii) may happen only with $\mathsf{negl}(\lambda)$ probability, as due to the signature scheme's EUF-CMA security, the client cannot send a message on the server's behalf to unnecessarily invoke the arbiter.

 Moreover,  it holds that $y'_{\st\mathcal{S},j} = 0$ because (a) the honest server never invokes the arbiter when the client's queries are well-structured and (b) due to the signature scheme's EUF-CMA security, the client cannot send a message on the server's behalf to unnecessarily invoke the arbiter.  Note that,  due to the binding property of the SAP commitment scheme, the client cannot change the payment parameters by convincing the contract to accept any  statement $\ddot{x}'_{\st cp}$ other than what was initially agreed  between the client and server (i.e., $\ddot{x}'_{\st cp}\neq \ddot{x}_{\st cp}$) except with a negligible  probability, $\mathsf{negl}(\lambda)$. Recall, according to Equation \ref{equ::what-server-recives-in-j}, in RC-PoR-P or RC-S-P protocol, the total coins the server should be credited for $j$-th verification is $coin_{\st\mathcal{S},j}=\frac{coin^{\st *}_{\st\mathcal S}}{z}+o\cdot(1-y_{\st\mathcal {S},j})-l\cdot(y_{\st\mathcal {S},j}+y'_{\st\mathcal {S},j})$. Therefore, given $y_{\st\mathcal{S},j}=y'_{\st\mathcal{S},j}=0$,  the server is credited $\frac{coin_{\st\mathcal S}^{\st*}}{z}+o$ coins for the $j$-th verification, with high probability. 

 Furthermore, as stated above, if the adversary invokes the arbiter, the arbiter with high probability outputs $I_{\st j}=\bot$ which yields  $y'_{\st\mathcal{C},j}=1$. According to Equation \ref{equ::arbiter}, the amount of coins the arbiter should be credited for $j$-th verification is $coin_{\st\mathcal{R},j}= l\cdot(y_{\st\mathcal {S},j}+y_{\st\mathcal {C},j}+y'_{\st\mathcal {S},j}+y'_{\st\mathcal {C},j})$. As shown above $y_{\st\mathcal{C},j}=y_{\st\mathcal{S},j}=y'_{\st\mathcal{S},j}=0$ and $y'_{\st\mathcal{C},j}=1$, which implies that with high probability,  $l$ coins are credited to the arbiter for the $j$-th verification if it is unnecessarily invoked by the adversary.  
 
 We now turn our attention to $\Big(M_{\st \mathsf{PoR}}(u^{\st *},k,{pp})\neq \sigma\ \wedge \ a=1\Big)$   which captures the case where the server accepts  ill-formed metadata. However, 
%
%
due to the security of the Merkle tree scheme that relies on the collision resistance of the utlized hash function, the probability this event happens is  $\mathsf{negl}(\lambda)$; therefore, with a high probability $a=0$. In this  case,  the server does not raise any dispute, instead it avoids serving the client. 

Next, we move on to  $\Big((Q_{\st \mathsf{PoR}}(\text{aux},k,{pp})\neq \hat{k}_{\st j}) \wedge  (b_{\st j}=1 \vee  y_{\st\mathcal{C},j}=0  \vee  coin_{\st\mathcal{S},j}\neq \frac{coin_{\st\mathcal S}^{\st*}}{z}+o  \vee  coin_{\st\mathcal{R},j}\neq l )\Big)$, i.e. the event that considers the case where the client provides an invalid query, and causes at least one of the following to happen:
\begin{enumerate}
\item convinces the server or arbiter to accept the query, i.e., $b_{\st j}=1$ or $ y_{\st\mathcal{C},j}=0$, or
\item makes the server or arbiter withdraw an incorrect amount of coins, i.e., $coin_{\st\mathcal{S},j}\neq \frac{coin_{\st\mathcal S}^{\st*}}{z}+o$ or  $coin_{\st\mathcal{R},j}\neq l$ respectively.
\end{enumerate}
Due to the correctness of the query-checking process, the probability that the server outputs $b_{\st j}=1$  is $0$.  
%
%
Note that, when the honest server rejects the query and raises a dispute, the arbiter checks the query and sets $y_{\st\mathcal{C},j}=1$. After that, due to  RC-PoR-P design, the client cannot make the arbiter set $y_{\st\mathcal{C},j}=0$ (unless it manages to modify the blockchain's content later on, but its probability of success is negligible due to the security of blockchain).  As already stated,  due to the binding property of the SAP commitment scheme and the EUF-CMA security of the digital signature scheme, the client cannot make the arbiter recognise the honest server as a misbehaving party with a probability significantly greater than $\mathsf{negl}(\lambda)$. So, with high probability $y_{\st\mathcal{S},j}=0$. Furthermore, since the arbiter  has identified a misbehaving party, it holds that   $y'_{\st\mathcal{C},j}=y'_{\st\mathcal{S},j}=0$.  The adversary may still try to  make them withdraw  incorrect amount of coins. To this end, in the coin transfer phase, it has to send a  different accepting statement  than what was initially agreed with the server. But, due to the binding property of the SAP commitment scheme, its  success probability is  $\mathsf{negl}(\lambda)$.  Hence with high probability, according to Equations \ref{equ::what-server-recives-in-j} and \ref{equ::arbiter}, for  $y_{\st\mathcal{S},j}=y'_{\st\mathcal{C},j}=y'_{\st\mathcal{S},j}=0$ and $y_{\st\mathcal{C},j}=1$, the server and arbiter are credited $\frac{coin_{\st\mathcal S}^{\st*}}{z}+o$ and $l$ coins respectively for  the $j$-th verification.  Furthermore, due to the liveness of the blockchain and the security and correctness of the smart contract (cf. Appendix~\ref{sec:Discussion-on-the-SAP}), the second security property of SAP holds, i.e., after the parties agree on the statement, an honest party can almost always prove to the verifier that it has the agreement. Thus, the adversary cannot block an honest server's  messages, ``pay'' and $\ddot{x}_{\st cp}$, to the smart contract in the coin transfer phase. \hfill$\dashv$

\begin{claim}\label{claim::RC-PoR-P-privacy} 
 The RC-PoR-P scheme with functions $F_{\st \mathsf{PoR}},M_{\st \mathsf{PoR}},E_{\st \mathsf{PoR}},$ $D_{\st \mathsf{PoR}},Q_{\st \mathsf{PoR}}$ preserves privacy (cf. Definition \ref{deff::RC-S-P-Privacy}), if SAP is secure and the symmetric-key encryption scheme is IND-CPA secure.
\end{claim}
\noindent\textbf{Proof of Claim~\ref{claim::RC-PoR-P-privacy}.} First, we show that the scheme guarantees the privacy of the service input. According to the associated experiment, let $\mathcal{A}_{\st 1}$ be an adversary that  given $pk, F,M,E,D,Q,z,pl$, outputs its choice of challenge files,  $u_{\st 0}$ and $u_{\st 1}$. Then, $\mathcal{A}_{\st 1}$ is provided with the vector of encrypted queries $\bm{c}^{\st *}$, the parties encoded coins $coin^{\st *}_{\st\mathcal S},coin^{\st *}_{\st\mathcal  C}$, the SAP commitments  $ g_{\st cp}, g_{\st qp}$, $\bm{\pi}^{\st *}$, the price list $pl$, and the sender's decision bit $a$ on the verification 
of the parameters and the metadata validity; given the aforementioned input, $\mathcal{A}_{\st 1}$ attempts to guess the challenge bit $\beta$. 

Due to  the hiding property of the commitment scheme, given  commitments $g_{\st qp}$ and $g_{\st cp}$ (stored in the blockchain as a result of running SAP), the adversary learns no information about the committed values (e.g. $o, l, pad_{\st\pi}$ and $\bar{k}$), except with $\mathsf{negl}(\lambda)$ probability. Moreover, given  price list $pl$,  $coin^{\st *}_{\st\mathcal{C}}$ and  $coin^{\st *}_{\st\mathcal{S}}$, the adversary learns nothing about the actual price agreed between the server and client, i.e., $(o,l)$, for each verification, due to Lemma \ref{lemma::encoded-coins}.  Also, since each proof $\bm{\pi}^{\st *}_{\st j}$ is encrypted with the IND-CPA secure encryption scheme and then padded, given $\bm{\pi}^{\st *}_{\st j}$ the adversary cannot tell  whether $\bm{\pi}^{\st *}_{\st j}$ is associated with $u_{\st 0}$ or with $u_{\st 1}$ with probability significantly greater than $\frac{1}{2}+\mathsf{negl}(\lambda)$. As each $c^{\st *}_{\st j}$ is an output of IND-CPA secure symmetric-key encryption and its size is fixed, it  leaks no information to the adversary.  The value of $a$ is also independent of $u_{\st 0}$  or $u_{\st 1}$, and only depends on whether the parameters and metadata are well-formed, so it leaks nothing about the choice of  input file $u_{\st\beta}$ and $\beta\in\{0,1\}$. Overall, the adversary $\mathcal{A}_{\st 1}$ cannot distinguish with a probability significantly greater than $\frac{1}{2}+\mathsf{negl}(\lambda)$ which file of its choice has been used as the server input. 

Next, we show that the scheme guarantees privacy of the service proof's status. For adversaries $\mathcal{A}_{\st 2}$, $\mathcal{A}_{\st 3}$ consider the experiment $Exp_{\st\mathsf{priv}}^{\st\mathcal{A}_2, \mathcal{A}_3}(1^{\st \lambda})$ defined in Definition \ref{deff::RC-S-P-Privacy}. Namely, $\mathcal{A}_{\st 2}$ is initially given $pk, F,M,E,D,Q,z,pl$ and outputs a file $u$. After client-side and server-side initiation steps have been honestly completed, $\mathcal{A}_{\st 2}$ acts as a client that creates potentially invalid queries that are provided to $\mathcal{A}_{\st 3}$ which, in turn, responds with potentially invalid proofs. For each $j$-th verification, the adversary $\mathcal{A}_{\st 2}$ or $\mathcal{A}_{\st 3}$ produces an invalid query or invalid proof, respectively, with probability $Pr_{\st 0,j}$  and a valid query or valid proof, respectively, with probability $Pr_{\st 1,j}$.  Finally, an adversary $\mathcal{A}_{\st 4}$, given $F,M,E,D,Q,\bm{c}^{\st *},coin^{\st *}_{\st\mathcal S},coin^{\st *}_{\st\mathcal  C}, g_{\st cp}, g_{\st qp}, \bm{\pi}^{\st *},pl, a$ is challenged to output a proof's verification bit $d_{\st j}$ for some $j$-th verification. Intuitively, the values $Pr_{\st 0,1},Pr_{\st 1,1},..., Pr_{\st 0,z},Pr_{\st 1,z}$ express the ``background knowledge'' that $\mathcal{A}_{\st 4}$ has on the behavior of $\mathcal{A}_{\st 2}$ and $\mathcal{A}_{\st 3}$.

We know that each encoded query $c^{\st *}_{\st j}\in\bm{c}^{\st *}$ is an IND-CPA encryption of a fresh PRF key that is always provided to the contract. So, $c^{\st *}_{\st j}$ leaks no information (except with $\mathsf{negl}(\lambda)$ probability), not even the query's status to the adversary. Furthermore, each padded IND-CPA encrypted proof $\bm{\pi}^{\st *}_{\st j}$ leaks no information (except with $\mathsf{negl}(\lambda)$ probability) and always contains a fixed number of elements. Besides, due to the hiding property of the SAP commitment scheme, the commitments $g_{\st cp},  g_{\st qp}$ leak no information about the committed values (e.g. $o, l, pad_{\st\pi}$ and $\bar{k}$), except with $\mathsf{negl}(\lambda)$ probability. In addition, the client-side and server-side initiation steps are carried out honestly in the experiment $Exp_{\st\mathsf{priv}}^{\st\mathcal{A}_2, \mathcal{A}_3}(1^{\st \lambda})$, which implies that $a$ is always $1$ and thus, it leaks no information about the proof and query validity. Finally, $pl$,  $coin^{\st *}_{\st\mathcal{C}}$ and  $coin^{\st *}_{\st\mathcal{S}}$ are generated independently and contain no information about the query's or proof's status. 

By the above, the input  $(F,M,E,D,Q,\bm{c}^{\st *},coin^{\st *}_{\st\mathcal S},coin^{\st *}_{\st\mathcal  C}, g_{\st cp}, g_{\st qp}, \bm{\pi}^{\st *},pl, a)$ does not provide $\mathcal{A}_{\st 4}$ with any more 
than $\mathsf{negl}(\lambda)$ advantage compared to any background knowledge $\mathcal{A}_{\st 4}$ may have on the behavior of $\mathcal{A}_{\st 2}$ and $\mathcal{A}_{\st 3}$. Therefore, for each $j$-th verification, the probability of $\mathcal{A}_{\st 4}(F,M,E,D,Q,\bm{c}^{\st *},coin^{\st *}_{\st\mathcal S},coin^{\st *}_{\st\mathcal  C}, g_{\st cp}, g_{\st qp}, \bm{\pi}^{\st *}, pl, a)$ guessing $d_{\st j}$ is maximized when $\mathcal{A}_{\st 4}$ exploits its background knowledge, so it is upper bounded  by $\mathrm{max}\{Pr_{\st 0,j},Pr_{\st 1,j}\}+\mathsf{negl}(\lambda)$. Overall, the probability of $\mathcal{A}_{\st 4}(F,M,E,D,Q,\bm{c}^{\st *},coin^{\st *}_{\st\mathcal S},coin^{\st *}_{\st\mathcal  C}, g_{\st cp}, g_{\st qp}, \bm{\pi}^{\st *}, pl, a)$ guessing $(d_{\st j},j)$ for some $j\in\{1,\ldots,z\}$ is upper bounded by
\begin{equation*}
\begin{split}
&\mathrm{max}\big\{\mathrm{max}\{Pr_{\st 0,1},Pr_{\st 1,1}\}+\mathsf{negl}(\lambda),\ldots,\\
&\quad\mathrm{max}\{Pr_{\st 0,z},Pr_{\st 1,z}\}+\mathsf{negl}(\lambda)\big\}\leq\\
\leq&\mathrm{max}\big\{\mathrm{max}\{Pr_{\st 0,1},Pr_{\st 1,1}\},\ldots,\mathrm{max}\{Pr_{\st 0,z},Pr_{\st 1,z}\}\big\}+\\
&\quad+\mathsf{negl}(\lambda)\leq\\
\leq&\mathrm{max}\big\{Pr_{\st 0,1},Pr_{\st 1,1},\ldots,Pr_{\st 0,z},Pr_{\st 1,z}\big\}+\mathsf{negl}(\lambda)=\\
=&Pr_\mathrm{max}+\mathsf{negl}(\lambda)\;,
\end{split}
\end{equation*}
so the privacy of the service proof is guaranteed.
\hfill$\dashv$\\

The security of the construction follows from Claims~\ref{claim::RC-PoR-P-SecurityAgainstMaliciousServer},~\ref{claim::RC-PoR-P-SecurityAgainstMaliciousClient}, and~\ref{claim::RC-PoR-P-privacy}.
\end{proof}

%% file: RC-PoR-P-without-Arbiter.tex
\section{Further Details on RC-PoR-P Without Arbiter's Involvement}\label{sec::RC-PoR-P-Without-Arbiter-}

As already stated, in the RC-PoR protocol, due to the efficiency of the arbiter-side algorithm we can delegate the arbiter’s role to the smart contract, SC.  To have the new variant of RC-PoR-P,  we need to amend the original RC-PoR-P’s protocol and definition. In the remainder of this section, we elaborate on how exactly the definition and protocol must change.

  \input{RC-S-P-without-Arbiter-Definition}

  \input{Recurring-Contingent-PoR-Payment-protocol}

  \input{x}

%% file: RC-S-P-without-Arbiter-Definition.tex

\subsection{Definition of RC-S-P Without Arbiter's Involvement}\label{sec::RC-S-P-Definition2}
There are cases, in RC-S-P schemes, where the third party arbiter's role  can be  efficiently delegated to a smart contract.  In this variant of the RC-S-P scheme, denoted by ${\text{RC-}\overline{\text{S}}\text{-P}}$, the arbiter's involvement is not needed anymore. The primary difference between RC-S-P and ${\text{RC-}\overline{\text{S}}\text{-P}}$ is the way a party pays to resolve a dispute. In particular, in RC-S-P, the party which raises a dispute does not pay the arbiter when it sends to it a dispute query. Instead, loosely speaking, the arbiter at coin distribution is paid by a misbehaving party. Whereas, in ${\text{RC-}\overline{\text{S}}\text{-P}}$, the party which raises a dispute and sends a dispute query to the contract, (due to the nature of  the smart contracts' platform) has to pay the contract, before the contract processes its query.  In this section, we show how the RC-S-P definition (presented in Section \ref{sec::RC-S-P-Definition}) can be  adjusted to  capture ${\text{RC-}\overline{\text{S}}\text{-P}}$. In the following, we highlight the main changes that should be applied to the RC-S-P definition. 

\begin{itemize}[leftmargin=4mm]

\item[$\bullet$] In Definition \ref{RC-S-P-def}: 
\begin{itemize}

\item [$\bullet$]  Three parties are involved; namely, client, server, and smart contract (so an arbiter is not involved anymore).

\item [$\bullet$]  Vectors $\bm{y}'_{\st\mathcal{C}}$ and $\bm{y}'_{\st\mathcal{S}}$ are not needed anymore. Because a misbehaving party, which unnecessarily invokes the contract, pays the contract ahead of time. Therefore, there is no need to keep track of unnecessary contract's invocation.

\item [$\bullet$] $\mathtt{RCSP}.\mathtt{resolve}(\cdot)$ is run by a smart contract.

\item [$\bullet$] $\mathtt{RCSP}.\mathtt{pay}(\cdot)$ outputs $({\bm{coin}}_{\st\mathcal C},{\bm{coin}}_{\st\mathcal S})$, so ${\bm{coin}}_{\st\mathcal{R}}$ is excluded from the output, as a third party arbiter plays no role anymore.
\end{itemize}

\item [$\bullet$] In Definition \ref{def::RC-S-P-Correctness}: only the above changes are applied to it.

\item [$\bullet$] In Definition \ref{deff::RC-S-P-SecurityAgainstMaliciousServer}:  the above changes are applied to the algorithms' syntax in the experiment. Moreover, the events are slightly modified, i.e. the amount of coins each party receives. For the sake of clarity and completeness, we state the entire modified  definition below. 

\begin{definition}[${\text{RC-}\overline{\text{S}}\text{-P}}$ Security Against Malicious Server]\label{deff::RC-S-P-SecurityAgainstMaliciousServer-2}  An ${\text{RC-}\overline{\text{S}}\text{-P}}$  scheme with functions $F,M,E,D,Q$ is secure against a malicious server for auxiliary information \text{aux}, if for any $z$ polynomial in $\lambda$,   any price list $pl$, every $j$ (where $1\leq j\leq z$), and any PPT adversary $\mathcal{A}$, the following probability is $\mathsf{negl}
(\lambda)$:

\[\Pr\left[{\small{
    \begin{array}{l}
  \hs   \mathtt{RCSP.keyGen}(1^{\lambda},)\rightarrow \bm{k}\\
  \hs   \mathcal{A}(1^\lambda,pk, F,M,E,D,Q,z,pl)\rightarrow u\\
  
  \hs   \mathtt{RCSP}.\mathtt{cInit}(1^\lambda, u,\bm{k}, z,pl)\rightarrow (u^{\st *},e,T, p_{\st\mathcal S},  \bm{y}, \\coin^{\st*}_{\st\mathcal C})\\ 
 \hs    \mathcal{A}(u^{\st *}, e, pk, z, T, p_{\st\mathcal S},\bm{y})\rightarrow (coin^{\st *}_{\st\mathcal S},a)\\
\hs 	\mathtt{RCSP}.\mathtt{genQuery}(1^\lambda, \text{aux},k,T_{\st qp})\rightarrow c^{\st *}_{\st j}\\
	
\hs      \mathcal{A}(c^{\st *}_{\st j},\sigma, u^{\st *},a)\rightarrow 
     (b_{\st j},m_{\st \mathcal{S},j},h^{\st *}_{\st j},\delta^{\st *}_{\st j})\\      
\hs \mathtt{RCSP}.\mathtt{verify}(\pi^{\st *}_{\st j},c^{\st *}_{\st j}, k,T_{\st qp})\rightarrow (d_{\st j},m_{\st \mathcal{C},j})\\
 \hs   \mathtt{RCSP}.\mathtt{resolve}(m_{\st \mathcal{C}},m_{\st \mathcal{S}},z, {\bm{\pi}}^{\st *}, {\bm{c}}^{\st *},pk, T_{\st qp})\rightarrow \bm{y}\\
 \hs   \mathtt{RCSP}.\mathtt{pay}(\bm{y},T_{\st cp},a,p_{\st\mathcal S}, coin^{\st *}_{\st\mathcal C},coin^{\st *}_{\st\mathcal S})\rightarrow ({\bm{coin}}_{\st\mathcal C},\\{\bm{coin}}_{\st\mathcal S})\\
   \hline
 \hs   \Big(F(u^{\st *}, \bm{q}_{\st j}, {pp})= h_{\st j}\ \wedge (coin_{\st\mathcal{C},j}\neq  \frac{coin_{\st\mathcal C}^{\st*}}{z}-o)\Big)\ \vee\\ 
\hs \Big(
F(u^{\st *},\bm{q}_{\st j}, {pp})\neq h_{\st j} \ \wedge\\ (d_{\st j}=1	\vee y_{\st \mathcal{S},j}=0 \ \vee \   {coin_{\st\mathcal{C},j}\neq \frac{coin_{\st\mathcal C}^{\st*}}{z}+l} ) \Big)\\
\end{array} } }  \right]\]

\noindent where  $\bm{q}_{\st j}\in D(c^{\st *}_{\st j},t_{\st qp})$, $\pi^{\st *}_{\st j}:=(h^{\st *}_{\st j}, \delta^{\st *}_{\st j}), h_{\st j}= D(h^{\st *}_{\st j},T_{\st qp})$, $\sigma\in e$, $m_{\st \mathcal{C},j}\in m_{\st \mathcal{C}}, m_{\st \mathcal{S},j}\in m_{\st \mathcal{S}}$,   $y_{\st \mathcal{S},j}\in \bm{y}_{\st \mathcal{S}}\in\bm{y}$, and $ {pp}\in T_{\st qp}$. 
\end{definition}

 \item[$\bullet$] In Definition \ref{deff::RC-S-P-Security-Against Malicious-Client}: similar to the previous point, only the algorithms' syntax (in the experiment) and the amount of coins each party receives changes. Below, we state the entire modified definition. 
 
 \begin{definition}[${\text{RC-}\overline{\text{S}}\text{-P}}$ Security Against Malicious Client]\label{deff::RC-S-P-Security-Against Malicious-Client-2}  An  ${\text{RC-}\overline{\text{S}}\text{-P}}$ scheme  wih functions $F,M,E,D,Q$ is secure against a malicious client for auxiliary information \text{aux}, if for any $z$ polynomial in $\lambda$,  every $j$ (where $1\leq j\leq z$), and any PPT adversary $\mathcal{A}$, the following probability is $\mathsf{negl}(\lambda)$:

\[\Pr\left[{\small{
    \begin{array}{l}
 \hs   \mathcal{A}(1^{\lambda}, F,M,E,Q,D,z,pl)\rightarrow (u^{\st *},\bm{k},e,T,  p_{\st\mathcal{S}}, \\coin^{\st*}_{\st\mathcal C},enc,\bm{y}, enc, pk)\\
   
\hs    \mathtt{RCSP}.\mathtt{sInit}(u^{\st *}, e, pk, z, T, p_{\st\mathcal S},\bm{y})\rightarrow (coin^{\st *}_{\st\mathcal S},a)\\
   
\hs 	\mathcal{A}(coin^{\st *}_{\st\mathcal S},a, 1^\lambda, \text{aux},k, T_{\st qp})\rightarrow c^{\st *}_{\st j}\\

  \hs    \mathtt{RCSP}.\mathtt{prove}(u^{\st *},  \sigma,  c^{\st *}_{\st j}, pk, T_{\st qp})\rightarrow (b_{\st j},m_{\st \mathcal{S},j},\pi^{\st *}_{\st j})\\
     
\hs  \mathcal{A}(\pi^{\st *}_{\st j}, \bm{q}_{\st j},k,T_{\st qp})\rightarrow (d_{\st j},m_{\st \mathcal{S},j})\\

 \hs   \mathtt{RCSP}.\mathtt{resolve}(m_{\st \mathcal{C}},m_{\st \mathcal{S}},z, {\bm{\pi}}^{\st *}, {\bm{c}}^{\st *}, pk, T_{\st qp})\rightarrow \bm{y}\\
\hs    \mathtt{RCSP}.\mathtt{pay}(\bm{y},T_{\st cp},a,p_{\st\mathcal S}, coin^{\st *}_{\st\mathcal C},coin^{\st *}_{\st\mathcal S})\rightarrow ({\bm{coin}}_{\st\mathcal C},\\{\bm{coin}}_{\st\mathcal S})\\
   \hline
\hs     \Big((M(u^{\st *},k, {pp})= \sigma  \wedge Q(\text{aux},k, {pp})= \bm{q}_{\st j})\ \wedge\\
  (coin_{\st\mathcal{S},j}\neq  \frac{coin_{\st\mathcal S}^{\st*}}{z}+o )\Big)\ \vee\\

 \hs  \Big(M(u^{\st *}, k,  {pp})\neq \sigma \wedge a=1\Big) \ \vee\\

\hs   \Big(Q(\text{aux},k,  {pp})\neq \bm{q}_{\st j}\ \wedge\\
\hs   (b_{\st j}=1 \ \vee \ y_{\st\mathcal{C},j}=0 \ \vee  {coin_{\st\mathcal{S},j}\neq \frac {coin_{\st\mathcal S}^{\st*}}{z}+o + l}) 
  \Big)\\
\end{array}} }   \right]\]

\noindent where $\bm{q}_{\st j}\in D(c^{\st *}_{\st j},t_{\st qp})$, $
D\in enc$, $\sigma\in e$,    $y_{\st \mathcal{C},j}\in \bm{y}_{\st \mathcal{C}}\in\bm{y}$, and ${pp}\in T_{\st qp}$.  
\end{definition}
\end{itemize}

Note that Definition \ref{deff::RC-S-P-Privacy} remains almost the same with a  minor change, that is vectors $(\bm{y}'_{\st\mathcal{C}},\bm{y}'_{\st\mathcal{S}})$ are excluded from the related algorithms input/output. 

 \begin{definition}\label{def::RC-S-P-Security-2} An ${\text{RC-}\overline{\text{S}}\text{-P}}$ scheme  is secure if it satisfies security against malicious server, security against malicious client, and preserves privacy (cf. . Definitions \ref{deff::RC-S-P-SecurityAgainstMaliciousServer-2},\ref{deff::RC-S-P-Security-Against Malicious-Client-2}, and \ref{deff::RC-S-P-Privacy}). 
 \end{definition}

%% file: Recurring-Contingent-PoR-Payment-protocol.tex

\subsection{Protocol For RC-PoR-P Without Arbiter's Involvement}\label{sec::R-PoR-P-protocol-Withohut-Arbiter}


Below we present the modifications that must be applied to the phases \ref{RCPoRP::Dispute-Resolution} and \ref{RCPoRP::CoinTransfer} of the RC-PoR-P protocol to allow the smart contract to play the role of the auditor. 
\begin{enumerate}[leftmargin=5mm]
\item[\ref{RCPoRP::Dispute-Resolution}.] \textbf{Dispute Resolution}. \label{RCPoRP2::Dispute-Resolution2}\\ $\mathtt{RCPoRP}. \mathtt{resolve}(m_{\st \mathcal{C}},m_{\st \mathcal{S}}, z, {\bm{\pi}}^{\st *}, \bm{q}^{\st *},  T_{\st qp})$

The phase takes place only in  case of dispute, i.e., when $\mathcal C$ rejects service proofs or $\mathcal S$ rejects the queries. 

\begin{enumerate}

\item $\mathcal S$  sends ${m}_{\st\mathcal S}$ and $\ddot{x}_{\st qp}$ to SC, at time $\texttt{K}_{\st 1}$, where $\texttt{K}_{\st 1}>\texttt{G}_{\st z,2} + \texttt{J}$

\item SC upon receiving ${m}_{\st\mathcal S}$ does the following a time $\texttt{K}_{\st 2}$. 
\begin{enumerate}

\item Checks the validity of  statement $\ddot{x}_{\st qp}$, by  sending it to the SAP contract which  returns   $1$ or $0$. If the output is $0$, then SC discards the server's complaint, $m_{\st\mathcal S}$, and does not take steps \ref{RCPoRP::check-v} and \ref{sc-check-server-claim}. Otherwise, it proceeds to the next step.

\item\label{RCPoRP2::check-v} Removes from $ {\bm{v}}_{\st\mathcal{S}}$ any element that is duplicated or not in the range $[1,z]$. It also constructs an empty vector $ {\bm{v}}$.

\item\label{sc-check-server-claim2}  For any element $i\in {\bm{v}}_{\st\mathcal{S}}$:
\begin{itemize}

\item[$\bullet$] Fetches the related encrypted query $\hat{k}^{\st *}_{\st i}\in \bm{q}^{\st *}$, and   decrypts it, $\hat{k}_{\st i}=\mathtt{Dec}(\bar{k},\hat{k}^{\st *}_{\st i})$.
 

\item[$\bullet$]Checks if the query is well-formed, by doing the same checks performed in step \ref{RC-PoR-P::checkquery} (of the RC-PoR-P).  If the query is rejected, then it increments $y_{\st\mathcal C}$ by $1$ and appends $i$ to $ {\bm{v}}$.

\end{itemize}
Let $\texttt{K}_{\st 3}$ be the time SC finishes the above checks.
\end{enumerate}

\item $\mathcal C$  sends $m_{\st\mathcal C}$ and $\ddot{x}_{\st qp}$ to SC, at time $\texttt{K}_{\st 4}$.

\item SC upon receiving $m_{\st\mathcal C}$, does the following at time $\texttt{K}_{\st 5}$. 
\begin{enumerate}
\item Checks the validity of  statement $\ddot{x}_{\st qp}$, by   sending $\ddot{x}_{\st qp}$ to the SAP contract which  returns  either $1$ or $0$. If the output is $0$, then SC discards the client's complaint, ${m}_{\st\mathcal C}$, and does not take steps \ref{sc-checks-client-claim}-\ref{the-contract-loop}. Otherwise, it proceeds to the next step. 

 \item\label{sc-checks-client-claim2} Ensures each vector   $ {\bm{m}}\in  {\bm{m}}_{\st\mathcal{C}}$ is well-formed. In particular, it verifies there exist no two vectors: $ {\bm{m}}, {\bm{m}}'\in  {\bm{m}}_{\st\mathcal{C}}$ such that $ {\bm{m}}[0]= {\bm{m}}'[0]$.  If such vectors  exist, it deletes the redundant ones from $ {\bm{m}}_{\st\mathcal{C}}$. This ensures no two claims refer to the same verification.  Also, it removes any vector $ {\bm{m}}$ from $ {\bm{m}}_{\st\mathcal{C}}$ if $ {\bm{m}}[0]$ is not in the range $[1,z]$ or if $ {\bm{m}}[0]\in  {\bm{v}}$. Note the latter check  (i.e., $ {\bm{m}}[0]\in  {\bm{v}}$) ensures $\mathcal C$ cannot hold $\mathcal S$ accountable if $\mathcal C$ has generated an ill-formed query for the same verification.   


\item\label{the-contract-loop2} For every  vector $ {\bm{m}}\in {\bm{m}}_{\st \mathcal{C}}$:
\begin{enumerate}
\item retrieves a rejected proof's details by setting   $j= {\bm{m}}[0]$ and $g= {\bm{m}}[1]$. Recall that $g$ refers to the index of a rejected proof in the proof vector which was generated for $j$-th verification, i.e., $ {\bm{\pi}}_{\st j}$.

 \item fetches the related encrypted query $c^{\st *}_{\st j}\in \bm{c}^{\st *}$   from SC and   decrypts it as $\hat{k}_{\st j}=\mathtt{Dec}(\bar{k},c^{\st *}_{\st j})$. It removes the pads only from $g$-th padded encrypted proof. Let $ {\bm{\pi}}'_{\st j}[g]$ be the result. Next, it decrypts the  encrypted proof, $\mathtt{Dec}(\bar{k}, {\bm{\pi}}'_{\st j}[g]))= {\bm{\pi}}_{\st j}[g]$.


%
%



\item identifies the misbehaving party as follows. 

\begin{itemize}

\item verifies $\hat{k}_{\st j}$ by doing the same checks done in step \ref{RC-PoR-P::checkquery} (of the RC-PoR-P). If the checks do not pass,   it sets  $I_{\st j}=\mathcal C$ and skips the next two steps; otherwise, it proceeds to the next step.

\item derives the related challenged block's index from $\hat{k}_{\st j}$, by computing $q_{\st g}=\big(\mathtt{PRF}(\hat{k}_{\st j},g)\bmod m\big)+1$.

\item verifies only $g$-th proof, by calling $\mathtt{PoR.verify}( {\bm{\pi}}_{\st j}[g] , $ $q_{\st g},$ $pp)\rightarrow  {\bm{d}}'$. If $ {\bm{d}}'[0]=0$, then it sets $I_{\st j}=\mathcal S$. Otherwise, it outputs $I_{\st j}=\bot$.

\item  if $I_{\st j}=\mathcal C$,  it increments $y_{\st\mathcal C}$  by $1$. If $I_{\st j}=\mathcal S$,  it increments $y_{\st\mathcal S}$  by $1$.  

\end{itemize}

\end{enumerate}
\end{enumerate}



Let $\texttt{K}_{\st 6}$ be the time that SC finishes all the above checks.

\end{enumerate} 

\item[\ref{RCPoRP::CoinTransfer}.] \textbf{Coin Transfer}. \hspace{-1.5mm}\label{RCPoRP::CoinTransfer2} $\mathtt{\mathtt{RCPoRP}}.\mathtt{pay}(\bm{y},T_{\st cp},a,p_{\st\mathcal S}, coin^{\st *}_{\st\mathcal C}, coin^{\st *}_{\st\mathcal S})$
\begin{enumerate}

 \item If SC  receives  ``pay'' message  at time $\texttt{T}_{\st 2}$, where $a=0$ or $coins^{\st *}_{\st\mathcal{S}}<p_{\st\mathcal{S}}$, then it sends $coin^{\st *}_{\st\mathcal C}$ coins to $\mathcal C$ and $coin^{\st *}_{\st\mathcal S}$ coins to $\mathcal S$. Otherwise (i.e., they reach an agreement), they take the following step. 

\item Either $\mathcal C$ or $\mathcal S$ sends ``pay'' message and  statement $\ddot{x}_{\st cp}\in T_{\st cp}$ to SC at time $\texttt{L}>\texttt{K}_{\st 6}$.
\item SC checks the validity of the statement by sending it  to the SAP contract that  returns  either $1$ or $0$. SC only proceeds to the next step if the output is $1$.
\item SC distributes the coins to the parties as follows:

\begin{itemize}
\item[$\bullet$]   $coin^{\st *}_{\st\mathcal C}-o\cdot(z-y_{\st\mathcal S})+l\cdot(y_{\st\mathcal S}-y_{\st\mathcal C})$ coins  to $\mathcal C$.

\item[$\bullet$] $coin^{\st *}_{\st\mathcal S}+o\cdot(z-y_{\st\mathcal S})+l\cdot( y_{\st\mathcal C}-y_{\st\mathcal S})$ coins to $\mathcal S$.

\end{itemize}

\end{enumerate}
\end{enumerate}

\begin{theorem}\label{theorem::RC-PoR-P-Without-Arbiter}
The ${\text{RC-}\overline{\text{PoR}}\text{-P}}$ protocol is secure, w.r.t. Definition \ref{def::RC-S-P-Security-2}, if PoR and SAP are secure and correct and the encryption scheme is IND-CPA secure. 
\end{theorem}
%
%


%% file: x.tex

\subsection{Proof of the RC-PoR-P Without Arbiter}

 To prove Theorem \ref{theorem::RC-PoR-P-Without-Arbiter}, we show that ${\text{RC-}\overline{\text{PoR}}\text{-P}}$ meets all security properties defined in Appendix \ref{sec::RC-S-P-Definition2}.  We start by proving that ${\text{RC-}\overline{\text{PoR}}\text{-P}}$ meets security against a malicious server.   The proof to some extent is  simpler to that  of RC-PoR-P against a malicious server (i.e., proof of Claim \ref{claim::RC-PoR-P-SecurityAgainstMaliciousServer})  as it does not involve any third party arbiter.

 \begin{lemma}

 If the SAP is secure, and the PoR scheme satisfies correctness and soundness, then ${\text{RC-}\overline{\text{PoR}}\text{-P}}$ is secure against a malicious server, w.r.t. Definition \ref{deff::RC-S-P-SecurityAgainstMaliciousServer-2}. 
 \end{lemma}

 \begin{proof}
 First, we  consider event  
  $$\Big(F(u^{\st *}, \bm{q}_{\st j},{pp})= h_{\st j}\ \wedge (coin_{\st\mathcal{C},j}\neq  \frac{coin_{\st\mathcal C}^{\st*}}{z}-o)\Big)$$
   that captures the case where the server provides an accepting  proof, i.e. PoR, but makes an honest client withdraw  incorrect amounts of coin, i.e., $coin_{\st\mathcal{C},j}\neq  \frac{coin_{\st\mathcal C}^{\st*}}{z}-o$. Note, in ${\text{RC-}\overline{\text{PoR}}\text{-P}}$ protocol, the total coins the client should receive after $z$ verifications is   $coin^{\st *}_{\st\mathcal C}-o\cdot(z-y_{\st\mathcal S})+l\cdot(y_{\st\mathcal S}-y_{\st\mathcal C})$. Since we focus on  $j$-th verification, the amounts of  coin that should be credited to the client for this verification is
  \begin{equation}\label{equ::what-client-recives-in-j-2}
   coin_{\st\mathcal{C},j}=\frac{coin^{\st *}_{\st\mathcal C}}{z}-o\cdot(1-y_{\st\mathcal {S},j})+l\cdot(y_{\st\mathcal {S},j}-y_{\st\mathcal {C},j})
     \end{equation}

As the  proof is valid, an honest client accepts it and does not raise any dispute. But, the server  would be able to make the client  withdraw incorrect amounts of coin, if it manages to either convince the contract that the client has misbehaved, by making the contract output  $y_{\st\mathcal {C},j}=1$ through the dispute resolution phase, or submit to the contract, in the coin transfer phase,  an accepting  statement $\ddot{x}'_{\st cp}$ other than what was agreed in the initiation phase, i.e., $\ddot{x}'_{\st cp}\neq\ddot{x}_{\st cp}$, so it can change the payments' parameters, e.g., $l$ or $o$. Nevertheless, it cannot falsely accuse the client of misbehaviour. As,  due to the security of SAP (i.e., the binding property of the underlying commitment scheme), it cannot convince the contract to accept different query's parameters other than what was agreed with the client in the initiation phase. In particular, it cannot persuade the contract to accept  $\ddot{x}'_{\st qp}$ such that $\ddot{x}'_{\st qp}\neq \ddot{x}_{\st qp}$, except with a negligible probability, $\mathsf{negl}(\lambda)$.  Furthermore, if the adversary provides a valid statement then, then due to the correctness of the PoR and query-checking
process, values  $y_{\st\mathcal {C}}$ and $y_{\st\mathcal {S}}$ are not incremented by $1$  in the $j$-th verification, i.e., $y_{\st\mathcal{C},j}=y_{\st\mathcal{S},j}=0$. Also,  due to the security of SAP (i.e., the binding of the used commitment scheme), the server cannot change the payment parameters by persuading the contract to accept any statement $\ddot{x}'_{\st cp}$ other than what was agreed initially between the client and server, except with a negligible probability $\mathsf{negl}(\lambda)$.  Therefore, according to Equation \ref{equ::what-client-recives-in-j-2}, the client is credited $\frac{coin_{\st\mathcal C}^{\st*}}{z}-o$ coins for the $j$-th verification, with a high probability.  We now move on to event 
  $$  F(u^{\st *},\bm{q}_{\st j}, {pp})\neq h_{\st j} \ \wedge$$ $$ (d_{\st j}=1	\vee y_{\st \mathcal{S},j}=0 \ \vee \  {coin_{\st\mathcal{C},j}\neq \frac{coin_{\st\mathcal C}^{\st*}}{z}+l} ) $$

     It captures the case where the server provides an invalid  proof however either persuades the client to accept the proof, or  persuades the contract to set $y_{\st \mathcal{S},j}=0$  or makes the client  withdraw incorrect amounts of coin, i.e.    $coin_{\st\mathcal{C},j}\neq \frac{coin_{\st\mathcal C}^{\st*}}{z}+l$.   Nevertheless, due to the soundness of PoR, the probability that a corrupt server can convince an honest client to accept invalid proof, i.e. outputs $d_{\st j}=1$, is negligible, $\mathsf{negl}(\lambda)$. So, the client detects it with a high probability and raises a dispute.  Also, the server may try to make the contract keep $y_{\st \mathcal {S},j}=0$. For $y_{\st \mathcal {S},j}=0$ to happen, it has to make the contract  recognise the client as the misbehaving party, i.e., makes the contract output $y_{\st \mathcal {C},j}=1$. In this case, the client's complaint would not be processed by the contract; therefore,  $y_{\st \mathcal {S},j}$ remains $0$.   Nevertheless, as we discussed above (due to the binding property of the commitment), the probability that the adversary makes the contract recognise the client as misbehaving is negligible, $\mathsf{negl}(\lambda)$. Therefore, with a high probability  $y_{\st \mathcal {S},j}=1$ and $y_{\st \mathcal {C},j}=0$, after the contract  is invoked by the client or  server. The adversary may  try to  make the client withdraw  incorrect amounts of coin, e.g., in the case where it does not succeed in convincing the client or contract. To do so,  in the coin transfer phase, it has to send a  different accepting statement  than what was initially agreed with the client. But, it would succeed only with a negligible   probability, $\mathsf{negl}(\lambda)$, due to the security of SAP, i.e., the binding property of the commitment.   So, according to Equation \ref{equ::what-client-recives-in-j-2}, the client is credited $\frac{coin_{\st\mathcal C}^{\st*}}{z}+l$ coins for the $j$-th verification, with a high probability. 
  \end{proof}

Next, we prove that ${\text{RC-}\overline{\text{PoR}}\text{-P}}$ satisfies security against a malicious client. The proof is also slightly simpler than that  of RC-PoR-P against a malicious client (i.e., proof of Claim \ref{claim::RC-PoR-P-SecurityAgainstMaliciousClient}) as it does not involve any third party arbiter. 

\begin{lemma}

 If SAP and the PoR's Merkle tree scheme are secure, then ${\text{RC-}\overline{\text{PoR}}\text{-P}}$ is secure against a malicious client, w.r.t. Definition \ref{deff::RC-S-P-Security-Against Malicious-Client-2}. 
\end{lemma}

\begin{proof}
  First, we consider event 
   $$(M(u^{\st *},k,{pp})= \sigma \  \wedge \ Q(\text{aux},k, {pp})= \bm{q}_{\st j})\ \wedge$$ $$
(coin_{\st\mathcal{S},j}\neq  \frac{coin_{\st\mathcal S}^{\st*}}{z}+o)$$

  It captures the case where the client provides  accepting metadata and query  but makes the server withdraw an incorrect amounts of coin, i.e. $coin_{\st\mathcal{S},j}\neq  \frac{coin_{\st\mathcal S}^{\st*}}{z}+o$. According to  ${\text{RC-}\overline{\text{PoR}}\text{-P}}$ protocol, the total coins the server should receive after $z$ verifications is $coin^{\st *}_{\st\mathcal S}+o\cdot(z-y_{\st\mathcal S})+l\cdot(y_{\st\mathcal C}-y_{\st\mathcal S})$. As we focus on  $j$-th verification, the amount of  coins that should be credited to the server for the $j$-th verification is
  \begin{equation}\label{equ::what-server-recives-in-j-2}
   coin_{\st\mathcal{S},j}=\frac{coin^{\st *}_{\st\mathcal S}}{z}+o\cdot(1-y_{\st\mathcal {S},j})+l\cdot(y_{\st\mathcal {C},j}-y_{\st\mathcal {S},j})
     \end{equation}

  Since the metadata and query are valid, an honest server accepts them and does not raise any dispute, so we have $y_{\st\mathcal{C},j}=0$.  The client however could  make the server  withdraw incorrect amounts of coin, if it manages to either convince the contract, in the dispute resolution phase,  that the server has misbehaved, i.e., makes the contract output $y_{\st\mathcal{S},j}=1$, or submit to the contract  an accepting  statement $\ddot{x}'_{\st cp}$ other than what was agreed at the initiation phase, i.e. $\ddot{x}_{\st cp}$, in the coin transfer phase. But, it cannot falsely accuse the server of misbehaviour, because due to the security of SAP (i.e., the commitment's binding property), it cannot convince the contract to accept different decryption key and pads' detail, by providing a different accepting statement $\ddot{x}'_{\st qp}$  (where $\ddot{x}'_{\st qp}\neq \ddot{x}_{\st qp}$), than what was initially agreed with the server, except with a negligible  probability, $\mathsf{negl}(\lambda)$.  So, with a high probability $y_{\st\mathcal{S},j}=0$. On the other hand, if the adversary provides a valid statement, i.e. $\ddot{x}_{\st qp}$, then due to the correctness of the PoR and query-checking process,  we would have $I_{\st j}=\bot$. Thus, due to the security of SAP and the correctness of the PoR and query-checking process,  we would have $y_{\st\mathcal{C},j}=y_{\st\mathcal{S},j}=0$ with a high probability.    Also,  due to the security of SAP (i.e., the commitment's binding property), the client cannot change the payment parameters by convincing the contract to accept any accepting statement $\ddot{x}'_{\st cp}$ other than what was initially agreed between the client and server (i.e. $\ddot{x}'_{\st cp}\neq \ddot{x}_{\st cp}$), except with a negligible probability, $\mathsf{negl}(\lambda)$. That means, according to  Equation \ref{equ::what-server-recives-in-j-2}, the server is credited $\frac{coin_{\st\mathcal S}^{\st*}}{z}+o$ coins for that verification, with a  high probability.  We now move on  to 
  $$\Big(M(u^{\st *},k,{pp})\neq \sigma\ \wedge \ a=1\Big)$$

    It captures the case where the server accepts ill-formed metadata. But, due to the security of the PoR's Merkle tree scheme (i.e., the collision resistance of the underlying hash function), the probability of the event happening is negligible,  $\mathsf{negl}(\lambda)$. So, with a high probability $a=0$; in this case,  the server does not raise any dispute, instead it avoids serving the client.    Next, we turn our attention   to 
     $$\Big(Q(\text{aux},k,{pp})\neq \bm{q}_{\st j})\ \wedge$$ $$ (b_{\st j}=1\ \vee \ y_{\st\mathcal{C},j}=0 \ \vee \ coin_{\st\mathcal{S},j}\neq \frac{coin_{\st\mathcal S}^{\st*}}{z}+o+l)\Big)$$
     
       It considers the case where the client provides an invalid query, but  either convinces the server or contract to accept it, or  makes the server   withdraw  incorrect amounts of coin, i.e. $coin_{\st\mathcal{S},j}\neq \frac{coin_{\st\mathcal S}^{\st*}}{z}+o+l$.  Due to the correctness of the query-checking process, the probability that the server outputs $b_{\st j}=1$ is $0$.  When the honest server rejects the query and raises a dispute, the contract checks the server's query and sets $y_{\st\mathcal{C},j}=1$.  Also, as discussed above (due to the commitment's binding property), the client cannot make the contract recognise the  honest server as a misbehaving party with a probability significantly greater than $\mathsf{negl}(\lambda)$. That means with a high probability $y_{\st\mathcal{S},j}=0$.  The adversary may still try to  make the server withdraw   incorrect amounts of coin (e.g., if the adversary does not succeed in convincing the server). To this end, at the coin transfer phase, it has to convince the contract to accept a  different  statement  than what was initially agreed with the server. However, due to the security of SAP, its  success probability is negligible, $\mathsf{negl}(\lambda)$.     Hence, according to Equation \ref{equ::what-server-recives-in-j-2}, the server is credited  $\frac{coin_{\st\mathcal S}^{\st*}}{z}+o+l$ coins for the $j$-th verification.   
 \end{proof}

In the following, we provide a lemma for  ${\text{RC-}\overline{\text{PoR}}\text{-P}}$'s privacy. For the lemma's proof, we refer readers to the proof of Claim \ref{claim::RC-PoR-P-privacy}. 
\begin{lemma} If SAP is secure and the encryption scheme is IND-CPA secure, then ${\text{RC-}\overline{\text{PoR}}\text{-P}}$ preserves privacy, w.r.t. Definition \ref{deff::RC-S-P-Privacy}. 
\end{lemma}

%% file: RC-PoR-P-complexity-table.tex

\section{A Table Summarizing RC-PoR-P Assymptotic Costs} \label{sec::RC-PoR-P-Assymptotic-Cost}

  \input{table-RCPoRP-cost}

 Table \ref{table::RCPoRP-Asymptotic-Cost} summarizes the RC-PoR-P's asymptotic costs of $z$ verifications, breakdown by parties. In the table, $\phi$ is the number of challenged blocks, $z'$ is the maximum number of complaints the client and server send to the arbiter,  $m$ is the number of  blocks in a file, $||{u^{\scriptscriptstyle*}}||$  is the file bit-size, and  $||\bm{\pi}^{\scriptscriptstyle*}||$  is the number of elements in the padded encrypted proof vector. 

%% file: table-RCPoRP-cost.tex

 \begin{table}[!ht]
\renewcommand{\arraystretch}{1.4}

\begin{minipage}{\linewidth}
\caption{ \small RC-PoR-P asymptotic complexity, of $z$ verifications, breakdown by parties.} \label{table::RCPoRP-Asymptotic-Cost} 
\scalebox{.78}{
\begin{tabular}{|c|c|c|c|c|c|c|c|c|}

   \hline
   


  \cellcolor[gray]{.9}{\scriptsize Phase} &\cellcolor[gray]{.9}{\scriptsize Party}&\cellcolor[gray]{.9}\scriptsize Computation  Cost&\cellcolor[gray]{.9}\scriptsize Communication Cost

  \\
\hline
\cellcolor[gray]{.9}\scriptsize  Client-side and Server-side Init. & \multirow{2}{*}{\rotatebox[origin=c]{0}{\cellcolor[gray]{.9}\scriptsize }} \scriptsize Client&\scriptsize$O(m)$&\scriptsize$O(||u^{*}||)$\\
     \cline{2-6}  
 \multirow{-2}{*}{\rotatebox[origin=c]{0}{\cellcolor[gray]{.9}\scriptsize  (i.e., outsourcing:  \ref{RCPoRP::Client-side-Initiation} and \ref{RCPoRP::Server-side-Initiation}) }}&\cellcolor[gray]{.9}\scriptsize Server&\scriptsize$O(m)$ &\scriptsize$O(1)$\\
     \cline{2-6}   
      
     \hline 
       
          \hline 
          
 \cellcolor[gray]{.9}  &\multirow{4}{*}{\rotatebox[origin=c]{0}{\scriptsize }}\cellcolor[gray]{.9}\scriptsize Client&\scriptsize$O(z \phi  \log_{\scriptscriptstyle 2}(m))$&\scriptsize$ O(z \log_{\scriptscriptstyle 2}(||u^{\scriptscriptstyle *}||))$\\
     \cline{2-6}
     
      \cellcolor[gray]{.9}  &\multirow{4}{*}{\rotatebox[origin=c]{0}{\scriptsize }}\cellcolor[gray]{.9}\scriptsize Server&\scriptsize $O(z \phi  \log_{\scriptscriptstyle 2}(m))$&\scriptsize$O(z  ||\bm{\pi}^{\scriptscriptstyle *}_{\scriptscriptstyle j}||)$\\
     \cline{2-6}
           \cellcolor[gray]{.9}  &\multirow{4}{*}{\rotatebox[origin=c]{0}{\scriptsize }}\cellcolor[gray]{.9}\scriptsize Arbiter&\scriptsize $O(z'  \log_{\scriptscriptstyle2}(m))$&\scriptsize$ O(1)$\\
     \cline{2-6}

    \cline{2-6}
\cellcolor[gray]{.9}\multirow{-4}{*}{\rotatebox[origin=c]{0}{\scriptsize The rest of phases (i.e., \ref{Billing-cycles-genQuery}- \ref{RCPoRP::CoinTransfer}) }}&\cellcolor[gray]{.9}\scriptsize Smart Contract&\scriptsize$O(1)$&-\\ 
 \hline
\end{tabular}  
}
\end{minipage}
\end{table}